\documentclass[fleqn,usenatbib]{mnras}

\usepackage{newtxtext,newtxmath}

\usepackage[T1]{fontenc}

\DeclareRobustCommand{\VAN}[3]{#2}
\let\VANthebibliography\thebibliography
\def\thebibliography{\DeclareRobustCommand{\VAN}[3]{##3}\VANthebibliography}

\usepackage{graphicx}   
\usepackage{epsfig}
\usepackage{epstopdf}
\usepackage{amsmath}	
\usepackage{array}

\newcommand{\hal}{H\ensuremath{\alpha}}

\newcommand{\hst}{\emph{HST}}
\newcommand{\msun}{\ensuremath{{M}_{\odot}}}

\newcommand{\kms}{km~s\ensuremath{^{-1}}}

\newcommand{\sersic}{S\'{e}rsic}
\newcommand{\sersicone}{S\'{e}rsic{1}}
\newcommand{\sersictwo}{S\'{e}rsic{2}}

\newcommand{\reff}{\ensuremath{r_\mathrm{e}} (kpc)}

\newcommand{\axisratio}{\ensuremath{b/a}}

\newcommand\Hunit{\ifmmode {\rm~km\ s}^{-1}\ {\rm Mpc}^{-1}
        \else ~km s$^{-1}$ Mpc$^{-1}$\fi}
\newcommand\ctssec{\ifmmode {\rm~count\ s}^{-1} \else ~count s$^{-1}$\fi}
\newcommand\ergsec{\ifmmode {\rm~erg\ s}^{-1} \else
        ~erg s$^{-1}$\fi}
\newcommand\funit{\ifmmode {\rm~erg\ s}^{-1}\;{\rm cm}^{-2} \else
        ~ergs s$^{-1}$ cm$^{-2}$\fi}
\newcommand\phflux{\ifmmode {\rm~photon\ s}^{-1}\;{\rm cm}^{-2}
        \else   ~photon s$^{-1}$ cm$^{-2}$\fi}
\newcommand\efluxA{\ifmmode {\rm~erg\ s}^{-1}\;{\rm cm}^{-2}\;{\rm
        \AA}^{-1} \else ~erg s$^{-1}$ cm$^{-2}$ \AA$^{-1}$\fi}
\newcommand\efluxHz{\ifmmode {\rm~erg\ s}^{-1}\;{\rm cm}^{-2}\;{\rm
        Hz}^{-1} \else ~erg s$^{-1}$ cm$^{-2}$ Hz$^{-1}$\fi}
\newcommand\cc{\ifmmode {\rm~cm}^{-3} \else cm$^{-3}$\fi}
\newcommand\FWHM{\ifmmode {\rm~FWHM} \else ${\rm~FWHM}$\fi}
\newcommand\Zsun{\ifmmode Z_{\odot} \else $M_{\odot}$\fi}
\newcommand\Lsun{\ifmmode L_{\odot} \else $L_{\odot}$\fi}
\newcommand\ltsim{\raisebox{-.5ex}{$\;\stackrel{<}{\sim}\;$}}
\newcommand\gtsim{\raisebox{-.5ex}{$\;\stackrel{>}{\sim}\;$}}

\newcommand\Kalpha{\ifmmode {\rm K}\alpha \else K$\alpha$\fi}
\newcommand\nh{\ifmmode N_{\rm H} \else N$_{\rm H}$\fi}

\newcommand{\Msun}{\ensuremath{\rm M_{\odot}}}

\title[HST study of spiral galaxy J2345-0449]{Unveiling the Bulge-Disc Structure, AGN Feedback, and Baryon Landscape in a Massive Spiral Galaxy with Mpc-Scale Radio Jets}

\author[Bagchi et al.]{
Joydeep Bagchi,$^{1}$\thanks{E-mail:joydeep.bagchi@christuniversity.in }
Shankar Ray,$^{1}$
Suraj Dhiwar,$^{2,3}$
Pratik Dabhade,$^{4,5}$
Aaron Barth,$^{6}$
\newauthor
Luis C. Ho,$^{7,8}$
Mohammad S. Mirakhor,$^{9}$
Stephen A. Walker,$^{9}$
Nicole Nesvadba,$^{10}$
Francoise Combes$^{11}$
\newauthor
Andrew Fabian,$^{12}$
Joe Jacob,$^{13}$
\\
$^{1}$Department of Physics and Electronics, CHRIST (Deemed to be University), Hosur Road, Bengaluru 560029, India\\
$^{2}$Inter-University Centre for Astronomy and Astrophysics, Pune 411007, India\\
$^{3}$Savitribai Phule Pune University, Pune 411007, India\\
$^{4}$Instituto de Astrof\' isica de Canarias, Calle V\' ia L\'actea, s/n, E-38205, La Laguna, Tenerife, Spain\\
$^{5}$Universidad de La Laguna (ULL), Departamento de Astrofisica,E-38206, Tenerife, Spain\\
$^{6}$Department of Physics and Astronomy, 4129 Frederick Reines Hall, University of California, Irvine, CA, 92697-4575, USA\\
$^{7}$Kavli Institute for Astronomy and Astrophysics, Peking University, Beijing 100871, China\\
$^{8}$Department of Astronomy, School of Physics, Peking University, Beijing 100871, China\\
$^{9}$Department of Physics and Astronomy, The University of Alabama in Huntsville, 301 Sparkman Drive, Huntsville, AL 35899, USA \\
$^{10}$Institut d’Astrophysique Spatiale, Centre Universitaire d’Orsay, F-91405 Orsay, France\\
$^{11}$LERMA, Observatoire de Paris, PSL Research University, College de France, CNRS, Sorbonne University, Paris, France\\
$^{12}$Institute of Astronomy, Cambridge University, Madingly Road, Cambridge CB3 0HA, UK\\
$^{13}$ Newman College, Thodupuzha 685585, Idukki District,
Kerala, India\\
}
\date{Accepted XXX. Received YYY; in original form ZZZ}

\pubyear{2023}

\begin{document}
\label{firstpage}
\pagerange{\pageref{firstpage}--\pageref{lastpage}}
\maketitle

\begin{abstract}

This study delves into the bulge-disc components and stellar mass distribution in the fast-rotating, highly massive spiral galaxy 2MASX~J23453268-0449256, distinguished by extraordinary radio jets extending to Mpc scales. Using high-resolution multi-wavelength Hubble Space Telescope (HST) observations and multi-parameter panchromatic spectral energy distribution (SED) fitting, we derive estimates for the star formation rate, total baryonic mass in stars, and warm dust properties. Our findings, validated at a spatial resolution of approximately 100 pc, reveal a pseudo-bulge rather than a classical bulge and a small nuclear bar and resonant ring, challenging conventional models of galaxy formation. Additionally, the lack of tidal debris and the  highly symmetric spiral arms within a rotationally supported stellar disc indicate a tranquil coevolution of the galactic disc and its supermassive black hole (SMBH). Significantly, the galaxy exhibits suppressed star formation in its center, potentially influenced by feedback from the central accreting SMBH with powerful radio jets. Detailed multi-wavelength studies of potential star-forming gas disclose that, while hot X-ray gas cools down in the galaxy's halo, new stars do not form in the center, likely due to feedback effects. This study raises questions about the efficient fueling and sustained collimated jet ejection activity in J2345-0449, underscoring the imperative need for a comprehensive understanding of its central black hole engine properties, which are presently lacking. The exceptional rarity of galaxies like 2MASX~J23453268-0449256 presents intriguing challenges in unraveling the physical processes responsible for their unique characteristics.

\end{abstract}




\section{Introduction}

Explaining the mechanism of galaxy formation in the Universe across cosmic eons in the framework of widely accepted   {\it Concordance Lambda Cold Dark Matter} ($\rm \Lambda$CDM) Cosmology,  is one of the  prime goals of  astronomy research.  Since  there  may be  several  competing  physical processes responsible for  the  observed  galaxy properties, identifying   and  relating them to   galaxy formation  models is an exceptionally important goal.  A  major mystery is how  galaxies convert their  baryons into stars and  why  do  some  galaxies are still  busy  forming  new  stars  while  others  have  virtually  shut down their  stellar  factory.  Observations show that the  Star Formation  Rate (hereafter SFR)  reached  it's  peak at `cosmic high noon' near  red shifts $\it{z} \rm = 1 - 2$  when galaxies were rapidly building up their  stellar masses by converting  cosmic baryons to stars and, subsequently,  over the next  few billions of years the 
star formation  went down drastically.  Eventually, by the  present  era,  most of the massive galaxies   have  stopped growing \citep[][]{2004MNRAS.351.1151B, 2015ApJ...801L..29R},  having entered  a  life  of apparent quiescence \citep[][]{2010ApJ...717..379B,2019MNRAS.488.3143B,2023MNRAS.518.4943D,2013MNRAS.428.3121M}.  Nevertheless,  for reasons still  unknown,  in  our  local universe, some  low-mass  galaxies  are still actively forming stars,  while the majority of  massive  galaxies with halo masses  $M_{halo} \gtsim 10^{12} \Msun$  display a  strikingly  low specific star formation  rate  (where  sSFR = SFR/stellar mass) compared to less massive galaxies (see reviews by \citet{2012NewAR..56...93A} and \citet[][]{2018ARA&A..56..435W}). 
 
Why this is so and  who  were the  progenitors of these massive local galaxies are still  open questions. It is likely that their formation history may go as  far back as  $ z \sim 6 - 10$ in  redshift  (about $\sim 0.9  - 0.5$ Gy after birth of Universe), in the era of re-ionization,  when their  ancestors  started  to grow in mass, as well as vigorously formed new stars from  cosmic baryons, infalling  and condensing around  their primordial dark matter halos. 
Puzzlingly, JWST has uncovered some extremely massive  (stellar mass $\rm \gtsim 10^{10} M_{\odot}$) galaxies existing even at very early epochs ($z \sim 7 - 9$; \citet[][]{2023Natur.616..266L}).
By the time when the Universe was about  three  billion years old, half of the most massive galaxies were in place as compact  proto-galaxies,  having already exhausted  their  fuel for star formation.  Of the rest half, about $25$ per cent  galaxies assembled before the peak of the cosmic star-formation,  while  $25$ per cent galaxies formed later \citep[][]{2014ARA&A..52..415M}. Observations reveal  that  at higher red shifts ($z \sim 4$) most of them  were  harbouring  intense nuclear starbursts and possibly  they ultimately grew into the most  massive local  galaxies seen today  through mergers with other galaxies \citep{2021JApA...42...59S}, and  by means of gravitational accretion  from  cold gas filaments of the cosmic web \citep[][]{2019Natur.572..211W}. 
Understanding how this remarkable transformation  happened  and what physical factors were responsible to bring about this change has been an enduring quest -  which  also brings  forth   some of   the most  widely  debated  key  questions in galaxy formation   (for reviews,  see \citet[][]{2020NatRP...2...42V} and \citet[][]{2015ARA&A..53...51S}).

In the above scenarios,  the Circum-Galactic Medium (hereafter CGM) of galaxies plays a  decisive  role in determining, at all cosmological epochs,  how galaxies acquire, eject, and  recycle their gas; which in turn  are the most important physical ingredients  in  determining  how the galaxies should evolve,  how much star formation will take place,  and eventually how much  of the baryonic gas  will  be  quenched  (prevented from forming stars),  or  recycled back to the galaxy.  Along  with the unseen, gravitationally dominant dark matter, the CGM is an extremely dynamic environment  reposing  a complex mix of  baryons existing  in  diverse  phases of  temperature, density and  metal content,  presenting an  extremely important, but  perhaps the  least understood  aspect of  galaxy formation (see the reviews \citet[][]{2017ARA&A..55..389T}, \citet[][]{2022PhR...973....1D} ). Among the many interesting and challenging questions that the  CGM  poses,  one of the most perplexing ones has been the so-called  the `missing baryons'  problem.  Indeed the term `missing baryons' is a misnomer, but it emphasises the  fact that even in  deep multi-wavelength observations of  galaxies, the total  detectable baryonic mass  falls far short (by $\sim$50 to 80 per cent) of the  cosmic baryon fraction  of $\rm \approx 0.15 - 0.17$ as expected in $\Lambda$CDM cosmology based on Big Bang nucleo-synthesis and Cosmic Microwave Background constraints \citep[][]{2009ApJS..180..330K,2009ApJS..180..306D}. A number of  theories and  computer simulations have tried to explain where do these baryons may reside and  why only a small fraction of the  baryons in galactic halos condense into stars, or  why by  present era ($\it{z} \rm = 0$)  the star formation efficiency (empirically defined as 
${f_{\star} = \frac{M_{\star}}{M_{halo}}~\frac{\Omega_{m}}{\Omega_{b}}}$)
 reaches a peak  of  about $\rm  20 - 30$ per cent for  typical $L_{\star}$ galaxies. Whereas for very massive galaxies (halo mass $\gtsim M_{h} \sim 10^{12} \rm M_{\sun}$), beyond the  peak, the $f_{\star}$  rapidly  declines to  only  $f_{\star} \sim 2 - 5$ per cent at $M_{h} \sim 10^{13} \rm M_{\sun}$ \citep[][]{2015MNRAS.446..521S,2010ApJ...717..379B,2019MNRAS.488.3143B,2013MNRAS.428.3121M}. It is likely  that  AGN (Active Galactic Nuclei) feedback  could be the  main agency responsible for  this steep decline in SFR of highly massive galaxies, but we need to understand if that is the only factor \citep[][]{2022A&A...659A.160B}.  On the other hand, a similar decline in SFR is seen on the opposite, lower mass end of the peak ($\ltsim M_{h} \sim 10^{12} \rm M_{\sun}$), which is mainly  attributed to supernova feedback, as it is more effective in less massive galaxies.

The promising agents of AGN feedback  are the  mass accreting  supermassive black holes (SMBH)  growing at the centre of galaxies \citep[][]{1998AJ....115.2285M} which  occasionally  turn into powerful engines  \citep[][]{1969Natur.223..690L}  that can potentially heat and even expel the  gas of their host galaxies, 
thereby  affecting  the  star formation process \citep{2008MNRAS.391..481S,2005Natur.433..604D,2017MNRAS.472..949B,2012NJPh...14e5023M}. It is now widely recognized that  in our local Universe, such  SMBHs of masses $\rm \sim 10^{6} - 10^{10} \Msun$ lurk in the nuclei of almost all massive galaxies (see  \citet[]{2013ARA&A..51..511K} for a review).  Moreover, AGN  feedback of  accreting BHs is likely to be more effective  in   massive galaxies because a remarkably  tight  coupling  exists  between the  mass of the black hole and the galaxy's  bulge  mass  or  with its'   total  stellar mass/luminosity, which  further suggests that; (a) mass of the central SMBH will be higher in  massive   galaxies  and  in  bulge dominated galaxies,  and  so  will be  its energetic  output and its  impact  on the surrounding media, (b) the star formation  rate  and the final mass of the  galaxy   somehow  has  been  symbiotically regulated  by the growing  black hole  over the cosmic eons, e.g., \citep[][]{1998AJ....115.2285M,2000ApJ...539L..13G,2004ApJ...604L..89H,2003ApJ...589L..21M,2009ApJ...698..198G,2019ApJ...876..155S}.  

The physical mechanism that links  the SMBH mass to the galactic bulge and disc properties is  still poorly understood, but AGN energy  driven feedback - either  via the mechanical power of the  narrow  radio jets \citep[][]{1984ARA&A..22..319B,2022JApA...43...97S}  or  through a broad  pressure-driven hot wind/outflow (and radiation flux) on the surrounding medium -  plays a major role in regulating the  growth  of the black hole and   may  affect the star formation rate and  growth of the host galaxy. In recent years we have learned    that  SMBHs are  the essential  ingredients  shaping the lives of  galaxies across  cosmic time  (excellently reviewed in  \citet[]{2013ARA&A..51..511K} and \citet[][]{2012NewAR..56...93A}).  Nevertheless,  there is  little understanding of exactly how the  black holes of AGN  are fuelled and how the radio jets or the hot outflows   deposit   energy to their surroundings; the energy that is  transferred to the different gas phases, and  what are the physical conditions  when  significant positive or negative feedback  on star formation rate  is produced (see reviews by \citet[][]{2012ARA&A..50..455F} and \citet{2015ARA&A..53..115K}).  Despite numerous efforts to observe  SMBH engines  in the  process of quenching star formation, conclusive evidence for such a process remains  elusive, particularly in the nearby Universe (\citet[][]{2018Natur.553..307M}).  However, the  strong impact of AGN feedback via   radio jets  on the hot intracluster medium (ICM)  is  observed directly through  X-ray observations of the  dark cavities and  non-thermal plasma  bubbles  found  around the central galaxies of  cool-core galaxy clusters  and  groups in X-ray images \citep[][]{2004ApJ...607..800B}.


Lastly, leading galaxy formation theories, e.g. \citep[][]{1978MNRAS.183..341W,1991ApJ...379...52W,2006ApJ...639..590F,2006ApJ...644L...1S} postulate a warm-hot  X-ray halo (or a `corona' first
mentioned by \citet[][]{1956ApJ...124...20S}) containing a large fraction  (at least $50$ per cent) of  cosmic baryons in highly massive  galaxies.  
To validate galaxy formation models and make a census of  all  cosmic baryons,  sensitive  X-ray or UV observations of circum-galactic halos are  necessary.
It has been proved very challenging to detect the  hot, baryon-filled  coronae  around  even in very luminous spirals and  ellipticals with our present telescopes due to sensitivity limitations.   Nevertheless, in past decade  using the  Chandra and XMM-Newton telescope, such coronae have  been robustly detected in  a small number  of   massive  galaxies, e.g., \citep[][]{2012ApJ...755..107D,2021MNRAS.500.2503M,
2015MNRAS.449.3527W,2011ApJ...737...22A,2016MNRAS.455..227A,2018ApJ...862....3B,2013ApJ...772...97B}, but  independent constraints on their temperature, density, and 
metallicity  from soft X-ray or UV spectroscopy is still lacking.  These  galaxies have provided compelling evidence for the existence of hot, low-metallicity atmospheres of gas that  could originate with accretion from the  IGM  and subsequent  heating to the virial temperature of the halo via accretion shocks. 
However,  the  fraction of baryons residing in the hot coronal phase and its dependence on stellar and  halo mass and 
how they are affected by the   AGN feedback  of the central black hole and the galactic star formation processes  are not  well determined  and  poorly  understood  (\citet[][]{2022PhR...973....1D}). 

 Evidently, to further the field  it  is necessary to  understand  how   SMBHs  have grown  over cosmic  time,  how do they  accrete   gas (both the rate of accretion, as well as the source of this gas) to  create their  relativistic jets and hot winds, and finally, their  impact on  the  surrounding  molecular gas and  the   star formation activity.  The  baryon  budget  and  star  formation  rate  are  the  two  vital  properties of galaxies that are significantly  affected  by  AGN    feedback  which needs much better understanding.  The quenching  of star formation in  massive galaxies is  one of the core predictions of the AGN feedback hypothesis, yet it is also one of its aspects that are very little understood (see \citep[][]{2017NatAs...1E.165H}). Some previous studies focus on gas removal in the host galaxies of very powerful AGN as  the main feedback mode 
\citep[][]{2021NatAs...5...13L}, however, in contradiction some observations show that  quasar-like outflows alone are not sufficient to understand AGN feedback: Neither do even the  most extreme outflows in very  powerful AGN result in a significant
reduction in star formation rate (e.g., \citep[][]{2015MNRAS.453..591S}), nor do many of the most frequent, medium to low-power 
AGN show strong outflow signatures. Thus, in spite of the  popularity of AGN feedback hypothesis, its central prediction – the direct  physical link between the energy injection by the AGN  and  star formation  remains a subject of several contradictions yet it provides  many new, vibrant ideas for explorations  \citep[][]{2017FrASS...4...42M}.  
 
Our prime focus is on a detailed exploration of active galaxies with massive black holes and jets, specifically examining their feedback impact on the interstellar medium and the extended baryonic matter of the host galaxy. We are particularly intrigued by the prospect of identifying clear counter examples—instances of powerful jets originating from an accreting supermassive black hole (SMBH) formed under highly unusual conditions. One such remarkable case under scrutiny here involves a Mpc-scale radio galaxy hosted by a luminous, quiescent massive spiral disc without a classical bulge, exhibiting no major signs of disturbances. Investigating such scenarios allows for a direct test of AGN feedback. Moreover, within this galaxy, we aim to uncover the precise physical factors influencing the launch of relativistic jets in powerful radio galaxies \citep[][]{1984ARA&A..22..319B}, marking the first instance of such knowledge. This unique and rare galactic configuration serves as an exceptional test-bed, offering valuable insights into the role of relativistic jets in the evolution of galaxies.

 
 Hereafter, in this paper we use redshift  $z = 0.07557$ and  $\Lambda$CDM   cosmological parameters;  $\rm H_{0} = 67.8$ km/sec/Mpc, 
$\Omega_{\rm m} = 0.308 $, and $\Omega_{\rm vac}= 0.692$. This results in an image scale of 1.48 kpc per arcsec. The luminosity
distance to J2345-0449  in this model  is 353.0 Mpc and the angular size distance is 305.2 Mpc.  Magnitudes are quoted in the AB system and  mass and luminosity are given in terms of solar mass (where $\rm M_{\odot}= 1.99 \times 10^{33}$ g) and solar luminosity (where $\rm L_{\odot} = 3.84 \times 10^{33}$ erg s$^{-1}$) where possible.

\begin{figure*}
\centering
\includegraphics[scale=0.55]{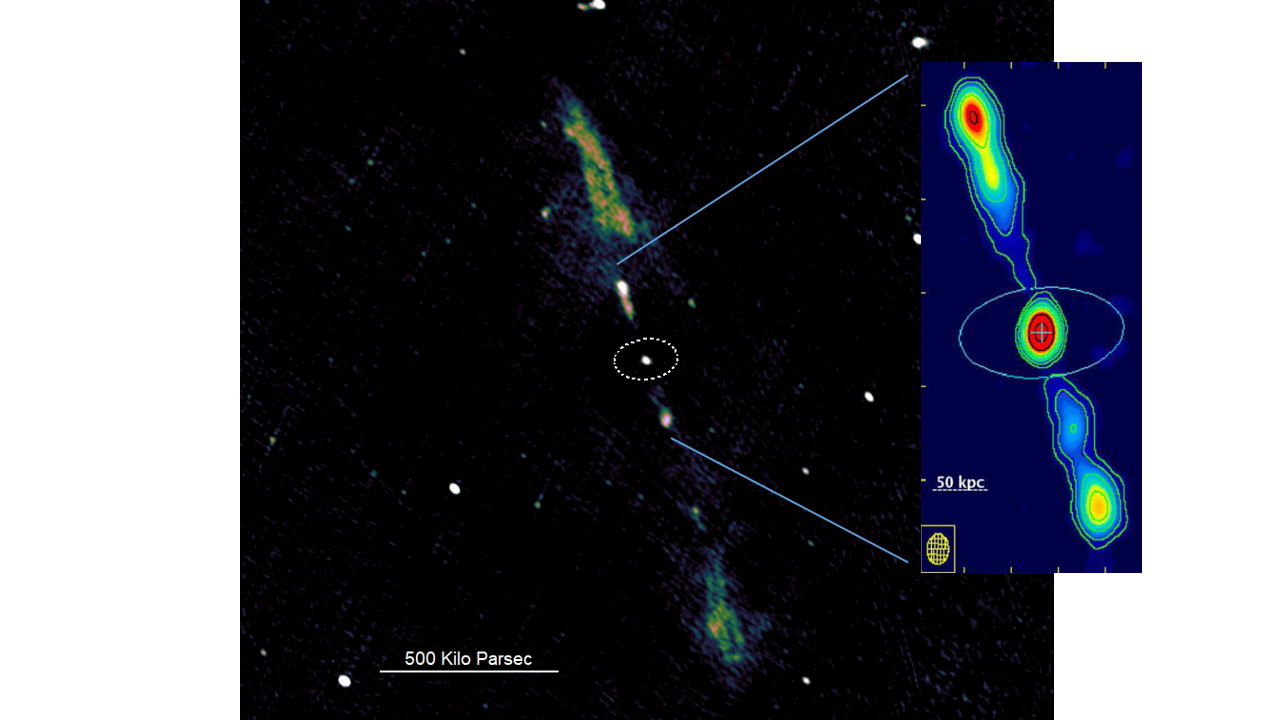}
\caption{GMRT 323 MHz  image of the
giant radio source hosted by the  spiral  J2345-0449.  The r.m.s. of fluctuations  at   noise floor is about $96 \, \mu$Jy/beam
and  the $3 \sigma$ surface-brightness detection threshold is about $300 \, \mu$Jy/beam. The major and minor axis of the synthesised elliptical beam are $10\farcs35 \times 7\farcs23$  at P.A. $45.24$
degree. The AGN core is detected at $\rm 5.5 \, m$Jy/beam brightness. Noteworthy is  the  rare 
occurrence of two nested pairs of  radio lobes. The extremely large extent of the  emission is evident with inner and outer  radio lobe  pairs extending  over $\sim387$ kpc ($\sim 4\farcm52$) and $\sim1.6$ Mpc ($\sim 19\farcm1$), respectively. 
The white dotted ellipse represents   an  outline of the optical galaxy zoomed by a factor of $\sim4$ for clarity. The scale bar represents 500 kpc. The inset  image shows  zoomed-in details of the inner radio-double
observed with VLA at 4.8 GHz and $ 19\farcs8 \times 13\farcs3$  restoring beam  at P.A. $178.5$ degree (lower left corner). Contour levels are (-0.1, 0.1, 0.2, 0.4, 0.8, 1.6, and 3.2 mJy beam$^{-1}$). The scale bar represents 50 kpc. The blue ellipse
represents zoomed outline of the optical galaxy. 
The Inner double  has an  edge-brightened FR-II  \citep[][]{1974MNRAS.167P..31F} morphology, being  fed by jets, but the outer lobes look peculiar; long, filamentous, and diffuse, 
possibly decaying relics of  jet activity that  had ceased millions of years ago.}
\label{GMRT}
\end{figure*}
\bigskip 
\begin{figure*}
\centering
\includegraphics[scale=0.35,angle=180]{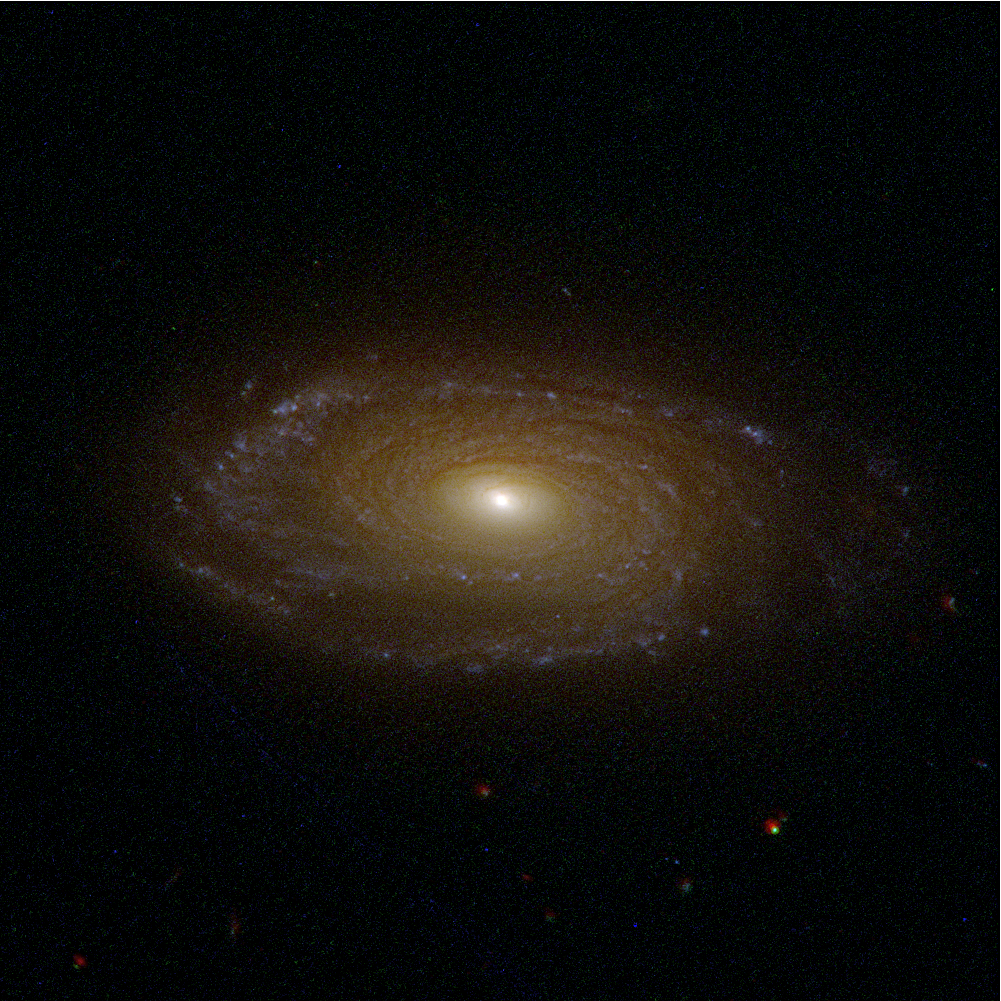}
\caption{A   tricolour image  of J2345-0449, constructed using the {\em Hubble Space Telescope}  UVIS  (F438W B-band and F814W I-band) and IR channel (F160W H-band) filter images. These images were taken with the   Wide Field Camera 3 (WFC3) instrument. The component images have been
superposed with individual intensities adjusted to obtain a natural colour balance. One 
can see   dark, winding dust lanes and  compact, blue star forming regions at the outer parts of the
galaxy. North is up and East is to the left.
}
\label{Fig1}
\end{figure*}


\noindent 

\section{2MASX J23453268-0449256: an  enigmatic spiral galaxy }

\bigskip

 The focus of present paper,  galaxy 2MASX J23453268-0449256 (hereafter  J2345-0449) at redshift $0.0755$. J2345-0449 is known to be an extremely luminous (optical r band: $M(r)$ =-23.26, near-IR K band: $M(K)$ = -26.15), isolated spiral galaxy  which  also hosts  one of the largest  and  most  powerful  radio jet  known  till now,  clearly powered   by  the   AGN  residing at the center of  a spiral galaxy disc  and   not   an  elliptical galaxy \citep[][]{2014ApJ...788..174B}. GMRT and VLA observations show that 
   this  giant radio  source  also  exhibits  a rarely  observed  `double-double'   radio  morphology;  viz. containing  a nested   pair of  smaller (active/young), inner  radio lobes,    and a second pair of   extended (relic/old),  diffuse  outer radio lobes,   with the  jet  axis (in projection) nearly   orthogonal to the host galaxy's  stellar  disc (Fig.~\ref{GMRT}).  The bipolar radio  emission  of  outer lobes can be  traced upto $\sim 1.63$ Mpc ($\sim 19\farcm1$) and inner lobes upto  $\sim387.2$ kpc ($\sim 4\farcm52$),  making it  an example of  class of  rarely seen  extreme sized  active galaxies, called the Giant Radio Galaxies  (GRG)  whose radio jets may  extend upto   $\sim 1$ Mpc or more \citep{2020A&A...642A.153D,2020A&A...635A...5D}, far beyond the  optical host. The  inner ``active” radio lobes are of edge-brightened, Fanaroff–Riley class II  morphology \citep[][]{1974MNRAS.167P..31F}, energized by the collimated jets shot out from the  central nucleus. The  outer lobes have peculiar long, filamentous, and diffuse morphology, possibly decaying relics of past  jet activity (Fig.~\ref{GMRT}). 

These bright and enormous set of radio jets and lobes  mark  J2345-0449  as a unique source even among  very  massive spirals and the rapidly growing examples of  GRGs in the local Universe, which  themselves  pose many important astrophysical questions \citep{2023JApA...44...13D}.
Undoubtedly,  this is a  very  unusual   occurrence,  since  for reasons still unknown such large scale, powerful radio jets (typically on 100 kpc - 1000 kpc scale)  are   exclusively  launched  from  bulge  dominated  elliptical  galaxies  hosting supermassive  black holes (e.g. \citet[][]{1964ApJ...140...35M,2004MNRAS.351..347M}) whereas the radio jet host of J2345-0449 is a spiral galaxy (Fig.~\ref{Fig1}). Consequently,  such  counter-intuitive   AGNs   challenge the standard paradigm of  growth and co-evolution of  massive black holes in   bulges of galaxies  \citep[][]{1998AJ....115.2285M,2000ApJ...539L..13G,2004ApJ...604L..89H,2003ApJ...589L..21M,2022ApJ...941...95W} and therefore,  they  are also of  profound interest as they may be the ideal  astrophysical models  of the central engines of  radio  galaxies and their feedback signatures on stellar and gaseous matter of  the galaxy.  


Finding  Mpc scale jets  in  a  spiral  galaxy is highly unusual in itself, but the  galaxy itself  is  equally remarkable. 
Its general  properties were discovered   by \citet[][]{2014ApJ...788..174B}  and further  investigated  by  \citet[][]{2015MNRAS.449.3527W}; \citet[][]{2021MNRAS.500.2503M} and \citet[][]{2021A&A...654A...8N}. The  WISE mid-IR colors suggest  that  some  mild  star formation is still occurring  within its  disc.  It is located in a sparsely populated galactic environment devoid of bright $L_{\star}$ galaxies. Having   $M_{r} = -23.26$ mag in SDSS r-band and  $M_{12} = -28.66$ in WISE 12\,$\micron$ band, it lies at the very top of the 
luminosity function of   red  spirals.   The galaxy’s  systematic  rotation speed  is very  high,  with maximum  flat rotation speed touching   $v_{rot} = 429\pm30$ km s$^{-1}$ at $r = 15$ kpc radius (corrected for inclination angle $i \approx 59$ deg.),  which  is  exceptionally large among  all  spirals.  Assuming Newtonian dynamics, this rotation speed implies a huge  mass  of  $\rm \sim 10^{13} M_{\odot}$  within the virial radius of $R_{200} = 450$ kpc,  which mostly is comprised of dark matter, thereby making J2345-0449 one of the  largest and most  massive galaxies  known in our local universe. \citet{2014ApJ...788..174B} found galaxy's  central velocity dispersion  is  $\sigma_{\star} = 351 \pm 25$ km s$^{-1}$,  much higher than normal Milky Way like spirals and close to the highest velocity dispersion found among nearby E and S0 galaxies (van den Bosch et al. 2012).    

As the  mechanism of radio-mode suppression of star formation in  massive galaxies is still  unknown and  there could be several ways to suppress star formation in  early type  galaxies, a powerful jet  emanating in a massive spiral disc offers an important alternate probe of jet-related feedback. J2345-0449 is one such galaxy, with its nested pair of extended radio lobes and a smaller scale restarted radio source that is  interacting with a molecular disc, injecting turbulence and affecting the star formation rate. Moreover, the large repository of  warm/hot baryons discovered  in the  circum-galactic  medium (CGM) of the  galaxy  and  the   cold molecular gas  content in the inner  disc  further highlight its  baryonic  budget and  feedback properties.  \citet[][]{2021A&A...654A...8N} detected  signatures of  AGN feedback on the inner CO gas ring  and  suggested possible  suppression of  star formation  on kpc scale at the center.  From VLT/MUSE spectroscopy \citet[][]{2023A&A...676A..35D} concluded that more than $93$ per cent of the stellar mass is very old having formed $\gtrsim$ 10 Gy ago in the disc and  detected only a few young 
star-forming regions that formed $\ltsim 11$ My ago. \citet[][]{2015MNRAS.449.3527W}   and   \citet[][]{2021MNRAS.500.2503M} using {\em Chandra} and {\em XMM-Newton} observatories  discovered  the  galaxy’s  extended  soft X-ray  (electron temperature $T_{e}  \sim 0.6$ keV or $6.9 \times 10^{6}$ K)  corona/halo of low metallicity ($Z \sim 0.1 Z_{\odot}$) gas,  which extends  at least to  r$\approx 160$  kpc galacto-centric radius,  making J2345-0449 one  of  just a few  spirals for  which  such  large, extended, massive ionized gaseous halos have been directly detected via X-ray imaging.  Extrapolating the  gas mass profile out to the virial radius ($R_{vir}\sim 450$ kpc) and accounting for the stellar and molecular gas masses of the disc, the baryon mass fraction  was found to be  $0.121 \pm 0.043$, close to the  universal baryon fraction  expected in LCDM cosmology \citep[][]{2009ApJS..180..330K,2015MNRAS.449.3527W,2021MNRAS.500.2503M}.


Ground based  (SDSS and CFHT) photometry  showed that 
J2345-0449 may  harbour a  flattened pseudo-bulge rather than a typical  ellipsoidal classical bulge,  has no  optically bright highly radiative AGN, and lacks  obvious  signs of mergers or interaction with any  massive neighboring galaxy \citep[][]{2014ApJ...788..174B}.  Such  rare, isolated and ultra massive galaxies ($ M_{200} \gtrsim 10^{13} \Msun$; \citealt[][]{2014ApJ...788..174B}, \citealt[][]{2015MNRAS.449.3527W}), located at the  upper end of galactic mass
spectrum are  highly valuable laboratories to
understand a  more general astrophysical mechanism, allowing us to isolate specific processes that are challenging to study in the dominant galaxy population due to the coexistence of various potentially competing mechanisms producing similar effects. Consequently, J2345-0449 becomes a distinctive laboratory for investigating crucial aspects of bulge-disc stars, relativistic jet formation, AGN feedback, and the co-evolution of supermassive black holes and galaxies. These aspects, often difficult to probe in early-type galaxies, come to the forefront in this exceptional astronomical specimen.



 In this paper we perform   a  detailed  multi-wavelength, high resolution  study of J2345-0449  with  Hubble Space Telescope (HST).  From  optical and near IR imaging made  with the  Wide Field Planetary Camera 3 (WFC3) providing a  significant  high spatial 
 resolution of   ($\sim$0\farcs05 to 0\farcs1  FWHM  or
 $\sim 50 - 100$ pc),  we  quantify  the  star  formation  rate and interstellar dust properties in the  disc. Furthermore, we unambiguously  resolve  the stellar and dust content organized in the  bulge and disc  components  and  determine  the bulge-disc  morphological   and  photometric parameters  accurately,   thereby vastly improving  on  the  previous  ground based studies.  We  also  estimate  the star formation  rate  in the disc and explore its implications in synergy with  some of  the   important  results from  our previous  investigations;  of cold molecular gas  content  and  its  dynamics  observed  via  the  Carbon Mono Oxide  (CO)  lines,   and   the    properties of  an  extended   soft  X-ray  halo/corona  detected  in  Chandra and deep  XMM-Newton observations. 
 
 We  present  the detailed,  best fitting spectral energy distribution (SED) model  of J2345-0449  using all available observations across  0.1  to  500\,  $\micron$ wavelength range. We decompose the SED into  star formation, ISM dust and  AGN  components.
 The SED fitting has been done using  two different software packages; CIGALE and MAGPHYS, for cross checking,  obtaining a number of key ISM and  stellar parameters of the galaxy, such as its  global  star formation rate, stellar mass, dust mass, dust temperature and dust luminosity,  gas mass, and the AGN fraction. 

 Finally, we examine the location of J2345-0449 on the  star formation  main-sequence and  other  well established scaling-relations governing  the  galaxies  in   local Universe with an aim to know how  this exceptional galaxy compares with other ordinary galaxies. We also  obtain a detailed census of all baryons in the disc and  the circum-galactic halo and derive  estimates of overall star formation efficiency and baryon fraction. Our  studies  enable  probing  the   environment of  the central   black  hole powering the powerful radio  jets and   investigate the  feedback action of  jets on  the  stellar  disc and  the surrounding circum galactic environment.
 
 These   multi-wavelength  data  set  and the   inferences   constitute  our   first   steps  towards understanding  the  origin  and  evolution of   this complex and enigmatic  galaxy.
  

 \footnotetext[1]{The NASA/IPAC Extragalactic Database (NED) is operated by the Jet Propulsion 
Laboratory, California Institute of Technology, under contract with the National Aeronautics and Space Administration.}

\begin{figure*}
\centering
\includegraphics[scale=0.6]{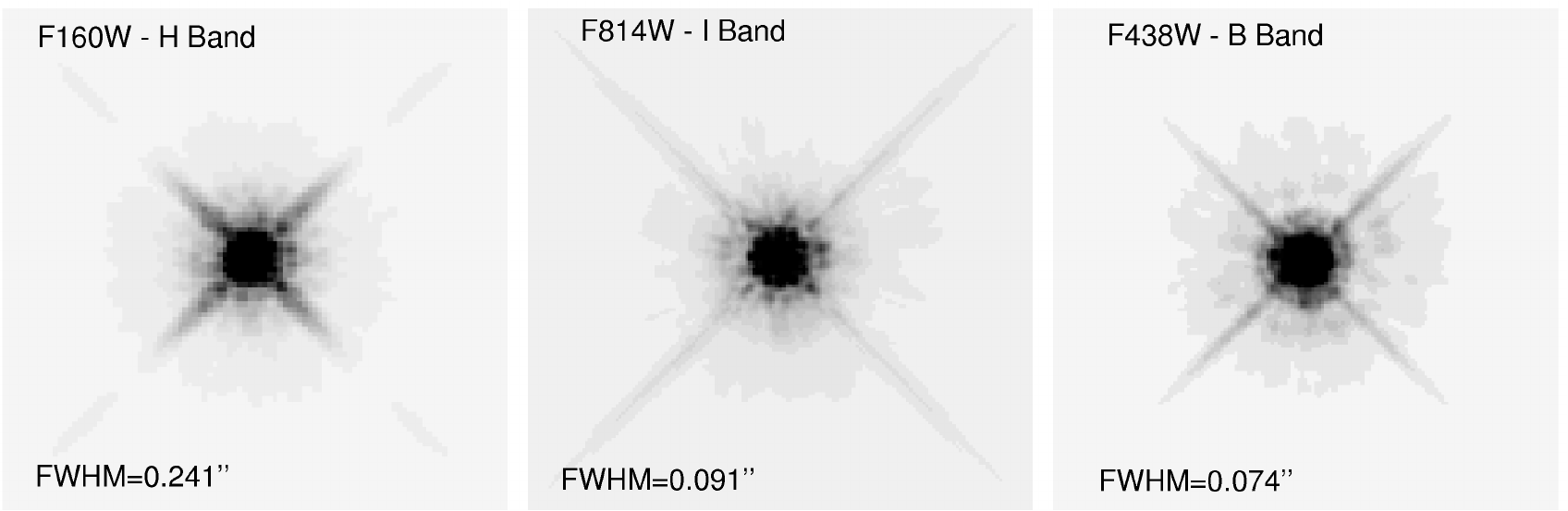}
\caption{Model PSFs in each HST filter used  in GALFIT models. The Full Width at Half Maximum (FWHM) of  PSF  is given in arcseconds at lower left corner
of each image. }
\label{PSF_images}
\end{figure*}

\begin{figure*}
\includegraphics[scale=0.8]{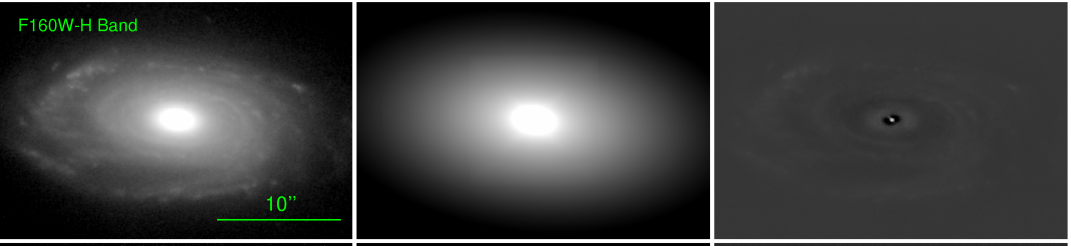}
\includegraphics[scale=0.8]{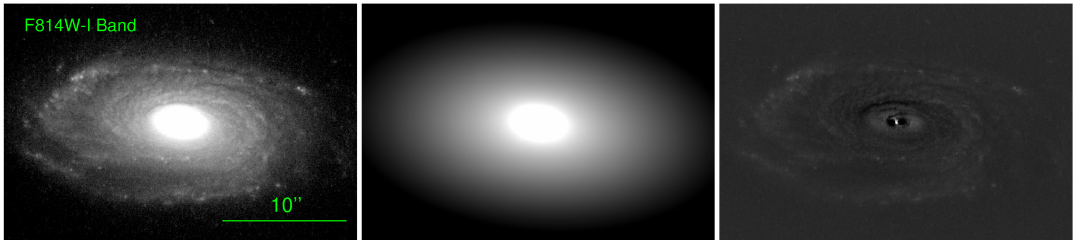}
\includegraphics[scale=0.8]{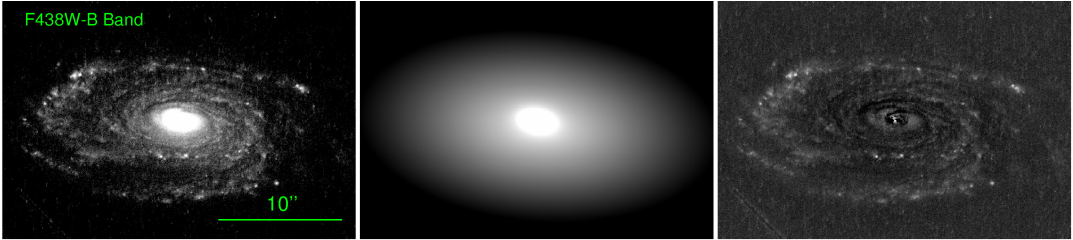}
\caption{ Model A. Grayscale images and GALFIT  modelling  of the inner $25'' \times 25''$ of spiral galaxy J2345 in 
all three \hst/WFC3 filters,  shown with log scale. Top row: H-band, middle row: I-band, bottom row: B-band.  Left image panel: the HST input image  of the galaxy, middle  image panel: the best fitting GALFIT model, right image panel: the residual image showing
the difference between the input and  the best fitting model image. Spiral arms 
were not fitted in the model. 
}
\label{set1}
\end{figure*}
\begin{figure*}
\includegraphics[scale=0.35]{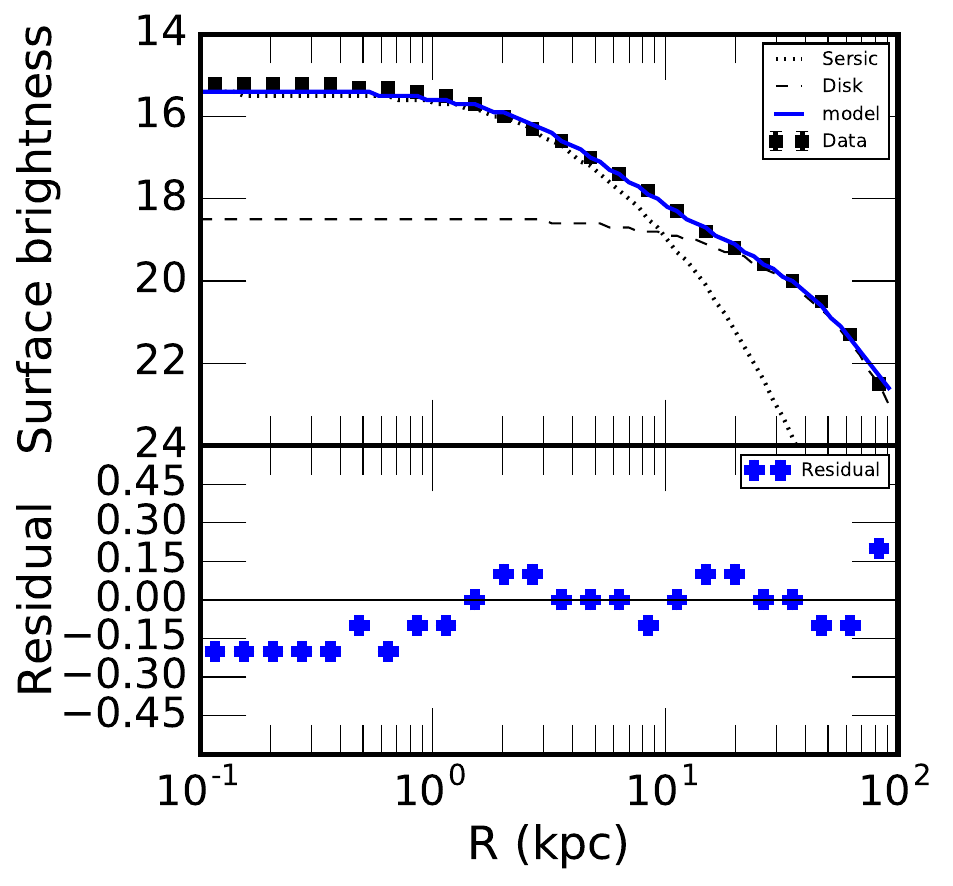}
\includegraphics[scale=0.35]{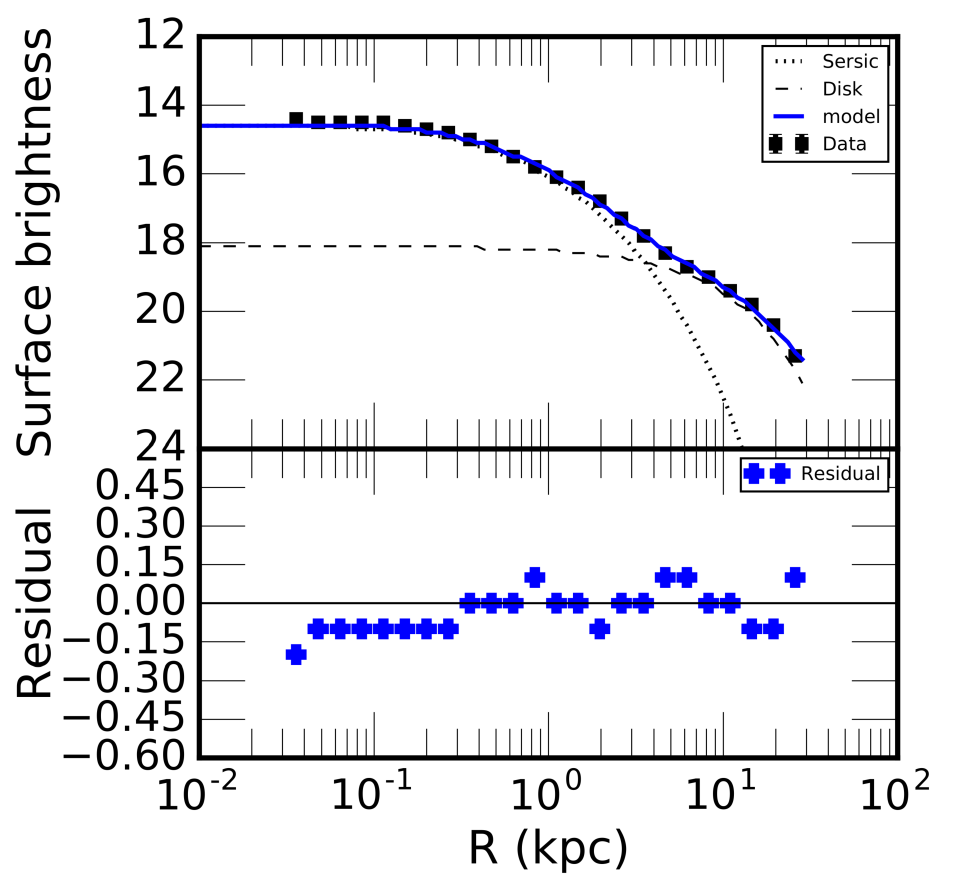}
\includegraphics[scale=0.35]{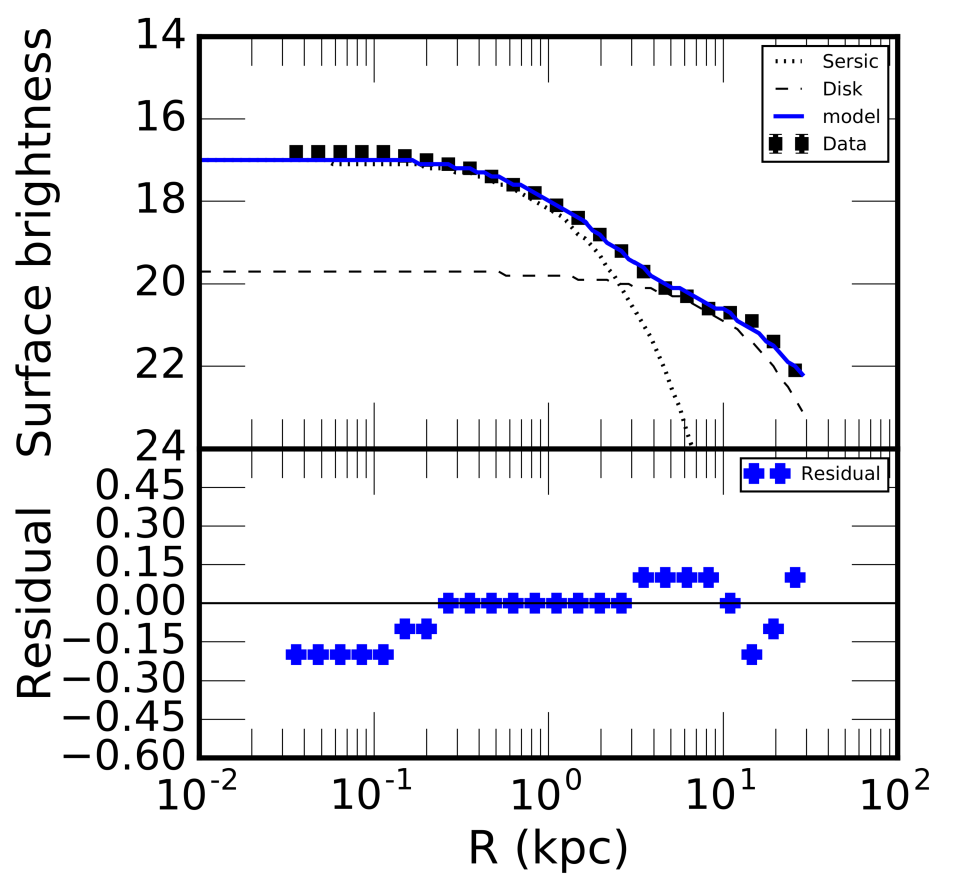}
\caption{Model A. Left to right: HST F160W, F814W and F438W filter  radial profiles of the surface brightness (mag arcsec$^{-2}$)  with \texttt{GALFIT} model (solid blue line) and \hst/WFC3 image data (black squares) as a function of radius (kpc), as well as the subcomponents of the fit
are plotted (see legend for various profiles). The data represent measurements along elliptical annuli of constant
ellipticity and  position angle as obtained from  best fitting parameters.
These radial profiles were obtained using the elliptical isophote fitting routine \texttt{Ellipse}. The parameters of fitting are tabulated
in Table~\ref{all_params}.}
\label{set1_profile}
\end{figure*}


\begin{figure*}
\includegraphics[scale=0.506]{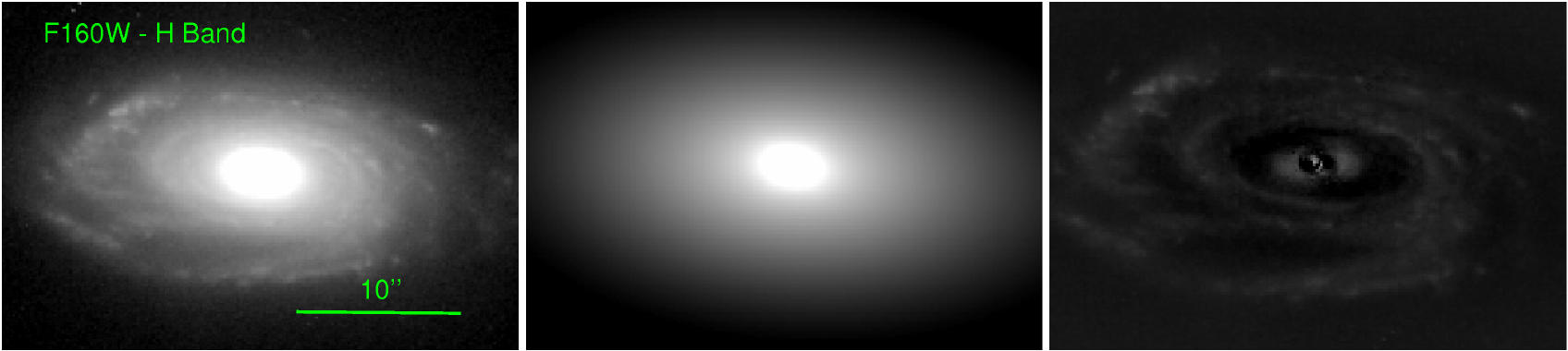}
\includegraphics[scale=0.8]{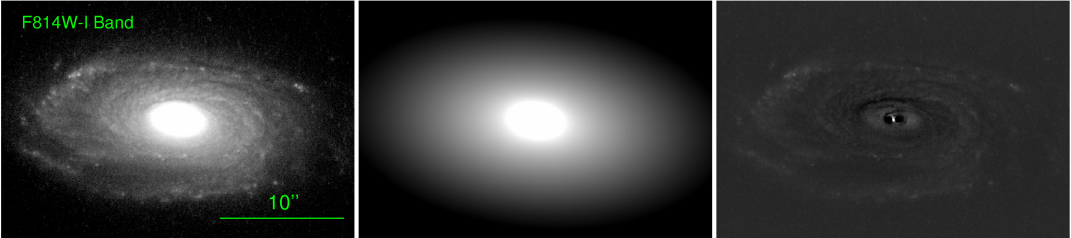}
\includegraphics[scale=0.8]{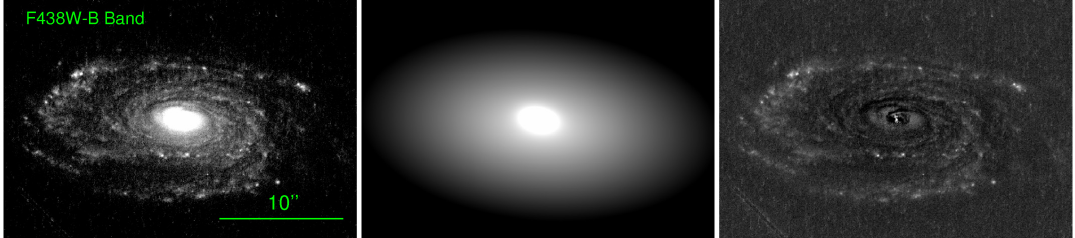}
\caption{Model B. Gray-scale images and GALFIT  modelling  of the inner $25'' \times 25''$ of spiral galaxy J2345 in 
all three \hst/WFC3 filters,  here shown with log scale. Top row: H-band, middle row: I-band, and bottom row: B-band.  Left image panel: the HST input image  of the galaxy, middle  image panel: the best fitting GALFIT model, and right image panel: the residual image showing
the difference between the input and the best fitting model image. Note that spiral arms were not fitted in the model. }
\label{set2}
\end{figure*}
\begin{figure*}
\includegraphics[scale=0.35]{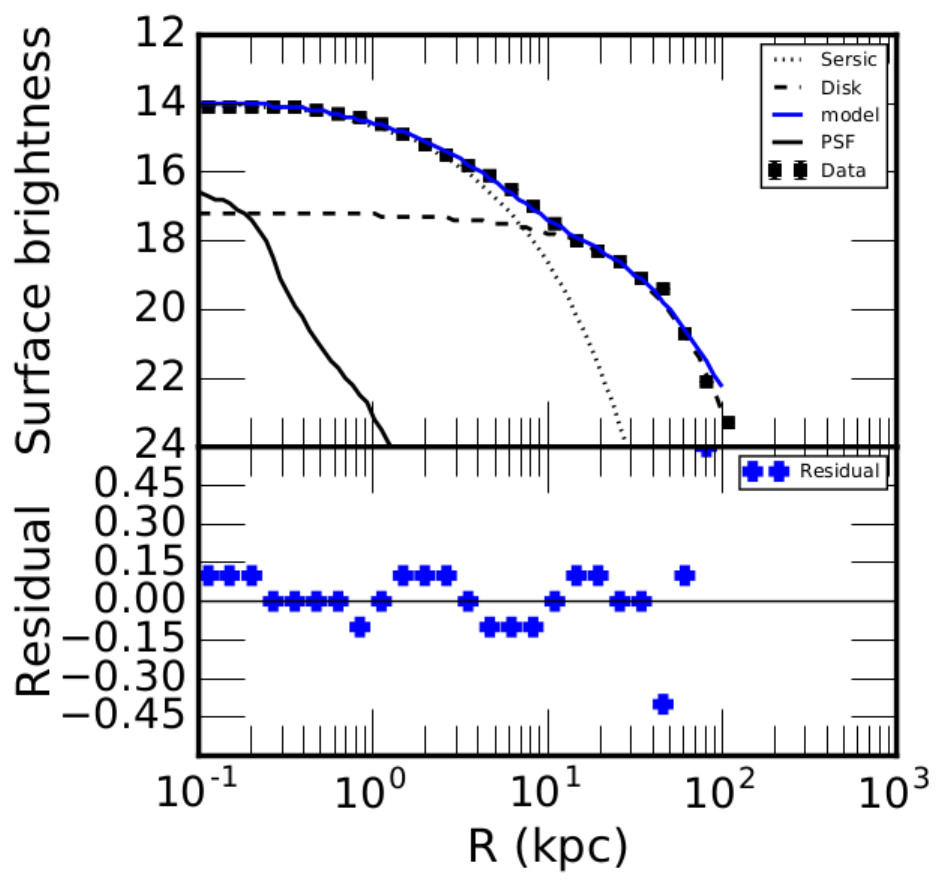}
\includegraphics[scale=0.35]{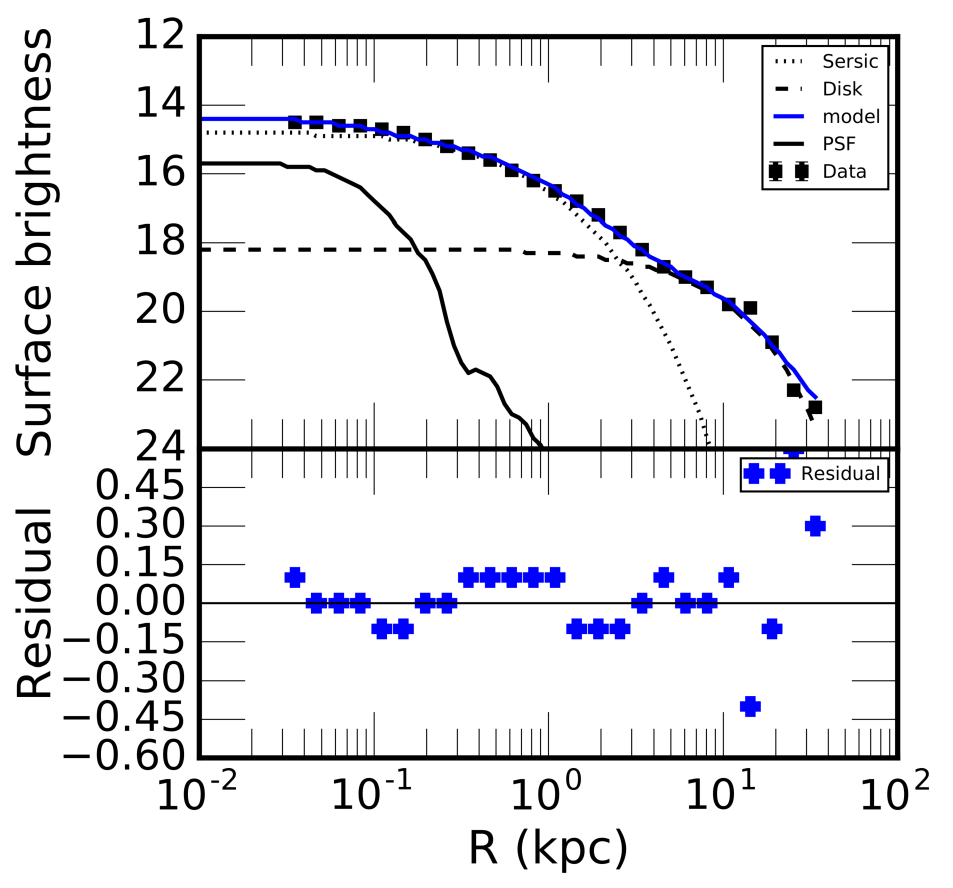}
\includegraphics[scale=0.35]{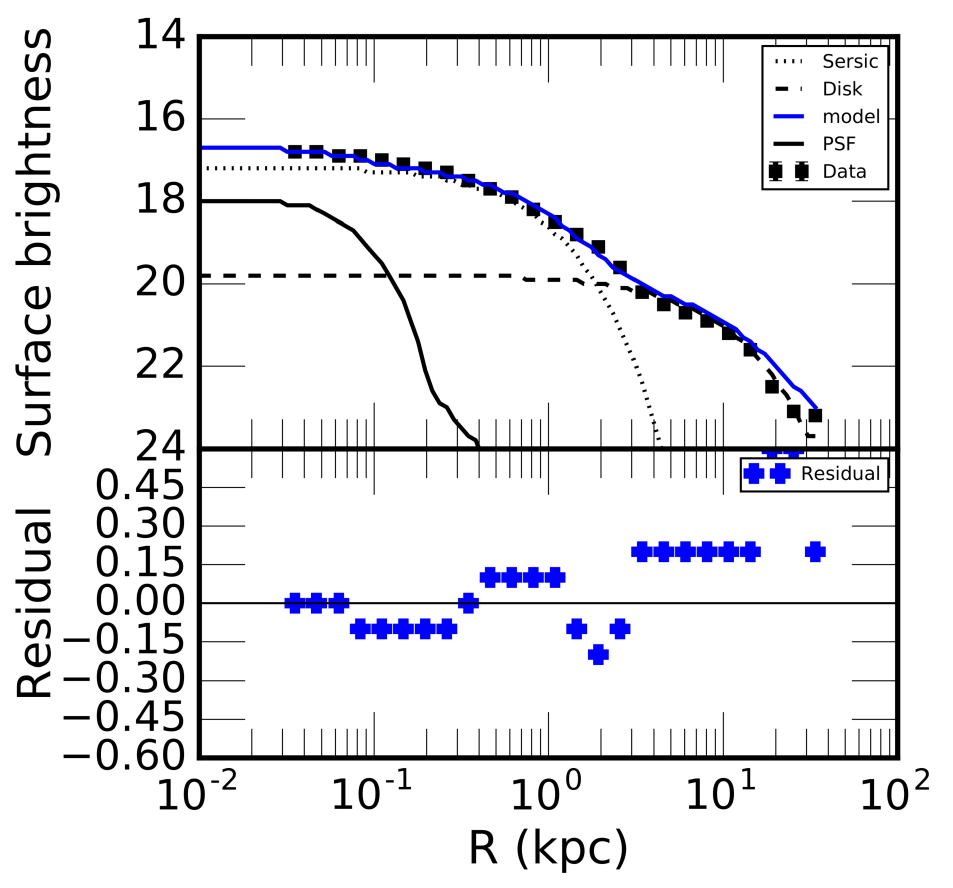}
\caption{Model B.  Left to right: HST F160W, F814W and F438W filter  radial profiles of the surface brightness  (mag arcsec$^{-2}$) with \texttt{GALFIT} model (solid blue line) and \hst/WFC3 image data (black squares) as a function of radius (kpc), as well as the subcomponents of the fit
are plotted (see legend for various profiles).  The data represent measurements along elliptical annuli of constant
ellipticity and position angle as obtained from  best fitting parameters.
These radial profiles were obtained using the elliptical isophote fitting routine \texttt{Ellipse}. The parameters of fitting are tabulated
in Table~\ref{all_params}.}
\label{set2_profile}
\end{figure*}

\begin{figure*}
\includegraphics[scale=0.8]{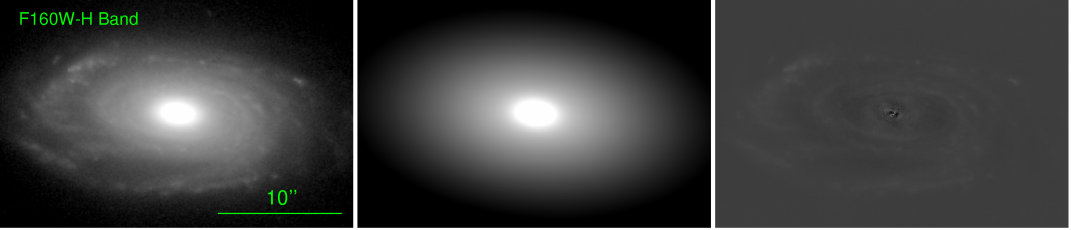}
\includegraphics[scale=0.8]{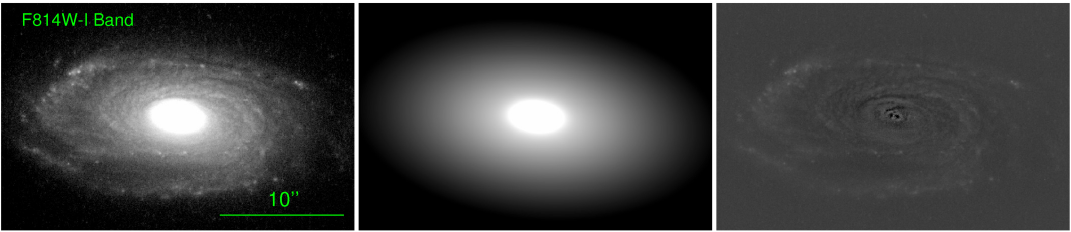}
\includegraphics[scale=0.8]{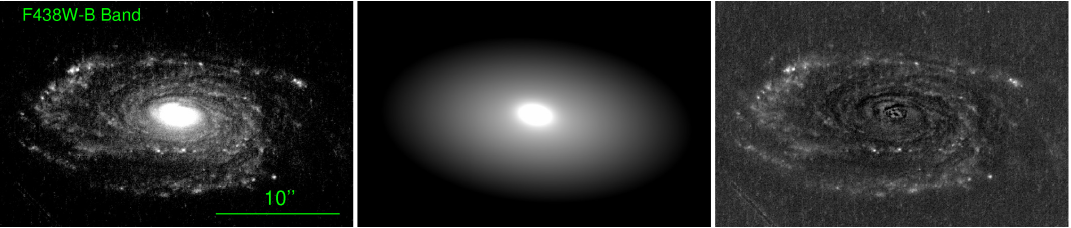}
\caption{Model C. Grayscale images and GALFIT  modelling  of the inner $25'' \times 25''$ of spiral galaxy J2345 in 
all three \hst/WFC3 filters,  shown with log scale. Top row: H-band, middle row: I-band, bottom row: B-band.  Left image panel: the HST input image  of the galaxy, middle  image panel: the best fitting GALFIT model, right image panel: the residual image showing
the difference between the input and the best fitting model image. Spiral arms 
were not fitted in the model.}
\label{set3}
\end{figure*}
\begin{figure*}
\includegraphics[scale=0.35]{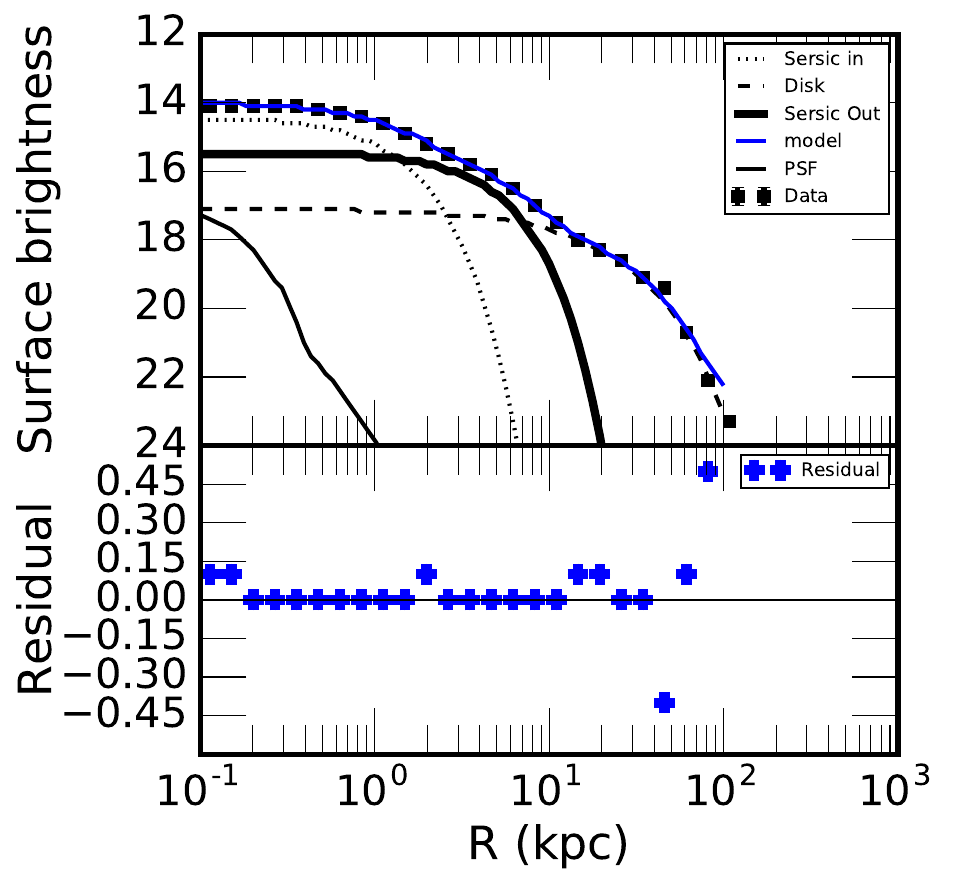}
\includegraphics[scale=0.35]{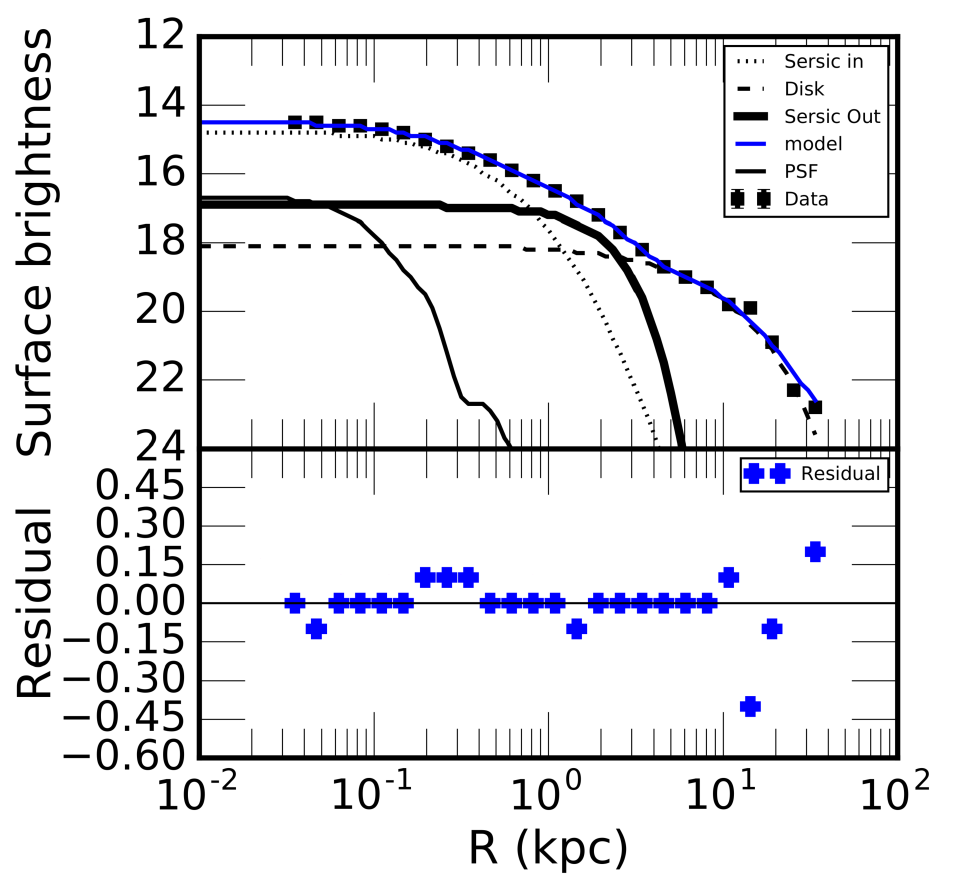}
\includegraphics[scale=0.35]{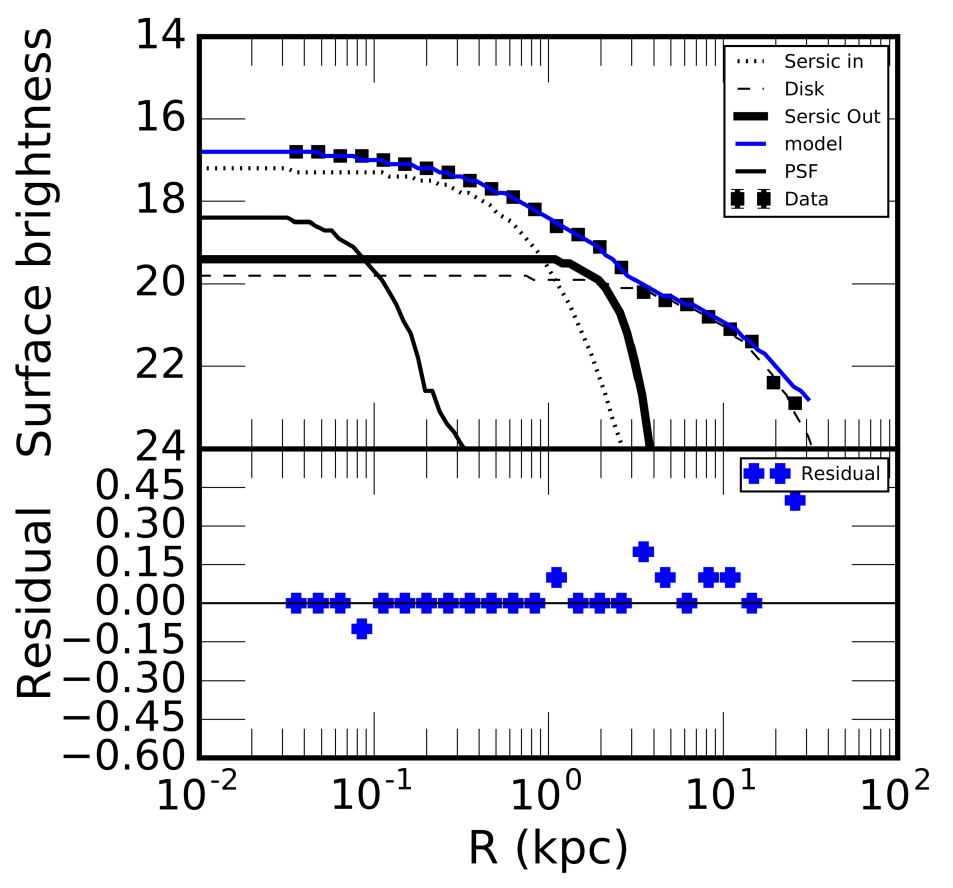}
\caption{Model C. Left to right: HST F160W, F814W and F438W filter  radial profiles of the surface brightness  (mag arcsec$^{-2}$) with \texttt{GALFIT} model (solid blue line) and \hst/WFC3 image data (black squares) as a function of radius (kpc), as well as the subcomponents of the fit
are plotted (see legend for various profiles). The data represent measurements along elliptical annuli of constant
ellipticity and position angle as obtained from  best fitting parameters. 
These radial profiles were obtained using the elliptical isophote fitting routine \texttt{Ellipse}. The parameters of fitting are tabulated
in Table~\ref{all_params}. }
\label{set3_profile}
\end{figure*}
\begin{figure*}
\includegraphics[scale=0.45]{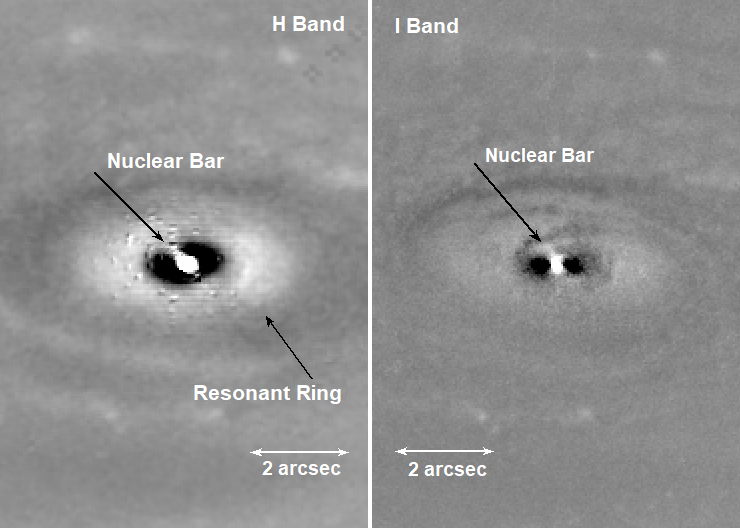}
\caption{Grayscale image of the innermost nuclear region of the galaxy showing the residual  of the difference between the input HST image  and  the best fitting model image (here we show the GALFIT model A). The left panel shows H-band and right one shows the I-band image. A small  nuclear bar and inner resonant ring of stars
are indicated by arrows. Note the I-band image is more affected by dust  extinction, particularly in the empty region surrounding the nuclear bar.}
\label{nuclear_bar}
\end{figure*}


\section{Description of HST Observations} 


Under  HST {\emph{GO}} program  14091 (PI: Aaron Barth) the spiral galaxy J2345-0449   was observed with WFC3  in three filter bands  taken over two  orbits on September 17, 2016. WFC3 offers a combination of broad wavelength coverage, wide field of view, and high sensitivity.  WFC3 is capable of direct, high-resolution imaging over the entire wavelength range  from 200  to 1700 nm by utilizing two optical/ultraviolet CCDs and  a near-infrared HgCdTe array. WFC3 has a range of wide, intermediate, and narrow-band filters, which provide unmatched capability to investigating a variety of new astrophysical phenomena  (\url{https://hst-docs.stsci.edu/wfc3ihb}).

During  observations, a series of three exposures were taken in each filter band, which are listed in Table.\ref{filters}.
These  exposures  were offset by fractional-pixel shifts relative to each  other using a three-point box dither pattern. The final images were constructed by combining these exposures, which allowed for the removal of cosmic-ray hits and hot  pixels. In the UVIS channel, we used the F438W  (\emph{B} band) and F814W (\emph{I} band)  filters. The UVIS channel CCD has a pixel scale 
of 0\farcs04 pixel$^{-1}$. At a distance of 323 Mpc, a single pixel on UVIS subtends about  56 pc. 
The  image  plane  point spread functions (PSFs) in Full Width at Half Maximum (FWHM) are;   0\farcs090 in F814W  and  
0\farcs0744  in F438W filters.  
Since J2345-0449 has effective radii of a few kpc (1.08$-$1.4\,kpc), WFC3 provides extremely good spatial resolution capability for a detailed study of its internal structure. 
In the IR channel, we used one wide band, F160W (\emph{H} band)  filter  and the  HgCdTe  detector array that  has somewhat coarser 
resolution than the UVIS, with  pixel scale  of 0\farcs13 pixel$^{-1}$ and  PSF  full width at half maximum  (FWHM) of   0\farcs241.  These  high resolution data are a key element of our overall  analysis, which  enables us to achieve our  science goals.

\subsection{High resolution HST imaging for accurate bulge-disc decomposition}

The exquisite spatial resolution offered by HST has some definite  advantages
over ground-based imaging.  For example,  for galaxies with a central bulge  and  strong light concentration,  a  surface brightness profile analysed in 
poor  seeing conditions may give the illusion of a  central light deficit. Atmospheric blurring  affects the fitted  light profile creating spurious  break/flattening on its slope. The  above effect is more pronounced when the true light concentration is on a scale comparable or below the seeing,  which can mislead one to  to believe on the detection of a depleted core, which can result from scouring of stars by mergers of binary super massive black holes. HST can largely mitigate this problem, while  is also  capable
of showing  a real depleted core if it  actually exists, which  can not be ruled out in  this galaxy. Moreover, the  simultaneous  
fits  over  a wide-wavelength range of HST
images makes the bulge-disc  decomposition more robust against any  spurious effects.

The galaxy’s favorable inclination to the line of sight ($ i = 59^{o}$) has enabled a direct view of the well-developed bright  spiral arms and dust features, as seen from the deep high-resolution optical images taken by HST presented here.


\section{HST Data Analysis, Photometric and SED Model Fitting}
 
 For analysis of HST data we have followed the same procedure as described in \cite{2015AJ....149..170C}.  The standard \hst\ calibration pipeline 
 was used for flat fielding, bias correction and dark current subtraction. In order to reject cosmic-ray hits and bad pixels 
 and to correct for the geometric distortion introduced by the optics of HST and WFC3, the three offset exposures for each  filter were combined into a final image using the PyRAF/STSDAS task \texttt{AstroDrizzle}, using a square kernel 
 with \emph{pixfrac}$= 1$, exposure time weighting, and sky subtraction turned off. In addition, as 
 part of the \texttt{AstroDrizzle} process, 
 we re-sampled the IR images onto the UVIS grid, resulting in a uniform pixel size of 0\farcs04 in all three bands.


We created  a  `natural'  looking   tri-colour  image  of  spiral galaxy J2345-0449 by  superposing  the HST WFC3 UVIS  
(F438W B and F814W I band) and  IR  (F160W H band)  monochrome images. The color balance was adjusted by eye to give
the most aesthetically pleasing effect. One can discern (Fig.~\ref{Fig1}) dark, winding dust lanes and  blue star forming regions along spiral arms providing  a detailed view of this enigmatic galaxy for the first time.


\begin{table}
    \centering
    \begin{tabular}{ccccc}
    \hline
        Filter & Channel & Central Wavelength & Band Width & Exp. time \\
        &  & \AA & \AA & seconds \\
        \hline
        F438W    & UVIS & 4326.5  & 197.30  & $3 \times 830$   \\
        F814W    & UVIS & 8029.5   & 663.33  & $3 \times 510$   \\
        F160W    & IR   & 15369.0 &2683.0 & $3 \times 199$  \\
        \hline
    \end{tabular}
    \caption{The central wavelength listed is the ``pivot wavelength'', a source-independent measure of the characteristic wavelength of the filter, and the band width is the ``passband rectangular width'', the integral with respect to wavelength of the throughput across the filter passband divided by the maximum throughput, as defined in the WFC3 Instrument Handbook.}
    \label{filters}
\end{table}


\subsection{Surface Brightness  Profile  Fitting}  

We used a composite of the \sersic \, \citep{1963BAAA....6...41S,1968adga.book.....S}  and  exponential functions
to model the galaxy's  light profile, where the former models the
bulge and the latter models the disc. In addition, to account for any   point-like excess light  at the  center, an additional PSF model was  used.  
A 2D model of the surface brightness profile was fit to each 
image using \texttt{GALFIT} version 3.0.4 \citep{2002AJ....124..266P,2010AJ....139.2097P}. 
For each band image, we specified the components of the model surface brightness profile to fit to the image 
(e.g., \sersic, exponential and  PSF), and supplied initial estimates for the model parameters 
(e.g., position of the center of the profile, total magnitude, effective radius). For simplicity  
we did not attempt to fit non-axisymmetric features like spiral arms, dust-lanes, bars or resonant rings in our models, which then should be visible in residual images.

A model image based on 
above parameters is produced and then convolved with the  corresponding  WFC3 PSF. Finally, the $\chi^{2}$ of
difference between the \hst input image and model image is  computed, and minimization of $\chi^{2}$ is achieved, yielding the best model parameters, using the Levenberg-Marquardt downhill-gradient algorithm. To properly weight each pixel in the $\chi^{2}$ calculation, we used the sigma image generated internally 
by GALFIT based on each input image. Although we used
the pipeline, we vary the initial conditions of different parameters to check whether the final results correspond to a global
minimum. We found that the final  results are stable against different
initial conditions.

The \sersic \, \citep{1963BAAA....6...41S,1968adga.book.....S} \ profile  is defined as,
\begin{equation}\label{eq:sersic}
    \rm{\Sigma(r) = \Sigma_{e} \, exp \left[-\kappa \, \left(\left(\frac{r}{r_{e}}\right)^{1/n} - 1\right)\right]}
\end{equation}

Here, $\Sigma(r)$ and $\Sigma_{e}$ are surface brightness at radius $r$ and $r_{e}$ respectively, $r_{e} $ is effective (half-light) radius, $n $ is the Sersic index, and $\kappa $ is a dependent variable coupled to $n$ such that $r_{e}$ encloses half of total brightness \citep{1999A&A...352..447C}. 

The exponential   profile is defined as,
\begin{equation}\label{eq:expdisc}
    \rm{\Sigma(r) = \Sigma_{0} \, exp \left(-\frac{r}{r_{s}} \right)}
\end{equation}
Here, $\Sigma_{0} = \Sigma(r)|_{r=0}$ and $r_{s} $ is disc the scale radius. 


For each filter,   simulated images of the \hst/WFC3 PSF were generated using 
version 7.5 of the \texttt{TinyTim} software package, which creates a model \hst\ PSF based on  the instrument, detector chip, detector chip position, focus and filter used in the observations. As the data have three exposures for each filter with sub-pixel offsets, we produced three versions 
of each PSF on a sub-sampled grid with sub-pixel offsets, using the same three-point box dither 
pattern as the \hst/WFC3 exposures to ensure that the model PSFs were processed in the same way as the \hst\ images.  

Each model PSF was convolved with the appropriate charge diffusion kernel in order to account for the effect of electrons leaking into neighboring pixels on the CCD. The PSFs were then combined and resampled onto a final grid with a pixel size of 0\farcs04  in all bands using \texttt{AstroDrizzle}, with the settings described in Section 2.0. The resulting simulated model PSFs are shown in Fig.~\ref{PSF_images}. 

\subsection{Bulge-Disc Decomposition and Best Fit Photometry}

We used a $600 \times 600$ pixel ($77\farcs0 \times 77\farcs0$), $1800 \times 1800$ pixel ($72\farcs0 \times 72\farcs0$) 
and $1800 \times 1800$ pixel ($72\farcs0 \times 72\farcs0$) fitting region centered on the brightest pixel in the H, B and I bands, 
respectively. The fitting region size is almost double as compared to the angular size of J2345-0449, whose  effective radii is  well under $1\farcs0$ (~1.4\,kpc). 

We experimented with different combination of functions available in \texttt{GALFIT} to fit to the WFC3 images. In first pass the simplest model A, with  one \sersic\,  bulge  and  one exponential disc function is used to model the light profile of the J2345-0449. The best fit models in  each filter  and the residual images are shown in Fig.~\ref{set1}. 
Interestingly in the model A residual images shown in Fig.~\ref{nuclear_bar}\, one can see
evidence for a small nuclear bar on $0.5\farcs$ ($\sim700$\, pc) scale, spiral shaped dust lanes winding inwards,  and a large resonant
ring surrounding the  nuclear bar. In addition due to higher dust extinction in I band, a dark disc shaped region surrounding the nuclear bar is more prominently visible in H-band image. We further discuss the implication of these structures below.

From Fig.~\ref{set1} it is evident that the  model A with one \sersic\ bulge plus an exponential disc is  promising but  not  enough to describe the 
innermost  light profile seen in the target galaxy. There is an excess,  compact light core and  a stellar halo and ring seen  at the central region of the galaxy which was not  accounted  for  by the simple two component  model. This excess light is not seen in  previous photometric decomposition of this galaxy \citep{2014ApJ...788..174B}, but only visible here due to the exceptionally high spatial resolution of HST. 

Next, in model B we  added  a PSF function to the simplest model of bulge and disc and  fit  this  model  to  data. In this case, the fit results in  improved  $\chi^{2}$, but even then  the residual images shows some  excess light and
halo around the central  AGN core region  (Fig.\ref{set2} right panels). 

Finally,   we  tried  a  more  complex  model C  with  more sub-components. We used a combination of   two \sersic \, functions; an inner 
\sersicone\ and another  outer \sersictwo\, , plus one exponential function  and  one  PSF function to fully account for the excess light at the center. This  composite  model provides very good fits to the 
F160W, F438W and F814W filter band images of this galaxy (Fig.~\ref{set3}) resulting in smallest
$\chi^{2}$.  The  three  fitted  models  A,B, and C  with  the derived structural parameters and their respective $\chi^{2}$ values  are  tabulated in  Table.\ref{all_params}.

We also subtracted  the best-fitted PSF,  \sersic \, and  the  exponential  disc  models from the  observed images  and searched for any   bar-like stellar component in the host galaxy of J2345-0449.  In addition, we  looked for the presence of a double or multiple nuclei as a sign of mergers. As  can be seen in the input data and the residual images, no evidence of  a   multiple nuclei is found in any of  the  HST bands, while a central nuclear bar and a stellar ring
is found (Fig.~\ref{nuclear_bar})

\subsection{Obtaining  Radial Surface Brightness Profiles}

We fit elliptical isophotes (ellipses of constant surface brightness) to both the \hst/WFC3 images and the PSF-convolved \texttt{GALFIT} model sub-components using the PyRAF/STSDAS routine \texttt{Ellipse} \citep{1993ESOC...47...27F}, 
which is based on methods detailed by \citet{1987MNRAS.226..747J}.  We also performed isophotal fits to the model PSFs in order to display 
their radial surface brightness profiles. Plots of the F160W, F814W and  F438W filter  radial profiles of the 
surface brightness for  spiral J2345-0449 and residuals of fits
are shown in Fig.~\ref{set1}, \ref{set2} and \ref{set3} panels showing the best fitting
profiles corresponding to models A, B and C
mentioned above. We caution that 1D radial profile plots is a powerful tool to visualize the goodness of photometric fits, however it is  not useful to visualize non-axisymmetric features of the 
surface brightness (like a stellar bar), which are best seen on 2D images. We further discuss  a non-axisymmteric nuclear bar structure below.

\subsection{A Small Scale Nuclear-bar and a  Resonant Ring}

Fig.~\ref{set1} and \ref{set2}  reveal in their right panels
-- that is their residuals from the H, I and B images -- a clear nuclear bar in the very center. This bar is
surrounded by a bright ring, forming a circumnuclear disc. The latter is then embedded in a two-arms spiral on larger scales,
which is somewhat flocculent, but still is the remains of a coherent density wave. There are clearly two
external spiral arms, which look coherent over 180 degrees (see Fig.~\ref{Fig1}). The details of this nuclear bar and ring are better seen in the residual images of
innermost region Fig.~\ref{nuclear_bar}. Bars typically have a radial profile characterized by a Ferrers function or a \sersic \, function with $n = 0.5$.  We note, one of the inner components of a double \sersic \, (plus one exponential function) GALFIT Model C indeed has  inner \sersic \, $n \approx 0.5$ (Table.\ref{all_params}); and the residuals seem to have removed the bar-like feature seen visually in the other models that did not include this extra component (Fig.~\ref{set3}).

The nuclear bar, oriented at PA $\sim 18$ degrees, has a diameter of $\sim 0.8$ arcsec in projection. Since it is located
near the minor axis, its intrinsic diameter is about  1.6 arcsec, or 2.4 kpc. This nuclear ring might be at the
origin of the slight velocity residual from disc rotation along a similar PA in the MUSE H$\alpha$ data cube (Fig.~\ref{spectrum} of \citet[][]{2023A&A...676A..35D}).
The nuclear bar is encircled by a nuclear stellar ring, of mean diameter of  $\sim3$ arcsec, or 4.4 kpc. The ring is clearly  seen due to the very dark regions devoid of any light (stars) in the regions east and west of the nuclear bar.
Since even in the NIR (H-band image) the region is very dark it is actually devoid of light and does not correspond  to highly dust obscured regions. With dark void regions on each side, and with a ring outside the bar, they form a  $\Phi$-shaped morphology.
 
 Bars are dynamical structures in galaxies that can generate a variety of resonant phenomena, including Lindblad Resonances (LR), due to their gravitational influence on the surrounding stellar and gas populations. Lindblad resonances are locations within a galactic disc where the rotational frequency of stars or gas matches the frequency of the gravitational potential induced by the rotating bar. These resonances occur at specific distances from the center of the galaxy and are associated with the spiral density waves generated by the bar.
Overall, the presence of a galactic bar can significantly influence the dynamics of stars and gas within a galaxy, and Lindblad resonances play a central role in mediating these interactions. Studying the connection between bars and resonances is essential for understanding the formation and evolution of galactic structure.
 
 From what has been learned with numerical simulations of embedded bar decoupling, a nuclear bar forms
 usually inside the resonant ring of a primary bar. The  ring is then both the inner Lindblad resonance (ILR) of
the primary bar, and the UHR (ultra-harmonic) resonance near corotation of the nuclear bar (e.g. \citet[][]{1996FCPh...17...95B}).
The common resonance occurs at the decoupling of a nuclear bar, embedded in a primary bar, to avoid chaos
in stellar orbits, and ensure more stability of both bars \citep[][]{1993A&A...277...27F}.

There is no primary bar seen at the present epoch in the J2345-0449 galaxy, but it might have been destroyed by the
 gas infall to the center (e.g. \citet[][]{2002A&A...392...83B}). Indeed, there has been a recent  AGN activity in the
 galaxy, as witnessed by the giant radio jets. A primary bar, supported by a secondary bar, is the most efficient
 dynamical mechanism to drive the gas to the center, and to fuel the AGN (e.g. \citet[][]{2014A&A...565A..97C}). The presence of a nuclear bar can affect how material is funneled towards the supermassive black hole, influencing its feeding rate and the resulting activity, such as the emission of powerful jets of particles and radiation. 
 
 This means  that in past a primary bar likely existed with an  ILR at 2.2 kpc radius, and then which  gave birth to the embedded nuclear bar,
and then it  got dissolved.  When gas has been driven in because of torques from the bar (due to strong tangential forces),
then the gas torques also the bar (from the equality of action and reaction), and gives its angular momentum
to the bar. This destroys the primary bar, as has been developed in models by \citet[][]{2002A&A...392...83B} and \citet[][]{2005A&A...437...69B}.

 We proposes the AGN at the origin of the radio jet has been highly active in the past; and consumed and/or ejected all the central molecular
 gas. There is no longer  cold gas left in the very center, but the disappeared primary bar might have played the major
 role in driving the gas to the center. Today we observe a large molecular ring, with CO emission (Fig.~\ref{ALMA_velocity}), with a mean
 radius of 6.6 arcsec, or 10 kpc \citep[][]{2021A&A...654A...8N}. This neutral gas ring could correspond to the OLR  (Outer Lindblad resonance) of the past primary bar.

The observed bulge looks more like a pseudo bulge, which comes usually from a bar evolution,
supporting that there existed a primary bar in the past, with a ring at its ILR. This also fits with the  position of the galaxy in the green valley, after a star formation episode corresponding to the primary bar driving the gas inwards.

\subsection{Spectral Energy Distribution (SED) Models}

\begin{figure*}
\centering
\includegraphics[scale=1]{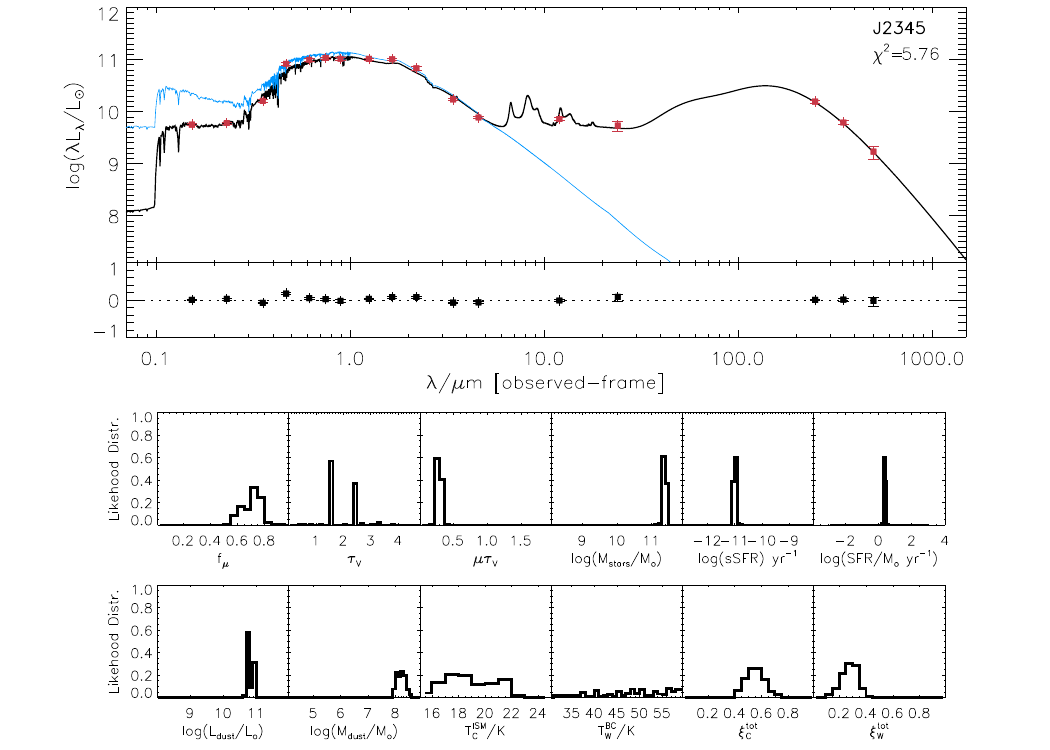}
\caption{ Upper figure: The rest-frame SED of J2345-0449, with observed photometric data points (red points) from the  UV to the far-IR.  
The black line is the best fit  {\fontfamily{qcr}\selectfont MAGPHYS} model  ($\ensuremath{\chi^{2} = 5.76}$ and  reduced  $\ensuremath{\chi^{2} = 0.98}$).
The blue line is the non-absorbed stellar model. The  
The bottom panel shows, the 
normalised residuals between the best-fit model and the observed data points. Central and lower  panels: The normalized likelihood distributions of  a few  best 
fit physical parameters for J2345-0449, derived from {\fontfamily{qcr}\selectfont MAGPHYS}  modelling. These are;  (1) fraction of total dust luminosity contributed by dust in the ambient (diffuse) ISM $f_\mu$; (2) average V band dust attenuation, $\rm \tau_{V}$;  
(3) total  effective V-band  absorption  optical  depth  of  the  dust  seen  by  young  stars inside birth clouds $\rm \mu\tau_V$; 
(4) stellar Mass, $ \rm log(M_{stars}/M_{\odot})$ ; 
(5) specific star formation rate, $ \rm log(sSFR) yr^{-1}$; 
(6) star formation rate, $\rm log(SFR/M_{\odot} yr^{-1})$;  
(7) total dust luminosity, $\rm log(L_{dust}/L_{\odot})$ ; 
(8) total dust mass, $\rm  log(M_{dust}/M_{\odot})$ ; 
(9) equilibrium temperature of cold dust in the ambient ISM,  $\rm T_{C}^{ISM}/K$;
(10) equilibrium temperature of warm dust in the ambient ISM,  $\rm T_{W}^{ISM}/K$
(11) fractional contribution by cold dust in the total infrared emission, $\rm \xi_C^{tot}$; 
(12)  fractional contribution by warm dust in  the total infrared emission, $\rm \xi_W^{tot}$
}
\label{Fig2}
\end{figure*}

For the SED fitting and to investigate the various  physical  parameters of J2345-0449,  such as the stellar mass,  star formation rate, dust mass, dust temperature, dust opacity, dust luminosity, and  contribution by AGN  radiation  to SED   etc., we have made use of the publicly available spectral energy distribution fitting code {\fontfamily{qcr}\selectfont MAGPHYS}  (Multi-wavelength Analysis of Galaxy Physical Properties) V 2.0 developed by \citet[][]{2010MNRAS.403.1894D} to fit the  available  multi-band photometric data over the wide  wavelength range from  912 \AA to  1 mm.
The input data  spans from  GALEX in  UV, SDSS in optical, 2MASS  in  the near-IR, WISE in the 
mid-IR and   Herschel in  the far-IR. The best fitting parameters and  SED model  fit on  photometric data points are tabulated in Table.~\ref{SED} and  shown  in Fig.~\ref{Fig2}, respectively. In this figure, The black 
line is the best fit model SED and the blue line is the unabsorbed optical stellar model. The bottom panel shows the residuals of  the difference between best-fit and the observed data points. The fit provides normalized likelihood distributions of best 
fit physical parameters which are given in Table.~\ref{SED} and  shown  in Fig.~\ref{Fig2}. These  parameters are as follows;  
(1) fraction of total dust luminosity contributed by dust in the ambient (diffuse) 
ISM $f_\mu$; (2) average V band dust attenuation, $\tau_{V}$;  (3) total  effective V-band  absorption  optical  depth  of  the  dust  seen  by  young  stars inside birth clouds $\mu\tau_V$; (4) stellar Mass, $\rm log$$(M_{\star}$$\rm /M_{\odot})$ ; 
(5) specific star formation rate, $ \rm log(sSFR/yr^{-1})$; 
(6) star formation rate, $\rm log(SFR/M_{\odot} yr^{-1})$;  
(7) total dust luminosity, $\rm log$$(L_{\rm dust}$$\rm /L_{\odot})$ ; 
(8) total dust mass, $\rm  log$$(M_{\rm dust}$$\rm /M_{\odot})$ ; 
(9) equilibrium temperature of cold dust in the ambient ISM,  $ T_{\rm C}^{\rm ISM}/K$;
(10) equilibrium temperature of warm dust in the ambient ISM,  $T_{\rm W}^{\rm ISM}/K$
(11) fractional contribution by cold dust in the total infrared emission, $\rm \xi_C^{tot}$; 
(12)  fractional contribution by warm dust in  the total infrared emission, $\rm \xi_W^{tot}$.

The goodness of the fit can be judged from the $\ensuremath{\chi^{2} = 5.76}$ with reduced $\chi_{\nu}^{2} = \chi^{2}/\nu = 0.98$ for $\nu = 6$ degrees of freedom (18 data points and 12 parameters for fitting), which indicates  a  very good fit. The best fit results of MAGPHYS are shown in Table.~\ref{Tab_SEDfit_magphys}. The error bars indicated are within $\rm 1\sigma$ confidence interval taken between the 16th to 84th percentile range of the likelihood distributions of each parameter.





\begin{figure*}
\centering
\centering
\includegraphics[scale=0.75,angle=0]{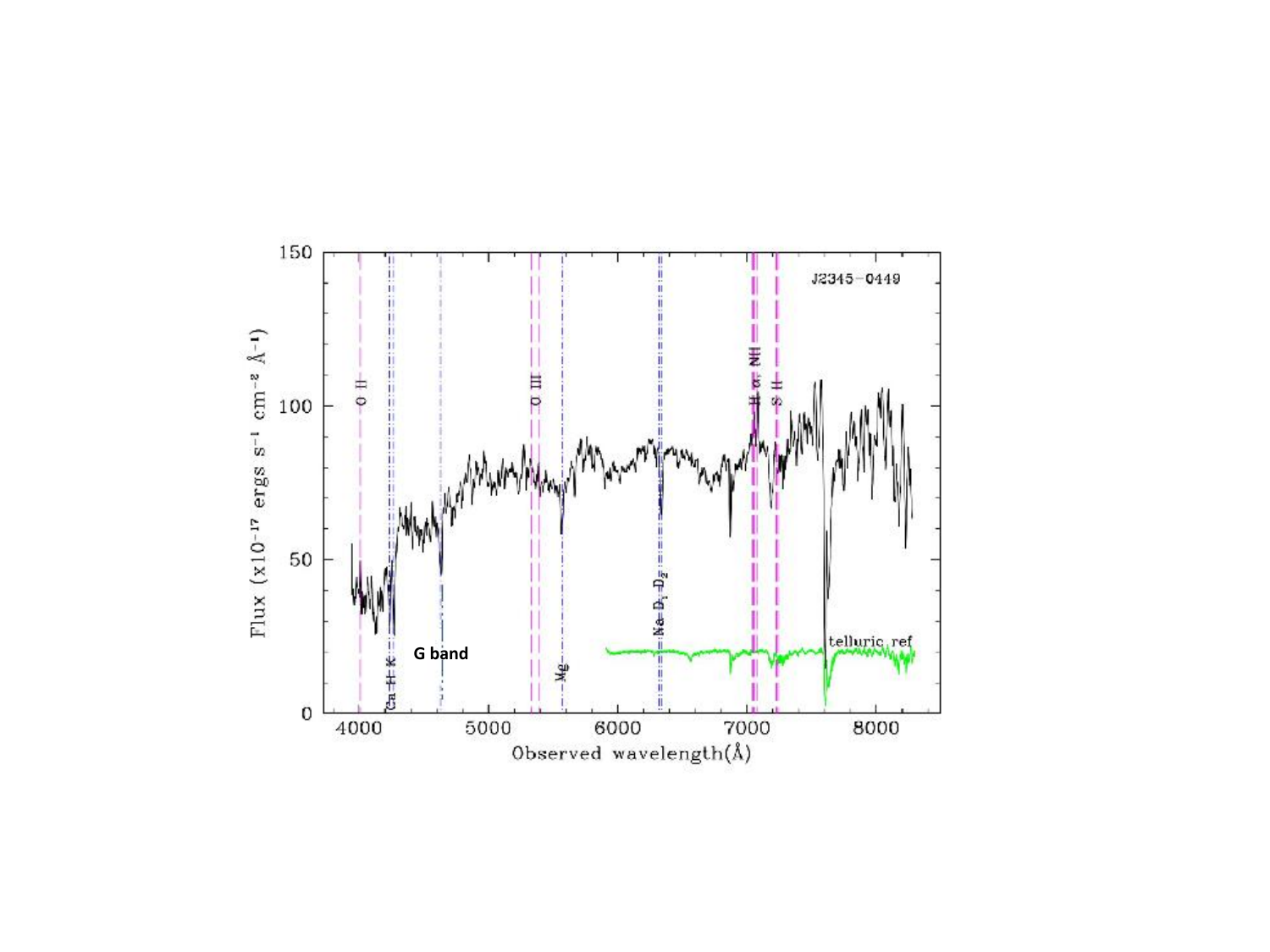}
\caption{Optical spectrum of J2345-0449 taken with IFOSC (IUCAA 
Faint Object  Spectrograph and Camera) on 
IUCAA 2m telescope, centered on the nucleus. We used two IFOSC  grisms, grism No. 7 and
grism No. 8, in combination with a 1.5 arcsec slit,  yielding a wavelength coverage of 3800 - 6840 \AA\,  and 5800 - 8350 \AA\,  and spectral resolutions of  300 kms$^{-1}$ and 240 kms$^{-1}$, respectively (\citet[][]{2014ApJ...788..174B} for more details). 
The absorption lines are marked by blue dot-dashed
lines, and the expected positions of the emission lines are marked with magenta
dashed lines. The red end of the spectrum is contaminated by  telluric features.
A telluric reference spectrum is shown in green at the bottom. 
Note the absence of any  prominent strong emission lines. Weak H$_{\alpha}$ and [NII] emission lines can be seen.}
\label{spectrum}
\end{figure*}
 
\section{Results and Discussion}

\subsection{On  the Bulge and Disc  Morphology and  Implications}

High-resolution bulge-disc modelling of  HST multi-wavelength images revealed that the \sersic\, index of the  bulge is  $n \approx 1$, which essentially implies that the galactic bulge  to be a pseudo-bulge having a disc-like exponential light profile rather than a  classical bulge (see Table~\ref{all_params}). However in H-band we find that the
\sersic\, index of the  bulge is  slightly larger (range of $n \approx 1.6 - 2.7$ in models A, B and C). Evidently this is consequent to the fact that in near IR light the dominant population are old  stars which have a larger scale height in the bulge compared to  blue stars in the disc. It is to be noted that,  empirically,  for pseudo-bulges $n< 2$ and classical bulges have $n > 2$, but the  division is not perfectly  bimodal  with  some degree of overlap  between the two classes \citep[][]{2009ApJ...696..411W,2008AJ....136..773F}. The \sersic \, index $n = 1$ represents a pure exponential disc like profile and $n = 4$  corresponds to the de Vaucouleurs elliptical galaxy like profile. While this is rather a simplified  division, more observations and analysis are  needed to investigate their physical details \citep[][]{2016ASSL..418..263G}.  Several previous  observations have found   that  bulges may exhibit fairly complex composite structures  with a mix of spherical/ellipsoidal, disc-like, or boxy/bar-like components, potentially showing a wide array of possible bulge morphology in the universe. Nevertheless, pseudo bulges are  always systematically flatter  than classical bulges and  appear more disc-like in  both their morphology and shape \citep[][]{2008AJ....136..773F}.

We find that  the  typical  scale length $r_{e}$ of the  pseudo-bulge  is  much smaller when compared to the disc $r_{d}$, which has   $r_{d} \approx 7 \times r_{e}$, (Table~\ref{all_params}), as one expects for pseudo bulges. To quantify the relative light distribution in bulge and disc components the bulge (B),  disc (D) and total (T) luminosity were derived along with  $\frac{B}{D}$ and $\frac{B}{T}$ measurements indicating bulge to disc and bulge to total (bulge+disc) luminosity ratios. It can be seen in Table~\ref{tab:bulge_disc_lum} that our object has a maximum $\sim 22$ per cent contribution to the total intrinsic light accounted for H-band and only $\sim 8.5$ per cent contribution in B-band. About $14-18$ per cent fraction of the  total  SDSS g-band or  r-band luminosity is contributed by the bulge. These  numbers again indicates that this  spiral  galaxy  has a pseudo-bulge  rather than a  classical  bulge  component.   Galaxies with classical bulges generally have a much more centrally peaked light profile, contain a higher fraction of total light, and their \sersic\, index is  larger ($n = 2-6$) than the pseudo-bulges \citep[][]{2008AJ....136..773F,2009ApJ...696..411W,2011ApJ...733L..47F,2022MNRAS.517...99R}.

Here we consider the  (B/T or B/D) ratio  an important parameter (notably both are strongly colour dependent) of galaxy structure that  is  also well correlated with several other key physical quantities relevant in studies of galaxy evolution; such as the morphology, kinematics, stellar mass and star formation rate et cetera. The B/T is also directly related to the  fraction of  total light and mass  contained in the bulge, which in turn  tightly  correlates  with the mass of the central supermassive black hole of the galaxy (see \citet{2013ARA&A..51..511K} and references therein).

The above results of bulge-disc decomposition strengthen  our previous  lower spatial resolution study  \citep{2014ApJ...788..174B} where we concluded that J2345–0449 possibly features a pseudo-bulge, 
whose small \sersic \, index and scale radius, and low  bulge-to-total light ratio make it similar to pseudo-bulges observed in other galactic systems (e.g. \citet{2009MNRAS.393.1531G}). Recent recognition of a  high fraction ($\sim60$ percent ) of  nearby spiral galaxies having no  kinematically  ``hot" central bulge challenges the conventional ideas of hierarchical galaxy formation; i.e. how have so many bulgeless pure disc galaxies come to exist, despite the galaxy mergers known to occur  \citep[][]{2009ApJ...696..411W,2008AJ....136..773F,2011ApJ...733L..47F}?. While
classical bulges are possible products of galaxy mergers
pseudo-bulges, seem to be still evolving, more likely as a result of their host galaxies’ slow internal evolution, which involves disc instabilities  and angular momentum transfer.

At present  very little is known about the formation mechanism and evolutionary path of such extremely massive disc galaxies, as they are found rarely, and moreover, not known to eject powerful relativistic jets on megaparsec scale, like those found in J2345-0449. Based on its structural and  photometric properties discussed above, here we  propose that the cosmic co-evolution of the galactic disc of J2345-0449 and the central SMBH driving the radio jets may have been primarily driven by external cold gas accretion and internal `secular' processes \citep[][]{2013seg..book....1K} instead of recent  major merger events. This is suggested by its  fast rotating exponential  disc-like  pseudo bulge, the  absence of close galactic neighbors and signature of tidal debris (e.g., stellar streams, plumes, or shells), and its stable, rotation supported massive stellar disc showing well-formed spiral arms.  
In a recent HST study of  spiral host galaxies showing double lobed radio jets  
\citet[][]{2022ApJ...941...95W}  found that they have exceptionally weak bulges, as judged by the low global concentrations, small global \sersic\, indices, and low bulge-to-total light ratios (median B/T = 0.13). With a median \sersic\, index of 1.4 and low effective surface brightnesses,  they are consistent with harboring pseudo bulges.

In  the late stages, dynamically,  a major merger of  galaxies  will  inevitably  create a  prominent classical bulge and  may exert  a strong perturbation of  the  central black hole's spin axis as it may be torqued by the gravitational 
influence of a second black hole nearby, and  eventually both  will merge in a single stable  black hole of a larger mass, emitting nano Hertz frequency gravitational waves \citep[][]{2023A&A...678A..50E,2023ApJ...952L..37A,2023ApJ...951L...6R,Xu_2023} during inspiral. 
On the contrary, in J2345-0449, a close alignment (within 10 degrees) of the inner and outer radio lobe pairs, and lack of  violent merger signs, implies that the  black hole’s spin axis, which in equilibrium state orients itself orthogonal  (either parallel or anti-parallel) to the inner  accretion disc spin direction because of the Lense–Thirring frame dragging effect \citep[][]{1975ApJ...195L..65B}, has remained stable (non-precessing),  at least between the two last episodes of jet triggering on timescales of  $\sim 10^{8}$ yr.  Thus, the jets are possibly  emitted from a stable gyroscope,   such as a rapidly spinning massive  object,  and present  theories provide a framework of jet formation  via a  magneto-hydrodynamic (MHD) plasma mechanism \citep[][]{1977MNRAS.179..433B,1982MNRAS.198..345M,2013MNRAS.436.3741P}.
All this suggests that the galactic disc and the  central black hole might have a quiet evolution together, without upheavals of major mergers. 


Recently \citet[][]{2019MNRAS.488..609I} studied evolution of  L-Galaxies  from the catalogues of merger-trees obtained by the Millennium \citep[][]{2005Natur.435..629S} and Millennium II \citep[][]{2009MNRAS.398.1150B} N-body simulations. They find that at early epochs ($z \approx 1$) pseudo-bulges are the main type of galaxies at high stellar masses $M_{\star} > 10^{10.5} \Msun$ while classical bulges and ellipticals are the main ones for $M_{\star} < 10^{10.5} \Msun$. While at low masses there is little evolution between $z = 1$ and $z = 0$, at high masses  by $z = 0$,   bulged ellipticals  are now the dominant galaxy population and pseudo-bulges have declined in number. This results from the hierarchical growth of structures: pseudo-bulges hosted in the most massive galaxies at high-z are subsequently destroyed by major mergers which turn  galaxies into pure bulges. However, since massive galaxies of $M_{\star} > 10^{10.5} \Msun$ only make up $\sim 4$ per cent of bright galaxies in the local volume, pseudo bulge galaxies and those with no bulge are the dominant type of bright galaxy by number in local Universe \citep{2009ApJ...696..411W,2008AJ....136..773F,2011ApJ...733L..47F}.
Contrary to the above findings, we find that J2345-0449 is a rare,   highly massive galaxy ($M_{\star} > 2 \times 10^{11} \Msun$  in local universe ($z = 0.07$) which apparently has managed to survive the  destruction of its'  pseudo bulge by violent mergers, as it might have evolved   predominantly through  secular,  disc driven processes -  the same  conclusion that  we arrived at from  other arguments. 


Using IUCAA telescope spectrum we obtained the best-fit values of stellar velocity dispersion $\sigma_{\star}$ of the central region of bulge from fitting the Mg b ($\lambda 5175$ \AA) line, after correcting for the redshift and instrumental broadening, which gave $\sigma_{\star} = 379(\pm25) \kms$ with the slit oriented along the major axis and $\sigma_{\star} = 351(\pm25) \kms$ along the minor axis. Both values are spatially averaged over 5 pixels (2.35 kpc) centered on the galactic nucleus \citep{2014ApJ...788..174B}. This gives an extremely large  dynamical mass of bulge $M_{\rm bulge} = 1.07 - 1.20 (\pm 0.14) \times 10^{11} \Msun $, roughly within one scale radius $r_{e} \approx 1.2$ kpc (We have further discussed the implication of this large bulge dispersion velocity and mass concentration below). The  massive stellar and gaseous disc of J2345-0449 is  clearly rotation supported, with ${(v_{\rm rot}/\sigma_{\star})}^{2} \approx 1.5$, where $v_{\rm rot}$ is the  rotational speed and $\sigma_{\star}$  is  velocity dispersion of stars) and  therefore,  galaxy disc should be dynamically stable against gravitational instability.

\begin{table}
    \centering
    \begin{tabular}{ccccc}
    \hline
    Band & $L_{\rm bulge}$ & $\rm L_{disc}$ & $\frac{B}{D}$ & $\frac{B}{T}$ \\
         & ($L_\odot$) & ($L_\odot$) & & \\
    \hline
    F438W &$4.10\times10^{9}$& $4.48\times10^{10}$&0.093&0.085\\
    F814W &$2.65\times10^{10}$& $1.11\times10^{11}$&0.240&0.193\\
    F160W &$1.50\times10^{11}$& $5.24\times10^{11}$&0.287&0.223\\
    \hline
    \end{tabular}
    \caption{Here, we show the bulge and disc luminosity along with  $\frac{B}{D}$ and $\frac{B}{T}$ measurements indicating bulge to disc and bulge to total (bulge+disc) luminosity ratios. It can be clearly seen that our object has a `pseudo-bulge' with  only  $\sim 20$ per cent contribution to the total intrinsic light in the I and H bands and only $\sim 8.5$ per cent contribution in B-band.}
    \label{tab:bulge_disc_lum}
\end{table}

\subsection{Stellar Mass, SFR, Dust  and  Gas Properties in the Disc}


\begin{figure*}
    \centering
    \includegraphics[width=0.49\textwidth]{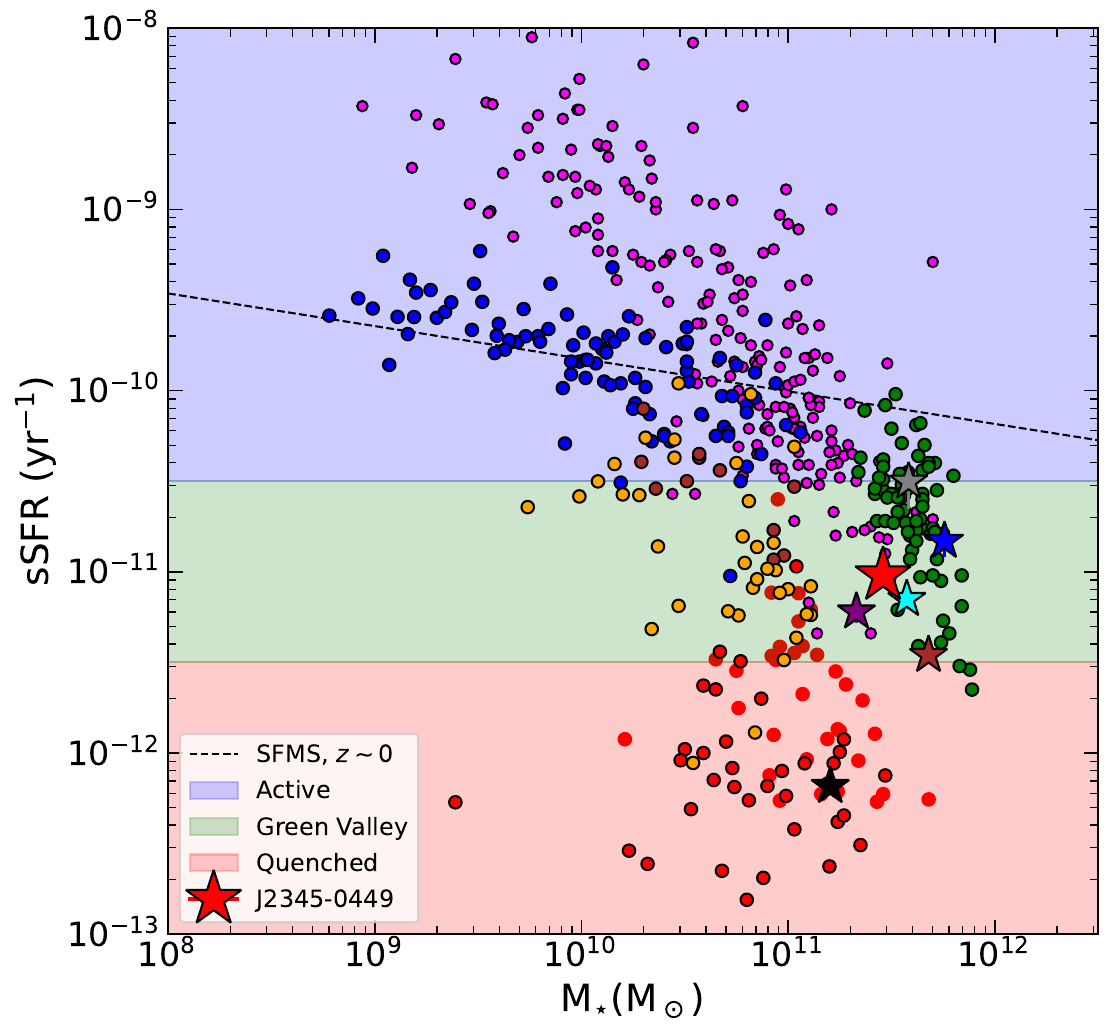}
    \caption{The variation of the specific star formation rates sSFR (= SFR/$M_{\star}$) of galaxies as a function of their stellar masses $M_{\star}$ is shown. The dotted black line is the star forming main-sequence (SFMS) 
      for redshift ${z}\sim 0$ \citep{2007A&A...468...33E}. The graph is divided into three parts; showing actively star forming region ($\rm sSFR > 10^{-10.5} yr^{-1}$, blue ), green valley region ($\rm 10^{-11.5} < sSFR < 10^{-10.5} yr^{-1}$, green) and quiescent region ($\rm sSFR < 10^{-11.5} yr^{-1}$, pink-red) (see \citealt{2016ApJS..227....2S}).  Miscellaneous galaxy samples of range of star formation are also plotted: (a) The highly star forming galaxies from \citet{2007ApJS..173..441H} (pink), (b) Super-spirals, lenticulars and post-mergers from \citet{2019ApJ...884L..11O} (green), (c) A sample of galaxies provided by \citet{2021A&A...648A..64K} including galaxies that are star forming (dark blue), mixed (orange), quiescent nuclear-ring (bright red), nearly retired (brown ) and fully retired (dark red ) are also shown for comparison. It can be seen that the position f our object J2345-0449 is well within the `green valley' region. Other than our object, the objects shown with star-symbols are six highly massive galaxies namely, UGC~12591 (black), NGC~1030 (brown), NGC~1961 (grey), NGC~4501 (blue), NGC~5635 (purple) and NGC~266 (cyan) from \citet[]{2022MNRAS.517...99R} and 
      \citet[][]{2023MNRAS.tmp.3539R}. 
      }
    \label{fig:ssfr_mstar}
\end{figure*}


 Structurally J2345-0449  shows  a prominent disc of  stars,  and interstellar  medium (ISM)  of   warm dust  along with a coplanar rotating  
cold, molecular (mainly H2 and CO)  gas  disc  in the  equatorial plane.  Based on our SED fitting models  (Table~\ref{Tab_SEDfit_magphys} and Table~\ref{Tab_SEDfit_cigale}) we  infer the following  properties of  the  ISM and  dust components;  

\citet[][]{2012MNRAS.419.2545R} did a study on dust properties and star formation history of the local sub millimetre detected galaxies, with a mixture of  early-type and passive spirals. The passive spirals are defined as  early-type galaxies with dust abundance and low SFR. They   found  the  mean dust luminosity  $\log M_{\rm dust} = 7.74 \Msun$ with $M_{\rm dust}$ in the range $\rm 10^{5} - 10^{8} \Msun$ for early-type galaxies,  and for passive spirals mean $\log M_{\rm dust} = 7.47$ with $M_{\rm dust}$ in the range $\rm 10^{6} - 10^{8}\Msun$.  From SED fitting we find $\log M_{\rm dust} =  8.1 \Msun$ (Table~\ref{Tab_SEDfit_magphys})  consistent with the results of \citet[][]{2012MNRAS.419.2545R},  with  dust mass for J2345-0449  falling at the highest  mass end of  passive spirals. The small  ratio  of dust to stellar mass $ M_{\rm dust}/M_{\star} \approx 6.3 \times 10^{-4}$ is typical  of  passive galaxies \citep[][]{2017A&A...602A..68O}, while the ratio of molecular gas mass (H2 mass  is $\rm 1.62 \times 10^{10} M_{\sun}$) to stellar mass is far higher at $0.085$ indicating a large,  quiescent (non star forming) repository of potential star forming material.

The estimated  integrated  stellar mass, $M_{\star}$ of the galaxy  is  $\rm 2.88 (+0,-0.79) \times 10^{11} M_{\odot}$  from  SED  model  {\fontfamily{qcr}\selectfont MAGPHYS}  and $\rm 1.53 (\pm 0.5) \times 10^{11} M_{\odot}$  from {\fontfamily{qcr}\selectfont CIGALE} model.  They are consistent with each other within errors. We may take  the  stellar mass of the galaxy  as the mean of two values at  $M_{\star} = 2.2 (\pm 0.7) \times 10^{11} M_{\odot}$.  In an earlier paper  \citet{2021A&A...654A...8N} quoted a value of  
$ M_{\star} = 3.6  \times 10^{11} M_{\odot}$ based on 2MASS J-band magnitude. This value  is higher than our SED fitted values but we note it is dependent on the assumed star formation history and  the mass-to-light ratio in J-band.

The global SFR computed  via two competitive SED fitting models; {\fontfamily{qcr}\selectfont MAGPHYS} and {\fontfamily{qcr}\selectfont CIGALE}, gives consistent results.  SFR and sSFR  from {\fontfamily{qcr}\selectfont MAGPHYS} is  $\rm SFR_{mag} = 2.45 \msun yr^{-1}$ and $\rm sSFR_{mag} = 1.09 \times 10^{-11} yr^{-1}$
while from {\fontfamily{qcr}\selectfont CIGALE} it is $\rm SFR_{cig} =2.38 \msun yr^{-1}$ and $\rm sSFR_{cig} = 1.55 \times 10^{-11} yr^{-1}$. In previous works \citet{2021A&A...654A...8N} measured SFR of $\rm SFR_{J} = 3.6 \msun yr^{-1}$ based on J-band magnitude and
\cite[][]{2015MNRAS.449.3527W} obtained $\rm SFR_{FUV} = 1.6 \msun yr^{-1}$ based on FUV flux of {\em GALEX}, bracketing our value.
These are global SFRs from the integrated flux values.  One should refer to \citet[][]{2023A&A...676A..35D}, who have discussed SFR surface density and observed properties of individual star-forming regions in the disc based on  VLT/MUSE spectroscopic observations.


J2345-0449 shows mild ongoing star-forming activity across the galactic disc, traced by \hal\, and [NII] emission lines in our slit spectrum. It is  also strongly detected by the Galaxy Evolution Explorer in the near (2316 \AA) and far 
UV (1539 \AA) bands, indicating a  starburst of  massive, hot  OB stars during the past $\sim100$ Myr. Interestingly,  UV or \hal \, signatures of  very  recent  ($< 100$ Myr old) stars are absent closer to the nucleus, which shows only stellar populations  older than $\sim 1$ Gyr in bulge,  while most of the ongoing star formation  in  blue stellar nurseries is occurring in the outer disc (cf. Fig.~\ref{Fig1} here and \citet[][]{2023A&A...676A..35D}).   
This requires  some physical process which will  rapidly
shut down star formation in the central pseudo-bulge region. One may interpret  this as a consequence  of heating and   expulsion of the  star-forming  medium 
by the energetic, episodic feedback of the central AGN, as revealed  by two pairs of  radio jets and lobes. 

\begin{figure*}
\centering
\includegraphics[scale=0.37]{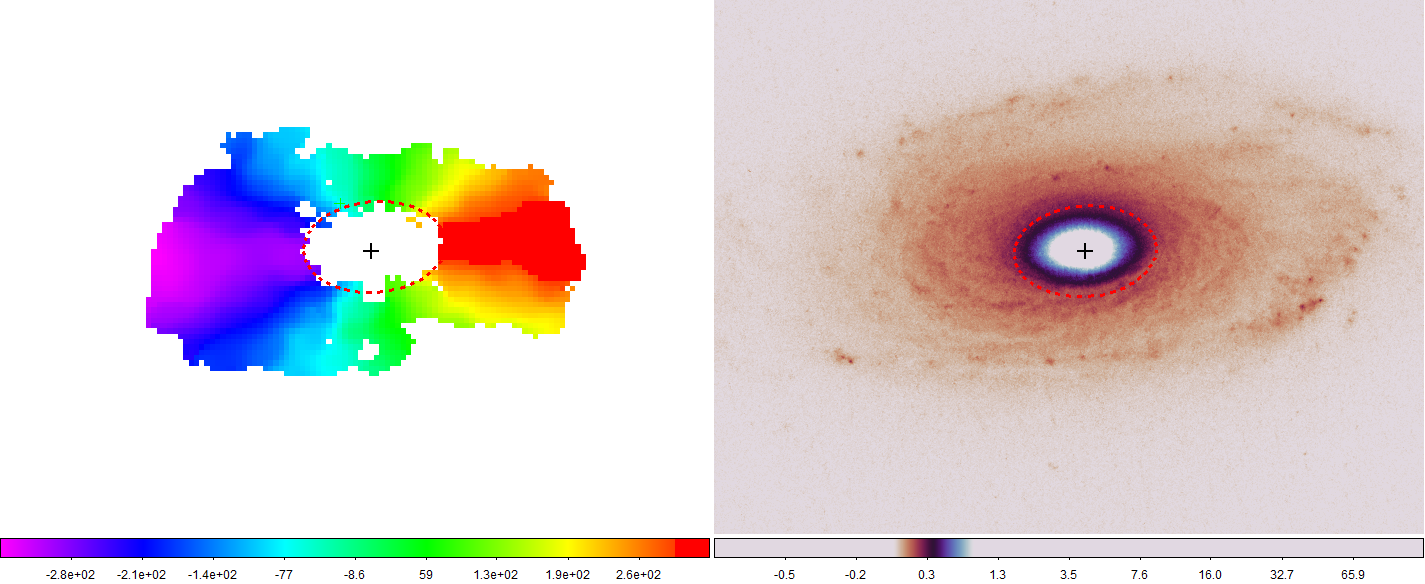}
\caption{ Left panel shows the gas kinematics map of $^{12}$CO(J = 1-0) molecular  line  transition   observed by ALMA in the
central region. The  velocity of  gas relative to kinematic center is represented by colors, ranging from violet/blue  for approaching and  orange/red for receding  velocity shown by the color bar at the bottom in km\,$\rm s^{-1}$. 
The kinematic center, located at a Right Ascension (R.A.
) of 23h 45m 32.708s and Declination (Dec.) of -04$^{o}$ 49$^{'}$
25.42$^{''}$, is represented by a  cross. This position is
$< 0.7$ arcsec from the compact 3.6 mJy/beam (about  $10 \sigma$ above noise) radio source detected in the VLA FIRST survey at 1.4 GHz as well as the
5 mJy/beam
point source on GMRT 323 MHz map.  No $^{12}$CO(J = 1-0) emission was detected in the central $2.8  \times 1.5 $ arcsec  from the nucleus, corresponding to a diameter of 4.2 kpc × 2.2 kpc (projected) above a flux limit of rms $= 332 \, \mu$Jy beam$^{-1}$. This CO
depleted  region is shown by a dotted ellipse superposed on the CO rotation map (left) and on the  HST I-band image (right).  The  diameter  of
CO depleted zone shown  is  about  2 times  the
effective diameter  of the stellar bulge in the HST  B and I band  images,  which is 
 $\rm 2 \times (2\, r_e) = 2 \times (2.2 - 2.6)$  kpc}.
\label{ALMA_velocity}
\end{figure*}

\subsection{Contribution of  the Active Galactic Nucleus (AGN) to SED and the Accretion State of the Black Hole}

For an accreting black hole the Eddington luminosity ($L_{\text{Edd}}$) is a critical limit for the luminosity that the AGN can  achieve before the radiation pressure from the emitted light becomes strong enough to counteract the gravitational force on the
infalling gas. It is given by the formula:

\begin{equation}
\label{eq:Edd_lumin}
L_{\text{Edd}} = \frac{4\pi G \, M_{\rm BH}\, M_{\rm p}\, c}{\sigma_{\rm T}} = 1.26 \times 10^{38} (\frac{M_{\rm BH}} {\Msun}) \,\, {\rm erg \, sec^{-1}} \,
\end{equation}

Here, 
$G$ is the gravitational constant,  $M_{\rm BH}$ 
 is the mass of the black hole, $M_{\rm p}$ 
 is the mass of proton, $c$ is the speed of light and $\sigma_{\rm T}$ is the Thomson scattering cross
 section.

 vigorous star formation produces significant UV emission, and  the  IR luminosity of dust is often interpreted as being directly proportional to the absorbed fraction of the energy from star formation. In addition, it is possible  that a   significant  radiative flux from the AGN  may  contribute to the SED as well as drive an outflow similar to QSOs. Active nuclei can  emit strongly in UV, and if embedded in  a dusty torus, may also  contribute to the  IR emission by  heating  dust in the torus and  clouds surrounding the AGN. In addition, 
an older stellar populations can also heat dust that is present in the  ISM of a galaxy, which can add to the total FIR emission. Simultaneous model fittings to the SED  provide  powerful insights into these processes.

In order to  investigate  all such  complex  astrophysical processes  we  used the {\fontfamily{qcr}\selectfont  CIGALE} V 0.9\footnote{https://cigale.lam.fr/}, a  Python-based SED fitting code \citep{2009A&A...507.1793N,2019A&A...622A.103B} to   constrain the   fractional contribution of 
AGN radiation to the total IR luminosity, represented  by  the   AGN-fraction parameter.  In addition,  CIGALE  also  simultaneously provides  several other parameters of  ISM dust and  star formation activity in the galaxy, similar to {\fontfamily{qcr}\selectfont MAGPHYS}. We have done SED fittings taking  fiducial  AGN-fractions   of  0 per cent, 
10 per cent  and  20 per cent. The AGN-fraction  is defined as the AGN  contribution to the total IR luminosity in $\sim 5 - 1000\, \mu m$ band.  In CIGALE, AGN-fraction explicitly takes into account three  emission components through a radiative transfer model; the primary source surrounded by a dusty torus, the scattered emission by dust, and the reprocessed thermal dust emission (more details can be seen in \citet[][]{2019A&A...622A.103B}). 

The results of  SED fitting with  varying  AGN contributions are shown in Table~\ref{Tab_SEDfit_cigale}. A  varying  degree of AGN  fraction  did not significantly (at  $> 99$ per cent confidence level) improve the reduced $\chi^{2}$  of the fit,  on the basis of a  Fisher-test. 
Therefore,  SED  fitting reveals  no evidence of  a (obscured or unobscured) radiatively efficient,  accretion  driven energy  source in the galaxy's SED.  Notably, the high excitation radio galaxies (HERGs) are fuelled at  high Eddington rate onto a  radiatively efficient standard accretion disc  by  cold gas, perhaps brought in through mergers and interactions, which is clearly absent in this galaxy. Rather, the AGN in J2345-0449  can be classified  as a  low-excitation radio loud AGN (LERG; e.g., \citet[][]{2012MNRAS.421.1569B,2014ARA&A..52..589H}), possibly  powered  by a radiatively inefficient, low Eddington ratio,  hot accretion flow of  gas from the  IGM  (know as the Bondi-Hoyle type accretion)  (\citet[][]{2014ARA&A..52..589H,2020NewAR..8801539H} and  references therein) on a super  massive black hole, which are  associated with AGN jets . In addition
In terms of  astro physical process  the gaseous fuel for the LERGs  is possibly derived from the hot X-ray halos surrounding massive galaxies, groups and clusters,  as part of a radio-AGN feedback loop.  Consistent with this  picture,  indeed  an extended  ($ \gtrsim 160$ kpc scale) soft X-ray  emitting halo of warm/hot baryonic matter is  robustly detected in this galaxy  \citep[][]{2015MNRAS.449.3527W,2021MNRAS.500.2503M}, whose implications are further discussed in next sections.  

Furthermore, in order to better constrain  the bolometric luminosity of the AGN we make use the first results from \citep{2014ApJ...788..174B} and  the   optical spectrum  taken on IUCAA 2m telescope of  central AGN region (Fig.~\ref{spectrum}), which
clearly falls into the LLAGN/LINER type with weak $H_{\alpha}$ emission of the nucleus \citep[][]{1998MNRAS.298.1035T,2012MNRAS.421.1569B,2010A&A...509A...6B}. From VLT/MUSE spectroscopy of J2345-0449 \citet[][]{2023A&A...676A..35D} also concluded the same.
The optical spectrum of J2345-0449 core is characteristic of an old stellar population.  The [OIII]$\lambda\lambda$4949 \AA + 5007 \AA\, emission  doublet  is not  detected  in  our  spectrum above  noise floor.  We use the empirical relationship of \citet[][]{2004ApJ...613..109H} to estimate the bolometric AGN luminosity limit by setting
$L_{\rm bol;AGN} = 3500 \times L_{\rm [OIII]\lambda5007}$. With the upper limit of  $L_{\rm [OIII]\lambda5007} < 1 \times 10^{40} \rm erg s^{-1}$  given by \citet{2014ApJ...788..174B}, this would correspond to  $L_{\rm bol;AGN} = 3.5 \times 10^{43} \rm erg s^{-1}$ and  an  Eddington ratio of $\log(\lambda_{\rm Edd}) = \log(L_{\rm bol;AGN}/L_{\rm Edd}) = - (2.6 - 3.6)$ for BH mass  range $\rm 10^{(8 - 9)} \msun$. 
\citet [][]{2023A&A...676A..35D} estimated $L_{\rm bol;AGN} = 3.3 \times 10^{43} \rm erg s^{-1}$ from [OIII]$\lambda\lambda$4949 \AA + 5007 \AA\ luminosity, quite close to our estimate.  However, both values should be considered upper limits because the fluxes are dominated by the  old stellar populations around the nucleus, which are the dominant gas heating source in the nuclear regions of J2345-0449 \citep[][]{2023A&A...676A..35D}.
In addition, the  absence of  broad  Balmer emission lines  in  our slit spectrum (mainly the $\rm H{\beta}$ line ) implies that J2345-0449 is not a Broad Lined Radio Galaxy (BLRG or a type-I AGN). Nevertheless, it is evidently  extremely radio loud for a  spiral  galaxy, which as a rule, do not  create  such large scale radio jets.  The integrated flux density and radio luminosity of J2345-0449 at 1.4 GHz are estimated at $S_{\rm 1.4} = 180.60 \pm 20.0$ mJy and $L_{\rm 1.4} = 2.5(\pm 0.3) \times 10^{31}$ erg s$^{-1}$ Hz$^{-1}$, respectively, which is typical of a low/medium powered  radio galaxy.

This  bolometric luminosity is also  indicative of  the  current rate of mass accretion  and  growth rate of the  black hole over time.  
The current rate of growth of the black  hole ($\zeta$) is governed by  the  mass accretion rate on  the  black hole and the  resulting  bolometric luminosity,

\begin{equation}\label{eq:bhar}
    {\frac{\zeta}{\rm M_{\odot} y^{-1}} = \frac {(1 - \epsilon) \, L_{\rm bol;AGN}}{\epsilon\, c^{2}}},
    \end{equation} 

where, c is the speed of the light and $\epsilon$ is the efficiency of the  accreted mass  to  radiative energy conversion (rest of the  accreted mass is swallowed by the BH). The value of $\epsilon$ is usually unknown, but some previous works  (e.g., \citet[][]{2020A&A...642A..65C,2021A&A...653A..32D}) find, $\epsilon = 0.1$,  which for J2345-0449  gives an upper limit to  mass accretion rate of $\zeta < 5.5 \times 10^{-3} \rm M_{\odot} yr^{-1}$, assuming lower limit of 
$L_{\rm bol;AGN}$ as above. At this fairly low accretion rate in the present epoch, assembling even  a lower mass  black hole of  $\rm 5 \times 10^{7} \msun$ would  take about a Hubble time, and much longer for a more massive black hole. This  implies that either the primordial `seed'  black hole was very massive initially  or the  mass  accretion rate was much higher  in the past than the  present  low rate, leading to  the  unusually rapid growth  of a SMBH in the galactic  nucleus, without necessitating major merger events.  

This is a promising  scenario for J2345-0449 because it's 
AGN  is presently in a low-excitation state (LERG/
LLAGN), showing extremely low radiative efficiency, but has a very
high kinetic luminosity in jetted outflows (a ratio  of Jet-power/bolometric luminosity $> 6$), strongly reminiscent of a hot Advection Dominated  Accretion  or the  Radiatively Ineficcient Accretion Flows (ADAF - RIAF,  \citet{1994ApJ...428L..13N},\citet{2014ARA&A..52..529Y}). In this accretion mode, because of rapid advection, compared to radiative cooling time, the radiative efficiency of accretion flow  is in  general lower than that of a standard thin accretion disc. This model also provides an important clue as to how the  radiatively inefficient, Mpc scale  kinetic  radio jets in J2345-0449  might be  launched  via a  low Eddington rate accretion state, as   observed in local  radio-loud galaxies  \citep[][]{2002ApJ...564..120H}.  Recently, in an extensive study of   Mpc scale  Giant Radio Galaxies (GRGs) \citet{2020A&A...642A.153D}  came to the conclusion that almost  every one of the GRGs in their sample is fuelled by an AGN  with low value of Eddington ratio ($\lambda_{\rm Edd} < 0.01$)   and    confirms to  the  low accretion rate, radiative inefficient ADAF/RIAF model. Interestingly, in a  parallel study  \citet[][]{2022A&A...660A..59M} showed that  average accretion rate of   Giant Radio Quasars (GRQs)  ($\lambda_{\rm Edd}$)  is roughly about 10  times higher than  Giant Radio Galaxies (radio jets of both  extend to similar Mpc scales, but GRQ jets have more kinetic power than GRGs), while their central black holes are  about equally  massive ($M_{\rm BH} \sim 10^{8}$ to  $10^{9}$ $\rm M_{\sun}$). This implies that powerful  Mpc scale radio jets  can be launched  even at  high rates of accretion  ($\lambda_{\rm Edd} \gtsim 0.1$) and  mass of the black hole is not  a  main deciding  factor. Recently \citet[][]{2024ApJ...963...91B} found strong evidence for jet activity in radio loud quasars triggered by
mergers. A corollary  of this result is that launch  of  radio jets   needs some  additional favorable or necessary conditions, like the  
 fast spin of BH  and interaction of the accretion disc with strong magnetic fields near the black hole,  possibly  launched via the  Blandford-Znajek  (BZ) or Blandford-Payne (BP) type mechanism \citep[][]{1977MNRAS.179..433B,1982MNRAS.198..345M,2013MNRAS.436.3741P}. In BZ model dynamically important horizon-scale magnetic fields near spinning Kerr black holes naturally produce powerful electromagnetic outflows \citep[][]{2021ApJ...910L..13E,2024ApJ...964L..25E} which may source the  Mpc scale jets.
These important studies provide  fundamental insight into the accretion disc-relativistic jet coupling process in  large dimension radio galaxies  and  quasars. 

In J2345-0449 the Inner lobes show FR-II  morphology, clearly being  fed by active jets, while outer lobes look peculiar; long, filamentous, and diffuse, 
possibly decaying relics of  previous jet activity that ceased millions of years ago. This makes the outer  lobes  excellent tracers of final stages  of the radio jet  activity (Fig.~\ref{GMRT}).
The kinetic power of radio jets is  estimated at $L_{\rm jet;kinetic} \approx 2 \times  \rm 10^{44} erg s^{-1}$ from GMRT 323 MHz  flux. 
As we have applied no correction for the (unknown)
energy loss in the  expansion of outer radio lobes, this is likely to be a lower limit. 
This jet power is much higher than $\rm L_{bol;AGN}$ and
therefore, any radiation driven feedback impact of AGN, for example a radiation pressure
driven outflow, akin to a quasar, would be at least about a factor 10 less  energetic than the kinetic power carried by the  radio jets.  

Thus, in this intriguing and rare spiral galaxy  we are witnessing a radio-mechanical 
(sometimes also called the maintenance-mode) energetic outflow in action wherein the AGN, despite its low Eddington  accretion rate 
and  low radiative output,  manages to produce large  mechanical energy  carried by the  energetic jets which may  couples efficiently  to the  circum-galactic atmospheric gas (e.g. \citet[][]{2002MNRAS.332..729C}).  In the absence of such  feedback heating, the hot X-ray emitting intra-cluster gas in  many clusters, groups and elliptical galaxies
would cool and rapidly form  stars, building much larger galaxies than are seen (see \citep[][]{2012ARA&A..50..455F,2012NJPh...14e5023M}).  Intriguingly,   J2345-0449  neither  originates in an elliptical galaxy (but in a spiral) nor it is embedded in a rich galaxy cluster or a group environment, and the working surface of the farthest 
 ends of radio jets  (the termination shocks) clearly extends well beyond the  galaxy's virial radius ($\sim 450$ kpc), they  do not  encounter a dense thermal plasma to inflate  X-ray dark non-thermal cavities or bubbles,  unlike  the cluster central
 BCGs of  elliptical morphology.   Nevertheless, the  radio-mechanical impact of  it's radio jets  on  the   circum-galactic ambient medium is inevitable, that  could  be  preventing  the  cooling  of its extended,  hot gaseous atmosphere strongly detected in deep X-ray observations \citep[][]{2021MNRAS.500.2503M}, thereby possibly  partially  quenching the new star formation in the  disc. As we show below, compared to the radiation pressure dominated quasar-mode feedback, the radio-mode mechanical feedback in this galaxy is possibly keeping it in the state of semi-quiescence  (in maintenance mode) rather than fully quenching  the star formation in the
 disc. Nevertheless, in early growth phase of the central supermassive BH,  when growing  at   super-Eddington/Eddington rate accretion, it may  have  exerted  strong  feedback affect via a hot wind on the molecular/atomic gas and   reduced the star formation rate on these very small scales. In  contrast,  UGC~12591 is one  such extreme mass, highly  bulge dominated  hybrid galaxy reported by \citet[][]{2022MNRAS.517...99R} which  has no  visible powerful radio jets observed  now, yet it has reached a  state of  virtual quiescence  showing extremely low star formation rate, which is  shown  plotted in  Fig.~\ref{fig:ssfr_mstar} along with J2345-0449 and  a many  other  galaxies of  various star formation states and  stellar mass ranges.

\subsection{An  Accreting  Supermassive Black Hole  Growing in a Bulgeless Environment}

At present very little is known of the growth and AGN activity of SMBHs in bulgeless spiral galaxies. 
Our  HST results above firmly establish  J2345-049 to be a  disc dominated spiral with unusual characteristics.  Despite lacking a classical bulge, this exceptional spiral  galaxy hosts a pseudo-bulge and  possibly a  massive, mass  accreting (radio-loud) central black hole of  minimal mass $M_{\rm BH} > 10^8$ $\rm \Msun$, which is  needed to launch the large  Mpc scale relativistic jets  at  low to medium   mass accretion rates, See   \citep[][]{2020A&A...642A.153D,2022A&A...660A..59M}. 


Here we  attempt to  estimate the mass of the central black hole using several established  scaling relations. Our present data do not provide strong constraints on $M_{\rm BH}$. One possible reason is  because  J2345-0449 lacks a classical bulge, whereas the black hole mass  is more  easily obtained from the tight $M_{\rm BH} - \sigma_{\star}$ correlation in bulge-dominated systems \citep[][]{2004ApJ...604L..89H, 2003ApJ...589L..21M,2009ApJ...698..198G}.  For this reason  the demographics of black holes in  pure disc galaxies are poorly determined.  However, there is  growing evidence that a  classical bulge is not essential for nurturing a massive black hole and some pseudo-bulge systems may also host fairly massive black holes of masses $\rm 10^{3} - 10^{7} M_{\odot}$ (\citet[][]{2011Natur.469..374K,2012A&A...540A..23S}). 
Therefore, these black holes must somehow attain masses substantially larger than the values implied by their bulge properties, possibly via a disc-driven, non-merger growth route. Possibly J2345-0449 is one such system, being an extreme member of such a population.  

The  (pseudo)bulge centric stellar velocity  dispersion value  on $\sim1$ arcsec scale  in J2345-0449 is $\sigma_{\star} = 351 \pm 25$ km s$^{-1}$ which  is much higher than normal Milky Way like spirals and close to the highest velocity dispersion found among nearby E and S0 galaxies, and is also significantly larger than that known for the majority of bulge-less disc galaxies on such a spatial scale (e.g., \citet[][]{2008MNRAS.386.2242H,2016ASSL..418..263G,2012A&A...540A..23S}), and compares well with that usually found for massive elliptical galaxies (e.g., \citep[][]{2004ApJ...604L..89H,2003ApJ...589L..21M,2009ApJ...698..198G}. This strongly hints at a large central concentration of mass, including a putative SMBH in this overall massive galaxy. Pseudo-bulges are products of secular evolution, where  instead of mergers, the internal dynamics of the galaxy drive the redistribution of material \citep[][]{2013seg..book....1K}. This evolution can strongly impact the feeding of the central black hole and influence its growth.

Using the dynamical mass of the (pseudo)-bulge  $\rm 1.07(\pm0.14) \times 10^{11} M{\sun}$, within one scale radius ($r_{e}$ = 1.25 kpc), we  calculate the  putative black hole mass $M_{\rm BH} = 2.54(\pm0.48) \times 10^{8} \rm \Msun$ using the black hole mass vs. bulge mass correlation found by Marconi and Hunt \citep[][]{2003ApJ...589L..21M}. This value is comparable to another
estimate of $M_{\rm BH} = 3.88(\pm 0.40) \times 10^{8} \rm \Msun$ obtained from the $M_{\rm BH} - \sigma_{\star}$ correlation of  \citet[][]{2008MNRAS.386.2242H}, applicable  for the pseudo bulges.  Another estimate using the Fundamental Plane (FP) relation of black hole activity  of \citep[][]{2009ApJ...706..404G} gives the black hole mass  $M_{\rm BH} = 5 (\pm0.5) \times 10^{8} \Msun$ using 5 GHz radio  luminosity of AGN $L_{\rm R} = 45 (\pm 5) \times 10^{38}$ erg sec$^{-1}$ and  2-10 keV X-ray luminosity $L_{\rm X} = 9.9 (\pm 1.5) \times 10^{40}$ erg sec$^{-1}$ \citep[][]{2014ApJ...788..174B}. However, if we use the  $M_{\rm BH} - \sigma_{\star}$  relation of \citet{2009ApJ...698..198G}, which is calibrated using a mixed sample of both ellipticals and spirals, with dynamically detected central black hole
masses and has an intrinsic rms scatter of $\epsilon = 0.44 \pm 0.06$, we obtain the black hole mass $M_{\rm BH} = 1.43 \times 10^{9} \Msun$.  Finally,  using K-band $M_{\rm BH} - L(\rm K)$ correlation of
\citet[][]{2007MNRAS.379..711G} we obtain $M_{\rm BH} = 1.22 \times 10^{9}  \Msun$. Both these latter BH masses are close to each other, yet are significantly larger than  the previous  mass  estimates. 
It is necessary to verify these rather large black hole mass values via more direct and robust methods. If confirmed it would imply that J2345-0449 hosts a SMBH as massive  as those found in giant elliptical
galaxies ($M_{\rm BH} \sim 10^{8 - 9} \Msun$), which is not expected at all for normal disc galaxies, and which will pose  a tension and raise various interesting questions about 
origin  and growth of  an accreting black hole in J2345-0449.

We note that some previous works have suggested (e.g., \citet[][]{2008MNRAS.386.2242H,2016ASSL..418..263G,2011MNRAS.412.2211G,2012A&A...540A..23S}) that pseudobulge and barred galaxies may host black holes that are about 3-10 times less massive than pure ellipticals  with the same $\sigma_{\star}$. It has also been proposed that their black hole masses do not correlate  well with the properties of galaxy discs or their pseudo bulges \citep[][]{2011Natur.469..374K}, and hence the growth process  of massive black holes in flattened galactic discs lacking prominent bulges is  still   puzzling and therefore a subject of ongoing deeper  investigations. Interestingly, 
\citet[][]{2013ARA&A..51..511K,2011Natur.469..374K} find that there is no convincing evidence of co-evolution of SMBH and the disc component  of  galaxies, i.e. feedback from BH accretion does not affect disc evolution and its stellar content. They  propose feedback from star formation is 
much more important in disc dominated systems than radio mode feedback. On the contrary, in J2345-0449, a predominantly disc dominated system, we find  that SFR is  very small (significantly quenched)  for its extremely large stellar and molecular gas masses,  and it clearly hosts a  super massive  black hole  launching  powerful radio jets, which clearly is located  in  the  nuclei  of a bulge-less  pure disc system,  contrary to the above idea.  This makes J2345-0449   a unique and exceptional  galactic  laboratory where  powerful AGN  energized radio jets seem to be directly  regulating the star formation activity in the surrounding  disc. 

\subsection{Molecular gas, star formation rate and comparison with other galaxies}

As many of the properties of this galaxy are rather extreme, like its huge mass, fast rotation speed and  unusual  `double-double'  episodic relativistic radio jets extending up to Mpc scale, it would be   very  informative to  investigate  their  relationship to the stellar and  baryonic mass  contents  of the galaxy,  and  contrast  that to  other local less massive galaxies.  The raw fuel for star formation in galaxies is the cold  molecular gas (e.g. \citet{2003RPPh...66.1651L}; \citet{Kennicutt98}), and to understand  galaxy evolution requires  understanding the various physical processes that are responsible  for the accretion and  depletion of  this gas in  different environments existing within the galaxy.  The  star-forming molecular clouds are predominantly made up of  molecular hydrogen  H2. As H2 lacks  a permanent dipole moment, therefore it is  difficult to directly observe at low gas temperatures. Instead   highly abundant CO is a good molecular gas tracer in the interstellar medium. The Doppler shift in the frequency of  any  such molecular line tracers, provides  immense information on  kinematics and gas flows. 

To compare the estimated SFR state of J2345-0449 with other disc galaxies we consider  the largest sample  of local SDSS galaxies on the  SFR - $M_{\star}$
plane \citep[][]{2004MNRAS.351.1151B}. The galaxies  that lie on the star formation main sequence (SFMS) are the active star forming (SF) ones, and those that have significant  offset from SFMS  are  the passive  or  quenched galaxies.  
To avoid the selection biases in the normalization, slope, and shape of the SFMS we use an   `objective definition' of the SFMS defined  by Renzini and Peng \citep[][]{2015ApJ...801L..29R} from SDSS galaxy sample \citep[][]{2004MNRAS.351.1151B}  that does not rely on a pre-selection of star forming or passive galaxies at all.  

The equation for  Renzini and Peng  derived best straight-line fit to the SFMS ridge line is given by (see Fig. 4 of \citet[][]{2015ApJ...801L..29R})

\begin{equation}\label{eq:sfms}
    \log(\frac{\rm SFR}{\rm M_{\odot} yr^{-1}}) = (0.76\pm 0.01) \, \log (\frac{M_{\star}}{\rm M_{\odot}}) - (7.64 \pm 0.02)
\end{equation}    
    
Therefore, for our estimated  $M_{\star} = 1.9 \, (\pm 0.5) \times 10^{11} \rm M_{\odot}$, the global SFR for J2345-0449, if it were to lie  on the $z = 0$ SFMS, should be  $\rm \sim 8.6 \, M_{\odot} yr^{-1}$ and $\rm sSFR = 4.5 \, \times 10^{-11} yr^{-1}$ as against estimated SFR $\rm (2.5 - 2.4) \Msun yr^{-1}$ and  sSFR of  $\rm (1.1 - 1.6) \times 10^{-11} yr^{-1}$. 
The range of values shown  above spans  between MAGPHYS and  CIGALE SED fitting estimates. Therefore we find here,  J2345-0449  shows a globally integrated SF quenching  by about a factor  4  when compared to the  $z = 0$  SFMS  of Renzini and Peng \citep[][]{2015ApJ...801L..29R}.  
We infer that although global SFR and sSFR for J2345-0449 both are significantly  below the  well established main  sequence, it's  star formation rate is  much above  the  fully quenched (spheroidal) galaxies, indicating a  drastic reduction, yet  not full termination  of  SFR. These kind of plots also reveal a  clear bimodality in  global and resolved realisations  of the star forming  properties of galaxies   around the
 main sequence \citep{2022A&A...659A.160B}. 

On  Fig.~\ref{fig:ssfr_mstar}  we see  the specific star formation rate (sSFR) of galaxies  versus $M_{\star}$.  When compared to $z = 0$ main sequence  of UV bright star forming galaxies of \citet{2007ApJS..173..441H}, J2345-0449  is placed  anomalously low in the region of  partially  quenched  `green valley'  galaxies. The graph is divided into three parts, showing actively star forming region ($\rm sSFR > 10^{-10.5} yr^{-1}$), green valley region ($\rm 10^{-11.5} < sSFR < 10^{-10.5} yr^{-1}$) and quiescent region ($\rm sSFR < 10^{-11.5} yr^{-1}$) (see \citealt{2016ApJS..227....2S}). Very interestingly, within the  'green valley' region we can also locate a  few  other most massive local galaxies (NGC 1030, NGC 1961, NGC 4501, NGC 5635 and NGC 266) of comparable high stellar
masses ($M_{\star} \gtrsim 10^{11} \rm M_{\odot}$). The position of J2345-0449 is  in stark contrast with the ultraviolet luminous galaxies  of \citet{2007ApJS..173..441H}  which show far more recent star forming activity for the same mass range. On the other hand, an  extraordinary massive and fast rotating  hybrid spiral  UGC~12591  is located in the lower-most region of highly quiescent 'fully quenched'  galaxies showing the lowest specific  sSFR of all \citep{2022MNRAS.517...99R}.  The fairly low sSFR of  all these very massive galaxies ( $M_{\star} >$ few\,$\times\, 10^{11} \rm M_{\odot}$) located in sparse environment  show  that  perhaps galaxy's high mass and  local  environment is    responsible to quench  star formation.  Presently very few  detailed studies of  cold molecular gas content of  galaxies falling    below the main sequence of star formation are available,  and a general  perception is that these systems are devoid of gas. However a detailed study of their molecular gas content is a key element for understanding the reason behind their remarkably low SFR.

What about local star formation rates  in the   disc of J2345-0449?
From resolved  $^{12}$CO(J = 1-0) line (rest frequency, 115.271 GHz) observations with  ALMA  \citet[][]{2021A&A...654A...8N,2023A&A...676A..35D} showed that J2345-0449 falls well below the standard Kennicutt-Schmidt relationship. The local star formation rates are  strikingly  lower by factors 50 to 75  from what is expected from the standard Kennicutt-Schmidt main-sequence relation \citep{Kennicutt98},  at the given gas mass surface densities, in  ordinary star-forming galaxies  with similar stellar and gas mass surface densities. Significant  offsets  towards lower  star-formation rate densities are also found from the Silk-Elmegreen law and the extended Schmidt law when taking into account the resolved stellar mass surface density in the disc \citep[][]{2021A&A...654A...8N}.
    

Previous  ALMA CO observations fully  resolve the molecular gas distribution  and kinematics, and thus it is able to distinguish between rotating discs, galaxy mergers, or gas inflows and outflows in J2345-0449.  $^{12}$CO(J = 1-0) line imaging and GALEX UV images both  show  a  low  rate star forming   disc  beneath the  disc-like pseudo-bulge. CO emission is sensitive to relatively low-density molecular gas, making it an excellent tracer of the bulk molecular material in star-forming regions. It is particularly useful for mapping the overall distribution and dynamics of molecular gas in galaxies. In J2345-0449
The  CO molecular gas is located in an annular ring as shown in Fig.~\ref{ALMA_velocity}. The dimensions of the molecular gas  ring are  $r_{\rm in} = 4.2$ kpc and $r_{\rm out} = 11.8$ kpc (projected) at the inner and outer (truncated) edge of the molecular disc, respectively, and  using the  deprojected rotation velocities, Keplerian values $v_{\rm rot,in} = 366$ km s$^{-1}$ and $v_{\rm rot,out} = 378$ km s$^{-1}$  are 
measured at these radii, respectively.
The kinematic center of CO molecular gas ring is located at position (J2000) RA= 23h45m32.708s, Dec = -04d49m25.42s, coincident with a faint ($\rm S_{115 GHz} = 1.3 \pm 0.34$ mJy) radio continuum point source, presumably the AGN. This position is also  within 0.5 arcsec of radio core detected on the GMRT 323 MHz and VLA FIRST 1.4 GHz maps. The position angle of CO  annular ring on the sky  is $94 \pm 2$ degree, and the inclination angle $i = 59 \pm 2$ degree, the same values previously also found for the stellar component by  \citet{2014ApJ...788..174B}. This is another indication that most of the molecular gas resides in the disc, and as measured from the CO line Doppler mapping, rotates largely unperturbed, in a highly  Keplerian fashion within the gravitational potential of the host galaxy. Moreover, no kinematic signature of rotation is detected along the minor axis, which is clear evidence for a planar, rotating molecular disc (see Fig.\ref{ALMA_velocity}).

Next we investigate the spatial correlation of  CO  emission ring and the central pseudo bulge component. As shown in Fig.~\ref{ALMA_velocity}, strikingly, no $^{12}$CO(J = 1-0) line emission is detected in the central $2.8 \times 1.5$ arcsec  from the nucleus, corresponding to a  diameter of 4.2 kpc × 2.2 kpc (projected) at a 
flux limit of rms $= 332 \, \mu$Jy beam$^{-1}$.  Summing  the CO line flux over all regions where CO line FWHM > 70 km s$^{-1}$   show that a total of $1.6 \times 10^{10} \Msun$ cold gas mass is being stirred up,  possibly by interactions with the compact radio nucleus (AGN) and inner jet, corresponding to about 75 per cent of the total molecular gas mass in the ring \citep{2021A&A...654A...8N}.  Here we find that  the spatial extent of CO  non-emission/depleted zone (Fig.\ref{ALMA_velocity})  is  about  2 times  the effective diameter  of the stellar bulge in the HST  B and I band  images,  which is $\rm 2 \times (2\, r_e) = 2 \times (2.2 - 2.6)$  kpc,  close  to the bulge size measured by \citet{2014ApJ...788..174B} from SDSS data. Therefore,  we  conclude that  the cold molecular gas is only associated with the inner disc of J2345-0449, while it clearly  avoids  the  central pseudo-bulge  where the   radio  AGN is located at  the very center. This reduced recent star formation in the innermost regions of the galaxy could be a manifestation of the   growth of the bulge itself. Alternatively this could be  a consequence  of  inside-out quenching of star formation as a result of supernova and/or AGN  feedback  which could  inject energy  to gravitationally unbind  star forming  gas from the galaxy's
center \citep[][]{2021MNRAS.508..219N};  a process most  effective  in  massive galaxies $M_{\star} > 10^{10.5} \Msun$ hosting SMBHs of masses $> 10^{8.2} \Msun$  accreting at low Eddington rates \citep{2020MNRAS.493.1888T}. In J2345-0449 we may be seeing   observational  signs of a similar AGN/SN feedback process at work,  but more detailed  investigations of cold molecular gas in future will provide a definite answer.

Recently \citet[][]{2020A&A...643A.111D}  reported   observations of  $^{12}$CO(J = 1-0) and  $^{12}$CO(J = 2-1)  transition emission  lines  from  the
repository of  cold molecular gas present in this and other giant radio galaxies (GRGs) using the IRAM 30m telescope.  Their global 
CO  abundance analysis  (single dish observations) suggest that  quenching does not require the total removal or  depletion of molecular gas, as many quenching models have previously assumed. Moreover,  cold molecular gas in the central   region   may  comprise the reservoir  for  fueling the central supermassive black hole. In J2345-0449 the molecular gas  (H2) mass was estimated at  $ M_{\rm H2} = 1.62 \times 10^{10} \Msun$, much smaller than the galaxy's  stellar mass, yet  is a significant repository of H2 which can  seed new stars.  On the other hand, the galaxy has a very long  global molecular gas  depletion  time   $t_{\rm dep} = M_{\rm H2 }/ \rm SFR  = 6.6$ Gy, for $M_{\rm H2} = 1.62 \times 10^{10} \Msun$ and $\rm SFR = 2.45 \Msun y^{-1} $. This seems an extreme value even for spirals, and it is the  highest of all  Giant Radio Galaxies studied so far \citep{2020A&A...643A.111D}.  But what is peculiar about  this  is  that  whereas  this galaxy contains a  massive reservoir of  cold molecular gas and dust ($M_{\rm dust} >  10^{8.1} \Msun$, Table~\ref{Tab_SEDfit_magphys}), it is  currently forming  new stars   at very  small rate. The exact cause  of  it  is still unclear but a good possibility is that this could be related to  suppression of  SF  due to strong  AGN feedback \citep{2021A&A...654A...8N}.

\subsection{The Baryon Landscape, Dark Matter Halo and the `Missing Baryon' Budget}

How did such ultra massive galaxies came to be? and what were the progenitors of these  giants  in the early universe? are the key open questions.  Recent detection of a hidden (in UV  to  near-IR bands) population of massive galaxies at $z >3$ detected by ALMA in sub-mm bands \citep[][]{2019Natur.572..211W} has provided new insights about origin of  these  elusive galaxies. Such dust laden galaxies contribute a total  star-formation-rate density ten times larger than that of equivalently  massive ultraviolet-bright  galaxies at $\it{z}$ > 3. Forming within the most massive dark matter haloes  of mass $\sim 10^{13} \Msun$, they  probably are the  progenitors of the present-day massive galaxies. Moreover  
JWST has recently found some extremely massive  (stellar mass $\rm \gtsim 10^{10} M_{\odot}$)  galaxies existing even at very early epochs ($z \sim 7 - 9$; \citet[][]{2023Natur.616..266L}). The infra-red and sub-mm detection of  an unexpected large population of  massive, evolved  galaxies in the early  Universe suggests that our present theories of  massive-galaxy formation may require substantial revision. With JWST \citet[][]{2023Natur.619..716C} found a similarly massive galaxy, which formed most of its stellar mass in  a remarkably
short interval of $\sim 200$ Myr before this galaxy quenched its star formation activity at $z \approx 6.5$, possibly via AGN feedback, when the Universe was about 800 million years old. They suggest that such galaxies are both  likely descendents of the highest-z sub-mm galaxies and quasars, and  probably the progenitors for the dense, ancient cores of the most massive local galaxies.

Much can be learned   by  exploring  the warm/hot gaseous coronae around massive  disc galaxies lacking  prominent bulges. Contrary to the bulged systems these galaxies are likely to be  much more direct tracers of the galaxy formation process governed  by  `secular' (isolated) disc based mechanisms, since they have not undergone  major mergers \citep{2004ARA&A..42..603K} which drastically alters them. On the other hand, due to their extreme stellar and halo mass and  possible  AGN feedback of  the central SMBHs the  baryons in these  galaxies may reside in  unusual  physical states  that are strongly  affected by these processes and may  participate in  much less star formation in the disc. Moreover, since the hot halo mass and  thermal X-ray emission of baryons both  scales with galaxy mass, it is highly informative  to   observe  deeply the  most massive spiral galaxies  in soft X-rays  to  achieve a detectable signal and make a good study.

Indeed, in J2345-0449  {\it XMM-Newton}  has robustly detected  a giant  halo of diffuse, soft  X-ray emission gas ($T < 1$ KeV)  which surrounds  the  galaxy and extends out to a  minimum radius of  160 kpc \citep[][]{2021MNRAS.500.2503M}, which is roughly 35 per cent of the virial radius  and $\sim20$ times the exponential disc half light radius. {\it Chandra} observation also  has  found the halo, but   less  extended  due to reduced  detection sensitivity of the telescope \citep[][]{2015MNRAS.449.3527W}.  The baryonic mass of the hot gas in the halo is very high,  about 3  times the total stellar mass of the galaxy (Table~\ref{tab:baryon_budget}).  This  remarkable  detection is  currently  limited  by the detection capability of X-ray telescopes, yet it is   the largest  repository of a  hot baryonic gaseous atmosphere observed in  any single galaxy so far.   Interestingly,  the X-ray emitting  gaseous halo is  not spherically  distributed but  forms  a  flattened  pancake  elongated parallel to the host galaxy's stellar disc \citep[][]{2021MNRAS.500.2503M}.

What could be the reason for this  elongated shape?  The  3-dimensional shape of dark matter halos around galaxies  and  galaxy groups or clusters  is  a  frontier topic of ongoing research, but  theoretical models (simulations; e.g. \citet[][]{2019MNRAS.484..476C,2021ApJ...913...36E,2023ApJ...958...44B}) and limited observations suggest \citep[][]{2002ApJ...577..183B} that they are not  perfectly  spherical  but  assume  triaxial spheroidal geometry,  with long axis  preferably oriented parallel to  longer axis of  the  central  galaxy or group.  This is consequence of the fact that the gravitational force that shapes dark matter halos are not perfectly  symmetrical, which governs their 3-dimensional  equilibrium figures in a rotating frame,  much  like a Jacobi ellipsoid. A Jacobi ellipsoid is a  triaxial ellipsoid that is in hydrostatic equilibrium under the influence of its own gravity and  rotation. This means that the pressure inside the ellipsoid is everywhere equal to the gravitational force acting on a small volume of fluid \citep[][]{1969efe..book.....C}.  Since the  X-ray halo  gas is essentially collisionless and likely to be  bound in the  gravitational potential of dark matter, it too  will  acquire a similar triaxial shape,  if not disturbed by major mergers and tidal forces. This is a simple model but this makes the X-ray halos  around  massive galaxies remarkably accurate tracers of dark matter  as  the DM halo shape is  a  result  of  initial density field anisotropy  and   the subsequent halo assembly history  (cold  filament  accretion and  mergers), and   the  galaxy halo shape statistics can be used to test and validate the $\Lambda$CDM structure formation paradigm. Future deep X-ray observations and theoretical  modelling  will  reveal if the   X-ray halo around J2345-0449 could be shaped by this mechanism,  directly related to its  currently unknown formation history.

Alternatively,  \citet[][]{2015MNRAS.449.3527W}  suggest that  pancake like shape of  the X-ray halo in J2345-049 possibly is due to  feedback by the energetic radio  jets which might  have disrupted or expelled hot gas  from an extended  region  ($\approx 100$ kpc)  above and below  the disc plane.  Although   still  unproven, If  confirmed by more  observations  and detailed  modelling,  this   will   be  an  extremely  unusual  scenario for a  spiral galaxy environment as  radio jet-mode feedback is   commonly seen in  AGN  hosted by  Brightest Cluster Galaxies (BCG) at the centers of  galaxy clusters  with  a cooling core, but  is extremely  rare to observe in an  isolated spiral galaxy like  J2345-0449  \citep[][]{2012NJPh...14e5023M,2012ARA&A..50..455F}.

A third  possibility is that the hot X-ray halo gas could have been  lifted up and  heated up by hot stellar winds   ejected by  multiple supernovae from high SFR in the past,  which has now  got suppressed due to  either  the  AGN feedback  or a morphological quenching by the massive DM halo \citep[][]{2009ApJ...707..250M}.  Such a massive DM halo  is also  needed to  stabilize the  extremely  fast rotating stellar disc. On the other hand, we find that a massive  stellar  spheroid/bulge formed by  mergers, which is most effective in morphological quenching, is  definitely  not present in J2345-0449, but   an  extended ($ > 160$ kpc radius), hot X-ray halo is  observable \citep[][]{2015MNRAS.449.3527W,2021MNRAS.500.2503M}.

\begin{figure*}
    \centering
    \includegraphics[width=0.8\textwidth]{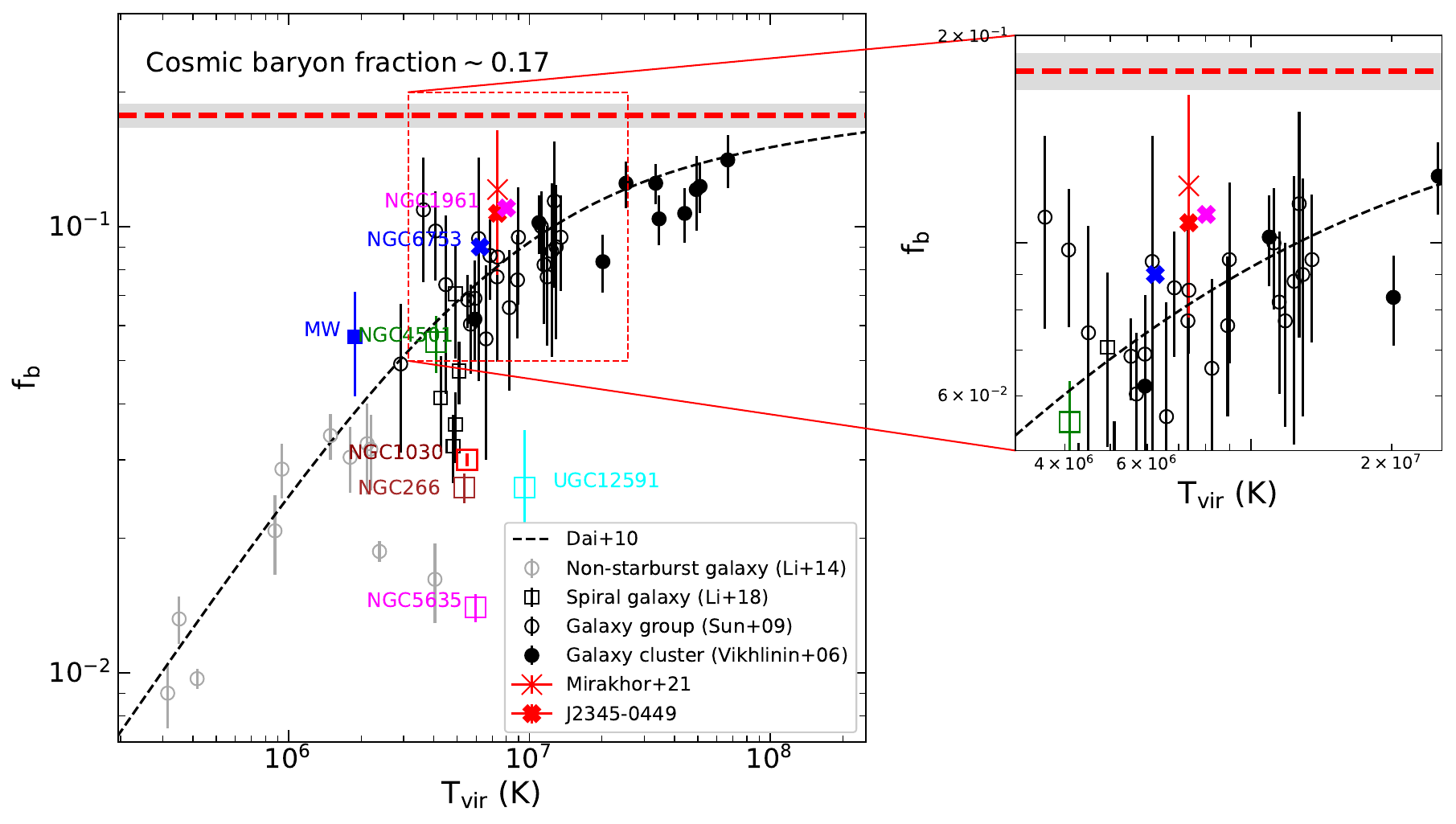}
    \caption{This plot shows the baryon fraction in relation to baryonic mass of different populations of galaxies, galaxy groups and galaxy clusters with the virial temperature $\rm T_{vir}\, (K)$  as the dependent variable. Non-starburst field galaxies are selected from \citet{2014MNRAS.440..859L}, spirals from    
    \citet{2018ApJ...855L..24L}, galaxy groups from \citet{2009ApJ...693.1142S} and galaxy clusters from \citet{2006ApJ...640..691V}. On top of these, we plot a handful of fast-rotating massive galaxies from \citet[]{2013ApJ...772...97B}{}, \citet{2016MNRAS.455..227A}, \citet[]{2022MNRAS.517...99R} and 
      \citet[][]{2023MNRAS.tmp.3539R}. The Milky Way galaxy \citep{2015ApJ...800...14M} is also shown for reference. The dotted line indicates the broken  empirical power-law model of \citet{2010ApJ...719..119D} which is given as $\rm f_{b} = [0.14(v_{rot}/440)]^{1.6}/[(1 + (v_{rot}/440)^{2})]^{1.5/2}$. For J2345-0449 we have
    used the baryon masses given in Table~\ref{tab:baryon_budget}. 
        The  cosmological  baryon fraction is shown with a shaded grey region of $\rm f_{b} = 0.17 - 0.19$ reflecting the uncertainty range of current
    measurements.} 
    \label{fig:fb_Vrot}
\end{figure*}

It would prove much informative if  the X-ray  halo  in J2345-0449 is  the  signpost of  initially  cold  gas  that has  fallen into the halo of the massive  galaxy,  and has  been shock heated and  virialized during galaxy structure formation. This can  be  shown   as follows:

(1) The gas  temperature of the  X-ray  halo is at the  right order of magnitude for being near to the virial temperature of the binding gravitational potential: 

\begin{equation}\label{eq:virtemp}
T_{\rm vir} =  4  \times  10^{5} \times \left(\frac{v_{\rm circ}}{100 \, \rm km \,s^{-1}}\right )^{2}\rm K,   
\end{equation}

which comes to $6.4 \times 10^{6}$ K (or 0.55 keV) for J2345-0449 for  a circular rotation speed of  $v_{\rm circ} \sim 400$ $\rm \kms$.  This  is about the same temperature as what  the  Chandra and XMM has measured in the hot halo region \citep[][]{2015MNRAS.449.3527W,2021MNRAS.500.2503M}. 

(2) If not a  coincidence, this implies that gas in
X-ray emitting halo  is in viral equilibrium  with a  massive dark matter halo of total mass $\rm few \times10^{13}\Msun$  as  suggested by  \citet{2012ApJ...755..107D}. This in turn implies the halo gas is possibly  shock heated after having been accreted  from the cold filaments of cosmic-web 
 beyond the  virial radius at $\gtsim 500$ kpc, rather than being recycled from the  disc. The thermal and metal line cooling time of the quasi-hydrostatic X-ray halo gas is  large,  in the range  $2.8 - 6.3$ Gyr for material within the cooling radius, and has the effective cooling rate of $\rm 0.15 - 0.21 \Msun\, yr^{-1}$ \citep[][]{2006ApJ...639..590F,2012ApJ...755..107D,2021MNRAS.502.2934K}. The average estimated cooling time is  much larger than the free-fall time, and therefore, it would require more than a Hubble time to build the entire  stellar mass of galaxy by condensation of cooling gas from halo, thus disfavoring such a possibility.  The  dynamical  state of the  circum-galactic  gas would be revealed  in future sensitive spectroscopic observations from the detection of the `Warm Hot Intergalactic Medium (WHIM)' in the soft X-ray lines at $\sim 10^{6}$ K, and via the  detection of  even cooler gas through the UV/EUV lines of  metals \citep[][] {2021ExA....51.1043S,2021ExA....51.1013N}.

The XMM-Newton detection of an extended, luminous, soft  Xray halo around the galaxy J2345-0449   is  a    prominent sign-post of  so called   `missing baryons' in  such massive galactic halos.  All prominent galaxy formation models predict such gaseous circum-galactic halos \citep[][]{1978MNRAS.183..341W,1991ApJ...379...52W,2006ApJ...639..590F,2006ApJ...644L...1S}.
The entire baryonic mass contained in the  X-ray halo  of J2345-0449 implies a baryon fraction near the cosmic average $\sim 0.15 - 0.17$ (\citet[][]{2009ApJS..180..330K,2009ApJS..180..306D}), which is very high for a  galaxy. 
In J2345-0449 the baryon to  DM mass density ratio appears to approach the cosmic value, if the baryonic mass is  extended upto the virial radius where the dark matter mass dominates.  Such a  massive  DM  halo  is  also needed  to gravitationally  stabilize the extremely fast rotating disc.  This  finding  does   not support  the possibility that in most massive galaxies in universe the  baryon shortfall is  due to the fact  that majority of  cosmic baryons never `condensed ' onto the  DM halo and thus  are invisible. On the contrary,  in J2345-0449 we find that  these    baryons actually  may  reside  in  the  circum-galactic halo   and thus it is a  very important  new result.

In Table~\ref{tab:baryon_budget} we have  shown  the  masses of different
baryonic components in the galaxy disc, which   are; the stellar mass, 
mass of the cold molecular (H2 and He) gas, and the interstellar dust mass. As typically done in many studies, we assume that the total disc gas mass $M_{\rm gas}$ is the sum of the atomic and molecular gas masses, corrected for the Helium contribution by multiplying by a factor of 1.36, i.e. $M_{\rm gas} = 1.36 \times [M_{\rm H2} + M_{\rm HI}]$. We have neglected atomic hydrogen mass M(HI), as it is not measured  and usually  is  a  small  fraction of total gas mass. In addition  we also list the  estimated  mass of hot x-ray halo gas, as detected  in XMM-Newton observations and the  model fitted 
total halo mass (dark matter plus baryons)  upto the virial radius, both as determined  by \citet{2021MNRAS.500.2503M}. The 

Combining all the baryon masses, we  have  determined a  total  baryon fraction of  $f_{\rm b}  =  0.11^{+0.04}_{-0.04}$, close
to the estimate $f_{\rm b} =  0.121^{+0.043}_{-0.043}$ reported by \citet{2021MNRAS.500.2503M}. Both these estimates are  close but  lower than  cosmological baryon fraction of  $0.15 - 0.17$ \citep[][]{2009ApJS..180..330K,2009ApJS..180..306D}. This is shown in Fig.\ref{fig:fb_Vrot} where  galaxy J2345-0449 with a baryon fraction of $\rm \sim  0.11^{+0.04}_{-0.04}$ (within the error bar of \citealt{2021MNRAS.500.2503M} who provided $f_{\rm b}$ as $\rm \sim 0.121\pm0.043$ upto the virial radius), falls above the  nearby region of the broken power-law  model  surrounded by a few other highly massive spiral galaxies and  galaxy groups. Note that, amongst the labelled massive spiral galaxies in the graph, we have  X-ray detection in the galactic halo for the objects J2345-0449, NGC~1961, NGC~6753, NGC~266 and UGC~12591 and  in effect  their hot gas mass upto the virial radius are also included in the total  baryon  fraction.

To further explore implications of the above  findings, we express the  
star formation efficiency of a galaxy,   $f_{\rm \star}$,  which can be expressed by \citep[][]{2019A&A...626A..56P},

\begin{equation} \label{eq:fstar}
    f_{\rm \star} = \frac{M_{\rm \star}}{M_{\rm halo}}~\frac{\Omega_{m}}{\Omega_{b}},
\end{equation}


with $f_{\star} = 1$ if all  baryons are converted to stars.
Now for J2345-0449 if we take  virial  mass $M_{\rm halo} = 1.07 \times 10^{13} \Msun$  as  calculated in \citet[][]{2015MNRAS.449.3527W,2021MNRAS.500.2503M} and  our estimated mean stellar mass $M_{\star} \approx (2.2\pm0.1)\times10^{11} \Msun$, we obtain $f_{\star} \sim 0.12\pm0.01$,  taking the universal   cosmic baryon fraction $\frac{\Omega_{b}}{\Omega_{m}} = 0.174$.  This  shows that  J2345-0449 is able to convert  only about $\rm 10$ per cent of  all  available  cosmic baryons to stars.  Another indicator of remarkably low efficiency of star formation in this galaxy is it's very large gas depletion time, given by  $t_{\rm dep} = M_{\rm gas}/ \rm SFR  \sim 10$ Gy, taking $M_{\rm gas} = 2.2 \times 10^{10} \Msun$ and $\rm SFR \sim  2 \, M_{\odot} yr^{-1} $. On the  contrary, the sample of most massive spirals with $M_{\star} = 1-3\times10^{11} \Msun$ of \citet{2019A&A...626A..56P} yields $f_{\star} \approx 0.3-1$ which shows their  larger  efficiency of star formation and their stellar baryon fraction is approaching the cosmological baryon fraction.  Although a very interesting result,  it is still too early to tell if these unusual  massive   spirals of   \citet{2019A&A...626A..56P} have  experienced the same evolutionary and star formation history  as the majority of normal spirals. On the contrary, J2345-0449  appears to be very  different   because of it's   very low $f_{\star}$, a large repository of  potentially  viable molecular gas,  and    because   most of the   `missing'  cosmic baryons  in this galaxy  appear to reside  within  the  dark matter halo inside the  virial radius in a  phase  of  warm/hot low  temperature ($\leq 1$ keV) gas,  but   unable to   turn  into  stars. 

These baryons may enable galaxies  to continue forming stars steadily for long periods of time and account for the elusive `missing  baryons' in galaxies in the local universe.  For some of the  most massive galaxies in local Universe,  like J2345-0449, UGC 12591, NGC 1961 etc.  we find just the opposite. This finding is broadly consistent with the constraints from  abundance matching techniques to
determine the typical galaxy stellar mass at a given halo mass. The simulations show that  peak efficiency of $\rm f_{\star}$ is always fairly quite low,  by $\rm {\it z} = 0$ reaching a peak  of  about $\rm  20 - 30$ per cent for  typical $L_{\star}$ galaxies  with halo mass $M_{\rm 200} \sim 10^{12} \Msun$. Beyond the peak $f_{\star}$ declines rapidly to  $f_{\star} \sim 2 - 5$ per cent for ultra high mass $M_{\rm 200} \sim 10^{13} \Msun$ \citep[][]{2010ApJ...717..379B,2019MNRAS.488.3143B,2013MNRAS.428.3121M}.  It is believed that AGN  feedback is one of the main ingredients required to bring this  decline in SFR, e.g. \citep[][]{2005Natur.433..604D,2017MNRAS.472..949B} and a cut off in the total stellar mass.

Next we  compare   the 0.5-2 keV band X-ray luminosity ($L_{\rm x}$) of all known   massive spiral  galaxies with  detected soft
X-ray halos.  We find that at $L_{\rm x} = 4.0\pm0.5 \times 10^{41}$ $\rm erg \, s^{-1}$ the  luminosity of  J2345-0449 is highest \citep[][]{2021MNRAS.500.2503M}, about 10 times that of  UGC~12591 \citep[][]{2012ApJ...755..107D}, NGC 6753 \citep[][]{2018ApJ...862....3B,2013ApJ...772...97B} and  significantly
higher than that of NGC 1961 \citep[][]{2016MNRAS.455..227A}.  The metal abundance of J2345-0449 halo is  low at $\rm \sim 0.1 Z_{\sun}$,  similar to  those found for UGC~12591, NGC~1961 and NGC 6753. This  metallicity is  consistent with baryon deficit,  low  metallicity  coronal gas  surrounding in the halo of  massive  galaxies shown in  EAGLE hydro simulations \citep[][]{2021MNRAS.502.2934K,2015MNRAS.446..521S}.  Although the
halo mass of UGC~12591, NGC~1961, NGC~6753 and J2345-0449 are similar, the higher X-ray luminosity and the largest  extent of J2345-0449 corona  (about 160 kpc, or 35 per cent of $\rm r_{200}$) is very  intriguing  which needs  better understanding. We point out  that J2345-0449  is the only one of those four galaxies  which hosts   large scale   radio jets. Therefore,  one may infer  it  is a consequence  of   powerful  Mpc scale radio jets and past supernovae  activity in this galaxy,  heating   and  expelling  the inner baryons upto a larger radius and thus preventing  gas cooling in the halo; although  a very  suggestive possibility,  more observational proof  is needed to robustly demonstrate  this scenario.



\begin{table}
    \centering
    \begin{tabular}{llr}
    \hline
    Parameter & Symbol & Value (in \Msun)\\
    \hline
   Stellar mass (this work) & $M_{\star} $ & $\rm 2.88^{+0.00}_{-0.79}\times 10^{11}$ \\
   Stellar mass (\citet{2015MNRAS.449.3527W})  & $M_{\star} $ & $\rm 4.60^{+0.1}_{-0.1}\times 10^{11}$ \\
    Molecular hydrogen mass & $M_{\rm H2} $& $\rm 1.62\times10^{10}$ \\
    Total gas mass (Including He)  & $M_{\rm gas} $& $2.20  \times 10^{10} $ \\
    Dust mass (this work) & $M_{\rm dust} $ & $\rm 1.21^{+0.44}_{0.21}\times10^{8}$ \\
    Hot halo gas mass upto 160 kpc radius & $M_{\rm hot,gas}^{160} $ & $\rm 1.15^{+0.22}_{-0.24}\times10^{11}$ \\
    Hot halo gas mass upto virial radius  & $M_{\rm hot,gas}^{r200} $ & $\rm 8.25^{+1.62}_{-1.77}\times10^{11}$ \\
    Total baryonic mass upto virial radius  & $M_{\rm baryon}^{r200} $ & $\rm 1.13^{+xx}_{-xx}\times10^{12}$ \\
    Total halo mass (dark matter + baryons) & $M_{\rm DM+bar} $ & $\rm 1.07^{+0.09}_{-0.09}\times10^{13}$ \\
    upto virial radius (450 kpc) &  &  \\
    Baryon fraction  (this work) & $f_{\rm b} $ & $\rm 0.11^{+0.04}_{-0.04}$ \\
    Baryon fraction  (\citet{2021MNRAS.500.2503M}) & $f_{\rm b}$ & $\rm 0.121^{+0.043}_{-0.043}$ \\
    \hline
    \end{tabular}
    \caption{In this table we specify the  mass of various baryonic components  and the dark matter mass of our target galaxy J2345-0449. The stellar mass and dust mass are from this work, second estimate of stellar mass is from \citet[][]{2015MNRAS.449.3527W}, molecular hydrogen mass is provided by \citet{2020A&A...643A.111D}, and  the hot X-ray halo gas mass and the  total halo mass (dark matter + baryons) is provided by \citet{2021MNRAS.500.2503M}. The estimated total halo mass within $r_{200}$,  $M_{\rm DM+bar} $ is  affected by systematic and statistical uncertainties, originating mainly from the unexplored properties  of the hot halo  beyond a radius of  160 kpc (see \citet{2021MNRAS.500.2503M}).}
    \label{tab:baryon_budget}
\end{table}

\section{Summary and Conclusions}

In the present work our research endeavour was directed  on the bulge-disc components and  stellar mass  distribution in the fast rotating, highly massive spiral galaxy J2345-0449 showing unusual radio jets
extending up to  Mpc scale.  Using the high-resolution multi wavelength  HST observations and  multi-parameter panchromatic  SED fitting  we   obtained  estimates of  star formation rate, the total baryonic  mass contained in stars and  derived  the warm dust properties, which were then connected to  the  cold molecular gas  distribution  based  on   previous ALMA and  IRAM  Carbon Monoxide spectral line observations.

 We confirm   down to $\sim100$ pc scale  spatial resolution, that this spiral galaxy features a pseudo-bulge instead of  a classical bulge. Search for non-axisymmetric structures in the AGN region revealed
a small nuclear bar 
surrounded by a bright ring, forming a circumnuclear disc.
This is consistent with the expectation as pseudo-bulges evolve from bars secularly. 
These have profound implications  for how this galaxy  formed and  how   the growth of  it's  mass  accreting  SMBH  happened in the past. 
Classical bulges form via more violent processes  such as hierarchical clustering, whereas the pseudo-bulges, seem to be still evolving, more likely as a result of their host galaxies’ slow internal evolution, which involves disc instabilities  and angular momentum
transfer. One important conclusion  from this study  is that the galactic disc and its SMBH  in J2345-0449  may  have
evolved together quietly (seculary), and  have not experienced the upheaval of some  recent  major  merger  events.  This suggestion is strengthened  by the  absence of any tidal debris (such as tails, shells, or faint plumes), the location of the galaxy in an isolated  galactic  environment having no nearby  massive galactic neighbors, and by the fact that the galaxy has highly-symmetric spiral arms  within a rotationally  supported stellar disc and  an  inner, highly  organized  Keplerian disc of  cold molecular gas.

Our detailed multi wavelength study  of  potential  star forming  gas reveal that,  while  hot gas cools down in  the massive  extended halo of this galaxy,  new  stars do not form in the center because of  possible  feedback from the central accreting supermassive black hole with powerful radio jets, which heats up the gas again or pushes it beyond the galaxy's reach. For most of the galaxies like J2345-0449 and a few others observed, the activity of the black hole seems to have put an  end  or drastically reduced  star formation but has not yet succeeded in  clearing them of all their cold gas.

Much more detailed followup work is required  and   will be taken up 
in  later  studies. The future studies will be directed to  understanding better how have such extremely massive, rotational supported disc  galaxies have
formed and are evolving, as it  depends on non-linear baryonic physics and complex interaction with the dark matter halos. Obviously, the most striking property that sets J2345-0449 apart from nearly all 
other spiral galaxies  is its large mass and unusual radio loudness, resulting in highly efficient ejection of relativistic jets on Mega parsec scales, possibly resulting from an unusual growth of a supermassive black hole in a bulge-less galaxy. The extreme rarity of such galaxies implies that whatever physical process had created such huge radio jets in  J2345-0449 must be very difficult to realize and maintain for long periods of time in most other spiral/disc galaxies. Thus, an important question is; What could be responsible for the efficient fueling and sustained  collimated  jet ejection activity in J2345-0449? The answer requires very detailed knowledge of the properties of its central 
black hole engine, i.e., its mass, spin,  and the nature of its accretion flow, namely the mass accretion 
rate and the magnetic field topology of the inner  regions of accretion disc, most of which is clearly lacking at the present moment.

In summary, our study sheds light on the unique characteristics of J2345-0449 and also raises new questions about the interplay between galactic structure, star formation, and the activity of the central supermassive black hole. Future investigations will likely focus on obtaining more detailed information to unravel the complexities of this intriguing galactic system.

\begin{table*}
    \centering
    \begin{tabular}{llrrrcc}
    \hline
     &  & F160W (H-band) & F438W (B-band) & F814W (I-band) & \\
     \hline
        {\bf Model A} \\ 
        (\sersic\ + Exponential disc)  					 & \sersic\,mag          & $13.92^{+0.05}_{-0.05}$ & $18.95^{+0.07}_{-0.07}$  & $16.04^{+0.08}_{-0.08}$ & \\
         					 & \sersic\,\reff        & $2.50^{+0.04}_{-0.04}$  & $1.07^{+0.03}_{-0.03}$   & $1.35^{+0.08}_{-0.08}$ & \\
         					 & \sersic\, n             & $2.69^{+0.03}_{-0.03}$  & $1.11^{+0.01}_{-0.01}$   & $1.67^{+0.06}_{-0.06}$ & \\
         					 & \sersic \, \axisratio    & $0.61^{+0.02}_{-0.01}$  & $0.61^{+0.05}_{-0.05}$   & $0.60^{+0.02}_{-0.01}$ & \\
							 & \sersic\,pa	          & $89.66^{+0.02}_{-0.02}$  &	$87.75^{+0.02}_{-0.02}$ & $90.87^{+0.02}_{-0.02}$	 & \\         					 
         					 & Expdisc mag           & $13.16^{+0.08}_{-0.08}$   & $16.44^{+0.10}_{-0.13}$  & $14.67^{+0.08}_{-0.07}$& \\
        					 & Expdisc scale radius ${r_{s}}$ (kpc)& $6.94^{+0.07}_{-0.07}$  & $8.19^{+0.03}_{-0.03}$   & $6.86^{+0.18}_{-0.13}$ & \\
         					 & Expdisc  \axisratio   & $0.52^{+0.02}_{-0.01}$     & $0.61^{+0.05}_{-0.05}$    & $0.60^{+0.02}_{-0.01}$ & \\
							 & Expdisc  pa	         & $96.03^{+0.03}_{-0.03}$    & $94.58^{+0.02}_{-0.02}$  & $95.03^{+0.06}_{-0.06}$	 & \\ 
         					 & Reduced $\chi^{2}$	         & 0.76                      & 1.34				  & 1.89					 &	\\
\hline\\
 	{\bf Model B} \\
 	(\sersic\ + Exponential disc + PSF)						 & \sersic\,mag & $14.28^{+0.05}_{-0.05}$ & $18.97^{+0.09}_{-0.09}$  & $16.09^{+0.03}_{-0.03}$& \\
         					 & \sersic\,\reff& $1.37^{+0.03}_{-0.02}$  & $1.08^{+0.08}_{-0.06}$   & $1.32^{+0.08}_{-0.08}$ & \\
							 & \sersic\ n   & $1.62^{+0.02}_{-0.03}$  & $1.03^{+0.01}_{-0.01}$   & $1.49^{+0.08}_{-0.08}$ &  \\
							 & \sersic\ \axisratio   & $0.61^{+0.02}_{-0.02}$  & $0.60^{+0.07}_{-0.07}$   & $0.60^{+0.01}_{-0.01}$ &  \\
							 & \sersic\ pa  & $89.71^{+0.02}_{-0.02}$  & $88.71^{+0.10}_{-0.10}$  & $90.93^{+0.03}_{-0.03}$& \\ 
         					 & Expdisc mag  & $13.05^{+0.07}_{-0.07}$ & $16.44^{+0.04}_{-0.04}$  & $14.66^{+0.03}_{-0.03}$& \\
        					 & Expdisc scale radius ${r_{s}}$ (kpc) & $6.71^{+0.04}_{-0.04}$  & $8.18^{+0.04}_{-0.04}$   & $6.77^{+0.07}_{-0.07}$ & \\ 
        					 & Expdisc \axisratio   & $0.53^{+0.01}_{-0.01}$   & $0.53^{+0.36}_{-0.03}$   & $0.54^{+0.01}_{-0.01}$ &  \\
							 & Expdisc pa           & $95.34^{+0.06}_{-0.06}$  & $95.03^{+0.06}_{-0.06}$  & $94.93^{+0.02}_{-0.02}$& \\         					 
         					 & PSF mag              & $18.22^{+0.01}_{-0.01}$ & $24.33^{+0.05}_{-0.05}$  & $21.25^{+0.01}_{-0.01}$& \\
         					 & Reduced $\chi^{2}$	       & 0.97						&1.326						&1.442					  &	\\
\hline\\ 
   	{\bf Model C} \\
   	(\sersicone\ + \sersictwo\ + \\ Exponential disc + PSF)							& \sersic\,1 mag        & $15.09^{+0.03}_{-0.03}$    & $19.68^{+0.01}_{-0.01}$   & $16.74^{+0.01}_{-0.01}$& \\
         							& \sersic\,1 \reff      & $1.58^{+0.02}_{-0.02}$     & $1.66^{+0.56}_{-0.56}$   & $1.75^{+0.04}_{-0.04}$ & \\
         							& \sersic\,1 n          & $0.66^{+0.01}_{-0.01}$     & $0.30^{+0.01}_{-0.01}$  & $0.53^{+0.08}_{-0.08}$ &  \\
									& \sersic\,1 \axisratio & $0.55^{+0.02}_{-0.02}$    & $0.48^{+0.07}_{-0.07}$   & $0.54^{+0.03}_{-0.03}$ &  \\
        							& \sersic\,1  pa   	     & $96.94^{+0.02}_{-0.02}$   & $92.01^{+0.10}_{-0.10}$ & $94.53^{+0.04}_{-0.04}$& \\
        							
        							& \sersic\,2 mag        & $15.21^{+0.03}_{-0.03}$    & $19.84^{+0.01}_{-0.01}$  & $17.22^{+0.01}_{-0.01}$& \\
         							& \sersic\,2 \reff      & $0.71^{+0.53}_{-0.05}$     & $0.61^{+0.02}_{-0.02}$   & $0.58^{+0.07}_{-0.07}$ & \\
        							& \sersic\ 2 n          & $2.45^{+0.03}_{-0.03}$    & $0.92^{+0.01}_{-0.01}$   & $1.33^{+0.01}_{-0.01}$ &  \\
         							& \sersic\ 2 \axisratio & $0.62^{+0.02}_{-0.02}$    & $0.70^{+0.07}_{-0.07}$   & $0.68^{+0.01}_{-0.01}$ &  \\
									& \sersic\,2 pa         & $80.45^{+0.14}_{-0.14}$   & $78.61^{+0.10}_{-0.10}$  & $82.35^{+0.09}_{-0.09}$& \\    					 
									
									& Expdisc mag                & $13.04^{+0.04}_{-0.04}$   & $16.43^{+0.05}_{-0.05}$  & $14.65^{+0.05}_{-0.05}$& \\
        							& Expdisc scale radius ${r_{s}}$ (kpc) & $6.00^{+0.06}_{-0.06}$    & $8.11^{+0.03}_{-0.03}$   & $6.47^{+0.08}_{-0.08}$ & \\ 
									& Expdisc \axisratio         & $0.54^{+0.02}_{-0.02}$    & $0.70^{+0.07}_{-0.07}$   & $0.68^{+0.01}_{-0.01}$ &  \\
									& Expdisc pa  		          & $94.57^{+0.02}_{-0.02}$   & $88.61^{+0.10}_{-0.10}$  & $82.35^{+0.09}_{-0.09}$& \\  
									& PSF mag                     & $23.17^{+0.02}_{-0.02}$   & $24.72^{+0.03}_{-0.03}$  & $22.22^{+0.02}_{-0.02}$& \\			 					       							& Reduced $\chi^{2}$	              & 0.85					  &1.314					 & 1.392				   & \\
\hline
    \end{tabular}
    \caption{The  best-fit bulge and disc photometric parameters for J2345-0449. 
    The parameters obtained in models A, B and C are listed here. 
    The error in each parameter as well as the  global minimum value of $\chi^{2}$  achieved in each model is also listed.  Foe each model the corresponding input images, best-fit  model images and the residual images  in all three bands are shown     in  Fig.~\ref{set1}, \ref{set2} and \ref{set3} while their 1D radial profiles are shown in 
    Fig.~\ref{set1_profile}, \ref{set2_profile} and \ref{set3_profile}
    \label{all_params}.}
\end{table*}

\begin{table*}
    \centering
    \begin{tabular}{lrrr}
    \hline
    Telescope and &   Wavelength ($\lambda_{\rm eff}$)  &  Flux  & Error \\ 
     channel/detector & (\micron) & (mJy) & (mJy)\\
    \hline
GALEX/FUV   & 0.15386                & 0.078 & 0.001 \\ 
 GALEX/NUV & 0.23157                & 0.127 & 0.002       \\
SDSS/u& 0.3551                 & 0.52  & 0.02          \\ 
SDSS/g& 0.4686                 & 3.53  & 0.20           \\
SDSS/r& 0.6165                 & 5.55  & 0.21           \\ 
SDSS/i& 0.7481                 & 7.31  & 0.17           \\ 
SDSS/z& 0.8931                 & 8.41  & 0.62           \\ 
2MASS/J& 1.2500                   & 3.34  & 0.52          \\
2MASS/H& 1.6500                  & 4.16  & 0.89           \\
2MASS/K& 2.17                   & 4.31  & 0.67           \\ 
WISE/W1& 3.4                    & 5.29  & 0.22           \\ 
WISE/W2& 4.6                    & 3.22  & 0.13           \\ 
WISE/W3& 12                     & 7.87  & 0.51           \\ 
WISE/W4& 24                     & 11.71 & 2.77           \\ 
HERSCHEL/SPIRE& 250                    & 352.7 & 17.8           \\ 
HERSCHEL/SPIRE& 350                    & 196.7 & 19.5           \\ 
HERSCHEL/SPIRE& 500                    & 77.3  & 21.4           \\
\hline
\end{tabular}
\caption{Observed  Multi-wavelength data used in SED fitting.
Radiative flux in different passbands (effective wavelength $\lambda_{\rm eff}$) is shown which is used for the SED fitting by CIGALE and MAGPHYS. 
The data is taken from \href{https://ned.ipac.caltech.edu/}{NED}
     and \href{http://cdsportal.u-strasbg.fr/}{CDS} (\citealt{2011Ap&SS.335..161B,2008AJ....136..735L,2020ApJS..249....3A, 2020yCat.8106....0H}). }
\label{SED}
\end{table*}

\begin{table*}
  \begin{center}
    \caption{SED best fitting  parameters from MAGPHYS. These parameters are
    explained in the text and shown in Fig.~\ref{Fig2}  where their likelihood
    distribution functions are also plotted.}
    \setlength{\extrarowheight}{2pt}
    \begin{tabular}{lccr|} 
    \hline
      \textbf{Best fit parameters}   &  Value              \\
       \hline
 sSFR  (${\rm yr^{-1}}$)        & $\rm (9.55^{+2.45}_{-0.00}) \times 10^{-12}$\\ 		      	
 $M_{\star}$   ( ${\rm M_{\odot}}$) 	 & $\rm (2.88^{+0.00}_{-0.79})\times10^{11}$	          \\
 $L_{\rm dust}$ (  ${\rm L_{\odot}}$)	 & $\rm (7.94^{+0.19}_{-1.91})\times10^{10}$  	           \\
  $M_{\rm dust}$ ( ${\rm M_{\odot}}$) 	 & $\rm (1.21^{+0.44}_{-0.21})\times10^{8}$  	           \\
 SFR   ( ${\rm M_{\odot} yr^{-1}}$)& $\rm 2.74^{+0.03}_{-0.33}$ 	            \\
 
 $\tau_{\rm V}$         	       & $\rm 2.48^{+0.00}_{-0.93}$                     \\
 $\mu\tau_{\rm V}$         	       & $\rm 0.37^{+0.00}_{-0.08}$                 \\ 
 
 $f_{\mu}$         	       		& $\rm 0.71^{+0.05}_{-0.03}$                   \\
 ${T_{\rm W}^{\rm BC}}$  (K)      	& $\rm 50.74^{+7.87}_{-7.83}$ 		           			\\
 ${T_{\rm C}^{\rm ISM}}$ (K)      & $\rm 20.41^{+1.03}_{-2.05}$		                  \\
 
 ${\rm \xi_{C}^{tot}}$     & $\rm 0.58^{+0.06}_{-0.05}$ 		                  \\
 ${\rm \xi_{W}^{tot}}$              & $\rm 0.27^{+0.06}_{-0.06}$   					\\
 \hline
 
    \end{tabular}
  \end{center}
  \label{Tab_SEDfit_magphys}
\end{table*}




\begin{table*}
\begin{center}
\caption{SED best fitting parameters obtained from CIGALE  with varying AGN fraction}
\label{Tab_SEDfit_cigale}
\begin{tabular}{|l|c|c|r|} 
\hline
\textbf{Fitted Parameter} & ${\rm AGN_{Frac}}=0\%$ &${\rm AGN_{Frac}}=10\%$ & ${\rm AGN_{Frac}}=20\%$ \\
 \hline
$M^{*}_{\rm old}$ (${\rm M_{\odot}}$)    & $1.53\times 10^{11}$  & $1.51\times 10^{11}$ & $2.30\times 10^{11}$      \\
$M^{*}_{\rm young}$ ( ${\rm M_{\odot}}$) & $2.28\times 10^{7}$  &  $2.16\times 10^{7}$ & $1.7\times 10^{7}$      \\
$L_{\rm dust}$ (${\rm L_{\odot}}$)	          & $3.36\times 10^{10}$  & $3.40\times 10^{10}$ &  $3.00\times 10^{10}$     \\
SFR  (${\rm M_{\odot} yr^{-1}}$)     & 2.38&  2.19&1.88  \\
$\tau_{V}$         	                  & 1.53      &  1.53     &1.71         \\
$\chi^{2}$                           & 5.10      & 5.63        & 5.86      \\
\hline
\end{tabular}
\end{center}
\end{table*}

\section*{Acknowledgements}

JB acknowledges the support from Department of Physics and Electronics, CHRIST (Deemed to be University),
Bangalore. SR gratefully acknowledges the support from IUCAA  under the student visiting program and funding by Indian Space Research Organisation (ISRO)  under 'AstroSat Data Utilization' project. SR is also supported by Department of Physics and Electronics, CHRIST (Deemed to be University), Bangalore.
 SD acknowledge support from the Indian Space Research Organisation (ISRO) funding under project PAO/REF/CP167.
LCH was supported by the National Science Foundation of China (11721303, 11991052, 12011540375, 12233001), the National Key R\&D Program of China (2022YFF0503401), and the China Manned Space Project (CMS-CSST-2021-A04, CMS-CSST-2021-A06). This research has made use of the data from {\it HST} Archive.   Part of the reported results is based on observations made with the NASA/ESA Hubble Space Telescope, obtained from the Data Archive at the Space Telescope Science Institute, which is operated by the Association of Universities for Research in Astronomy,  Inc., under  NASA  contract  NAS 5-26555.  This research has made use of  NASA's  Astrophysics Data  System, and of the  NASA/IPAC  Extragalactic Database  (NED) which is operated by the Jet  Propulsion Laboratory, California Institute of Technology, under contract with the National Aeronautics and Space Administration.
We have used images and results from SDSS and funding for SDSS has been provided by the Alfred P. Sloan Foundation, the participating institutions, the National Science Founda- tion, and the U.S. Department of Energy’s Office of Science. 
We thank the Ned Wright's Javascript Cosmology Calculator for several cosmological calculations (\citet{2006PASP..118.1711W}). 
\section*{Data Availability}
The HST data used in this paper are publicly available at the Hubble Legacy Archive (https://hla.stsci.edu)
and the Mikulski Archive for Space Telescopes (https://archive.stsci.edu).  All the retrieved  data used in the analysis are tabulated in Table.~\ref{filters} and Table.~\ref{SED} which are publicly available. The GALAFIT code used in photo metric bulge-disc modelling  and  SED fitting routines MAGPHSY and CIGALE  are publicly available at https://users.obs.carnegiescience.edu/peng/work/galfit/galfit.html, http://www.iap.fr/magphys/, and https://cigale.lam.fr. For any further help the authors can be contacted.



\bibliographystyle{mnras}
\bibliography{References} 








\bsp	
\label{lastpage}
\end{document}